\documentclass{jfm}

\usepackage[percent]{overpic}
\usepackage{graphicx}
\usepackage{epstopdf, epsfig}
\usepackage{graphicx}
\usepackage{dcolumn}
\usepackage{bm}
\usepackage{color}
\usepackage{array}
\usepackage{amsmath,amssymb,stmaryrd} 
\usepackage{hyperref}
\usepackage{soul}
\usepackage{multirow}
\definecolor{blue3}{rgb}{0, 0.1770, 0.3410}

\hypersetup{
	colorlinks=true,
	linktoc=all,
	linkcolor=gray2,
	citecolor=gray2,
}

\newcolumntype{M}[1]{>{\centering\arraybackslash}m{#1}}

\newcommand\ie{i.e.\ }

\definecolor{blue2}{rgb}{0, 0.4470, 0.7410}
\definecolor{red2}{rgb}{0.8500, 0.1250, 0.0480} 
\definecolor{orange2}{rgb}{0.8500, 0.3250, 0.0980} 
\definecolor{yellow2}{rgb}{0.9290, 0.6940, 0.1250}
\definecolor{purple2}{rgb}{0.4940, 0.1840, 0.5560}
\definecolor{green2}{rgb}{0.4660, 0.6740, 0.1880}
\definecolor{ltblue2}{rgb}{0.3010, 0.7450, 0.9330}
\definecolor{dkred2}{rgb}{0.6350, 0.0780, 0.1840}
\definecolor{gray2}{rgb}{0.4, 0.4, 0.5}
\definecolor{gray3}{rgb}{0.65, 0.65, 0.65}

\def\drawline#1#2{\raise 2.0pt\vbox{\hrule width #1pt height #2pt}}

\def\solid{\drawline{8}{1}\nobreak\ }

\shorttitle{Resolvent-analysis-based design of separation control}
\shortauthor{C.-A. Yeh and K. Taira}

\title{Resolvent-analysis-based design of airfoil separation control}

\author{Chi-An Yeh\corresp{\email{cy13d@my.fsu.edu}} \and Kunihiko Taira }

\affiliation{Department of Mechanical Engineering, Florida State University,
Tallahassee, FL 32310, USA}

\begin{document}

\maketitle

\begin{abstract}

We combine three-dimensional (3D) large-eddy simulations (LES) and resolvent analysis to design active separation control techniques on a NACA 0012 airfoil.  Spanwise-periodic flows over the airfoil at a chord-based Reynolds number of $23,000$ and a free-stream Mach number of $0.3$ are considered at two post-stall angles of attack of $6^\circ$ and $9^\circ$.  Near the leading edge, localized unsteady thermal actuation is introduced in an open-loop manner with two tunable parameters of actuation frequency and spanwise wavelength.  For the most successful control case that achieves full reattachment, we observe a reduction in drag by up to $49\%$ and increase in lift by up to $54\%$.  To provide physics-based guidance for the effective choice of these control input parameters, we conduct global resolvent analysis on the baseline turbulent mean flows to identify the actuation frequency and wavenumber that provide high energy amplification.  The present analysis also considers the use of a temporal filter to limit the time horizon for assessing the energy amplification to extend resolvent analysis to unstable base flows.  We incorporate the amplification and response mode from resolvent analysis to provide a metric that quantifies momentum mixing associated with the modal structure.  By comparing this metric from resolvent analysis and the LES results of controlled flows, we demonstrate that resolvent analysis can predict the effective range of actuation frequency as well as the global response to the actuation input.  Supported by the agreements between the results from resolvent analysis and LES, we believe that this study provides insights for the use of resolvent analysis in guiding future active flow control.

\end{abstract}


\section{Introduction}
In aerodynamic applications, flow separation can cause detrimental effects such as stall.  Flow separation can also intensify the pressure fluctuation and cause structural fatigue.  For these reasons, suppression of flow separation over aerodynamic bodies has been an area of focus for the flow control community \citep{Joslin&Miller:2009}.  Active flow control, which requires steady or unsteady input of external energy, is capable of adapting to a wide range of operating conditions. It has the advantage over passive control strategies whose performance can degrade in off-design conditions.  For separation control, in particular, unsteady forcing has demonstrated its enhanced capability of reattaching the flow and enhancing aerodynamic performances \citep{Zaman:JFM1989,Wu:JFM1998}.  Consequently, attempts have been made to investigate the control effect of different unsteady forcing frequencies \citep{SeifertPack:AIAAJ1999,Glezer:AIAAJ2005}.  A range of flow responses to forcing frequency were reported by conducting parametric studies of separation control \citep{Amitay:AIAAJ2002}.  However, the characterization of global frequency response of the separated flow lacks quantitative support from theoretical analyses.  Moreover, detailed knowledge of effective frequency range for unsteady separation control remains limited. 

\citet{Greenblatt:PAS2000} provided an overview on the use of periodic excitation for separation control. They suggested that the fundamental mechanism for suppression of separation lies in the excitation of the Kelvin--Helmholtz instabilities in the shear layer forming from the separated flow.  The seminal work of \citet{BrownRoshko:JFM1974} pointed out that the formation of spanwise coherent structures due to these instabilities are the main driving force for the momentum mixing and entrainment.  Clearly, leveraging the shear-layer instabilities has been an important strategy to suppress flow separation \citep{Joslin&Miller:2009}.  As such, the knowledge on the instability and receptivity of the separated flow is crucial to guide the design of active separation control.  

For the study of hydrodynamic instability, a variety of approaches have been summarized by \citet{SchmidHenningson:2001} and \citet{Theofilis:ARFM2011}.  One traditional approach for analyzing instability seeks a modal representation for infinitesimal perturbations about an equilibrium base state, \ie a solution to the Navier--Stokes equations.  Such an approach forms an eigenvalue problem for the global instability modes and emphasizes on the spectrum of the linearized Navier--Stokes operator \citep{Barkley:JFM1996,Sipp:JFM2007,Liu:JFM2016,Sun:JFM2017,Taira:AIAAJ2017}.  Inherently, it characterizes the asymptotic long-time behavior of the perturbations in the flow.  Complementing this traditional approach, the nonmodal approach addresses flow instability by seeking an energy measure for the time-evolving response of the flow \citep{Schmid:ARFM2007}.  The nonmodal approach either forms an initial-value problem that examines the transient energy growth over a finite-time window \citep{Schmid:JFM2004}, or investigates the energy amplification from a forcing to the harmonic response \citep{Trefethen:1993Science,FarrellIoannou:PoF1993,Jovanovic:JFM2005}.  The latter path is closely related to receptivity analysis \citep{Goldstein:ARFM1989,Choudhari:TCFD1993}, and has built the foundation for the resolvent analysis extended for turbulent flows.

With the recent developments, resolvent analysis has become a valuable approach to investigate the frequency response of a fluid-flow system.  Resolvent analysis concerns the pseudospectrum of a linear operator \citep{TrefethenEmbree:2005}.  It provides particularly valuable insights when the linear operator is nonnormal, which is encountered in shear-dominated flows \citep{SchmidHenningson:2001}.  \citet{Trefethen:1993Science} conducted such an analysis on laminar Poiseuille flows.  They showed that the perturbation energy can exhibit significant transient growth due to the nonnormality of the operator.  This growth can depart from the linear regime and cause subcritical laminar-turbulent transition.  For a nonnormal operator, a linear mechanism of pseudoresonance can also result in a large resonant behavior to forcing even when the forcing frequency is far from the spectrum (eigenvalues) of the operator.  \citet{MckeonSharma:JFM2010} extended the resolvent analysis for turbulent flows.  The challenge in formulating the analysis for turbulent mean flow stems from the nonlinear terms of finite-amplitude perturbations.  In their framework, these nonlinear terms are treated as an internal forcing, yielding a linear relationship between the retained nonlinearity and the harmonic flow response.  The linear relationship describes an input-output process that takes place through the resolvent operator constructed about the statistically stationery turbulent mean flow.  By examining the characteristics of the resolvent operator, they captured the coherent structures in wall-bounded turbulence, revealing scalings for length and velocity that are in agreement with experimental measurements.  Following this resolvent formulation, similar approaches have been undertaken in numerous studies \citep{Moarref:JFM2013,Beneddine:JFM2016,Gomez:JFMR2016}.

Resolvent analysis, as an input-output analysis, gives knowledge of energy amplification as well as the associated structural response to the perturbation over a range of frequencies.  Such knowledge is crucial in designing active flow control, because both amplification and response structure provide insights on identifying the unsteady forcing that takes minimal energy to change the mean flow.  Applying this analysis  to turbulent flows, our study aims to provide theoretical support to examine the flow responses under unsteady forcing and to develop a predictive tool for identifying the range of effective actuation frequencies.  In this study, we conduct an active flow control effort combining LES and resolvent analysis on flows over a canonical airfoil.  Over the airfoil, the control input is introduced in the form of local periodic heat injection near the leading edge. We parameterize the actuation frequency and spanwise wavenumber in this numerical effort.  Our choice of the thermal actuator is motivated by the energy-based actuators that have become widespread in active flow control, such as nanosecond pulse driven dielectric barrier discharge plasma actuators \citep{Little&Samimy:AIAAJ2012} and thermoacoustic actuators \citep{Yeh:AIAA2015}.  These energy-based actuators have a sheet-like arrangement with no moving parts, which facilitates surface-compliant installation without occupying any internal space or adding significant weight.  The thermal actuator setup used in the present study models the thermoacoustic and plasma-based actuators at a fundamental level \citep{Bin:JAP2015,ChaeKim:AIAA2017}.


\begin{figure}
	\includegraphics[width=1.0\textwidth]{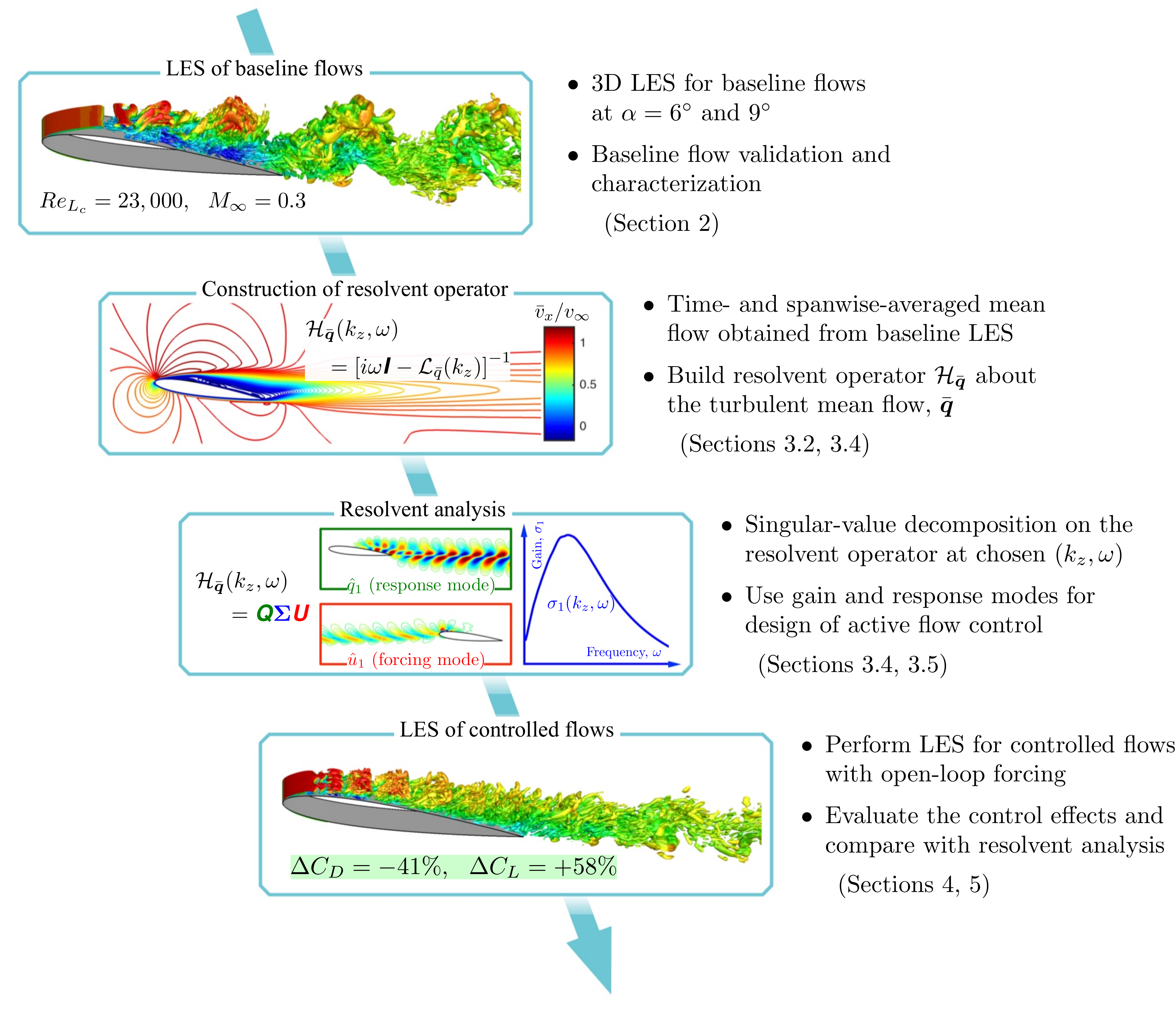}
	\vspace{-0.3in}
	\caption{\label{fig:intro} Roadmap of the present study.}
	\vspace{-0.1in}
\end{figure}

A roadmap of this study is provided in figure \ref{fig:intro}.  Starting in section \ref{sec:Comp_setup},  we perform the baseline flow simulations at two post-stall angles of attack.  The baseline flows are validated and characterized.  With the turbulent mean flow obtained from the baseline LES, the global resolvent operator is constructed about the time- and spanwise-averaged mean flow at a specified wavenumber-frequency combination in section \ref{sec:ResolventAnalysis}.  Resolvent analysis performs a singular value decomposition (SVD) of the discrete resolvent operator to determine the forcing modes, response modes and the associated amplification (gain).  The amplification as well as the modal structures are characterized over the Fourier space, as to obtain physical insights to the potentially effective range of actuation frequencies and wavenumbers for active flow control to suppress flow separation.  In section \ref{sec:Control}, we present the LES results of over 250 controlled cases using open-loop actuation with different actuation frequencies and wavenumbers.  The control effects are quantified and compared to the prediction of resolvent analysis on the mean baseline flows.  We comment on the agreements and limitations on the usage of resolvent analysis for design of active flow control are commented in section \ref{sec:Resovent_vs_CtrlLES}.



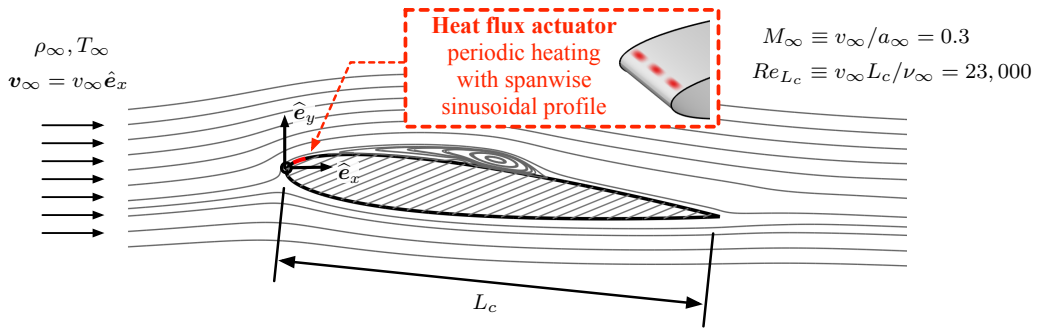
\begin{figure}
\begin{center}
	\vspace{0.3in}
	\begin{overpic}[width=0.85\textwidth]{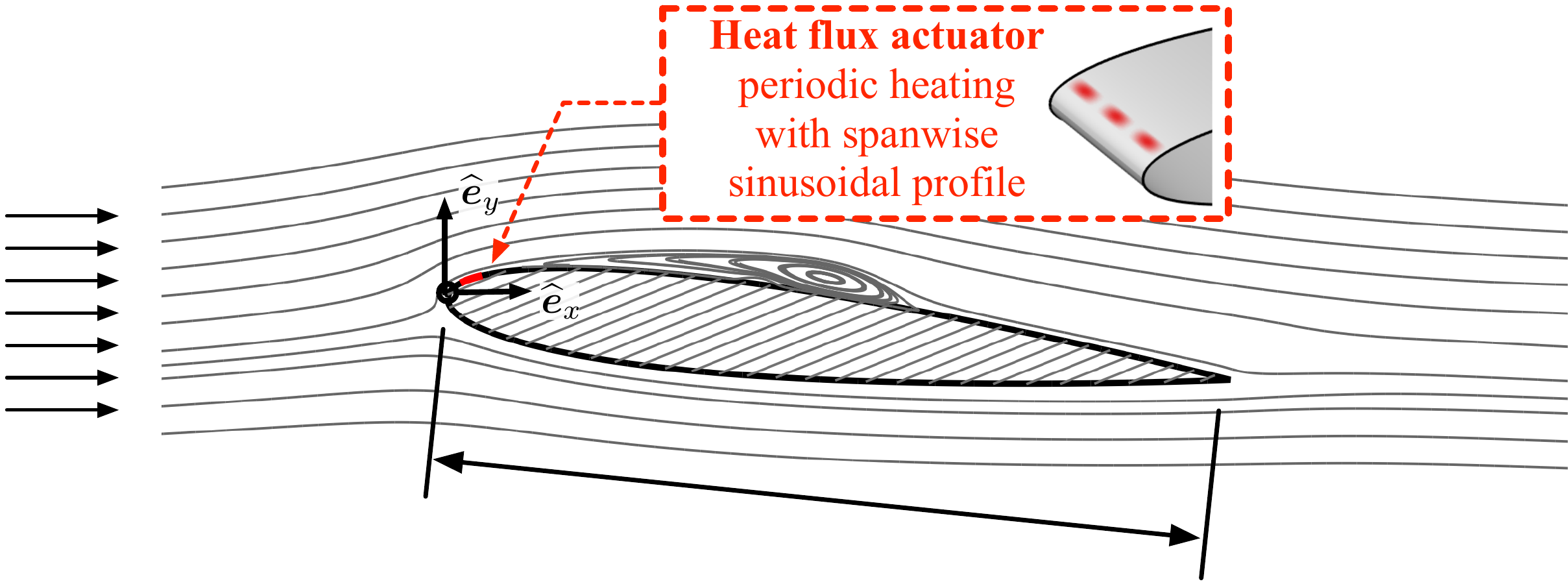}
			\put(-0.5, 32){\indexsize$\rho_\infty,T_\infty$}
			\put(-3.5, 28){\indexsize$\boldsymbol{v}_\infty = v_\infty \hat{\boldsymbol{e}}_x$}
			\put(83.4, 33){\indexsize$M_\infty \equiv v_\infty/a_\infty = 0.3$}
			\put(82.2, 29){\indexsize$Re_{L_c} \equiv v_\infty L_c/\nu_\infty = 23,000$}
			\put(50, 2.5){\indexsize$L_c$}
	\end{overpic}
	\vspace{+0.12in}
\caption{\label{fig:setup} The problem description: separated flow over a NACA 0012 airfoil (shown for $\alpha = 6^\circ$) at free stream Mach number $M_\infty = 0.3$ and chord-based Reynolds number $Re_{L_c} = 23,000$. }
\end{center}
\end{figure}

\section{Problem setup}
\label{sec:Comp_setup}

\subsection{Problem description}
We consider separated flows over a NACA 0012 airfoil at two angles of attack of $\alpha = 6^\circ$ and $9^\circ$ for a moderate chord-based Reynolds number $Re_{L_c} \equiv v_\infty L_c/\nu_\infty = 23,000$ and a free stream Mach number $M_\infty \equiv v_\infty/a_\infty = 0.3$, as shown in figure \ref{fig:setup}.  Here, $v_\infty$ is the free-stream velocity, $L_c$ is the chord length, $a_\infty$ is the free-stream sonic speed, and $\nu_\infty$ is the kinematic viscosity.  To perform active flow control, a thermal actuator is placed across the span near the leading edge.  This actuator introduces oscillatory heat flux at a prescribed frequency and spanwise profile as an open-loop actuation input.  The details of this thermal actuator will be discussed in section \ref{sec:ActuatorModel}.

\subsection{Simulation setup}

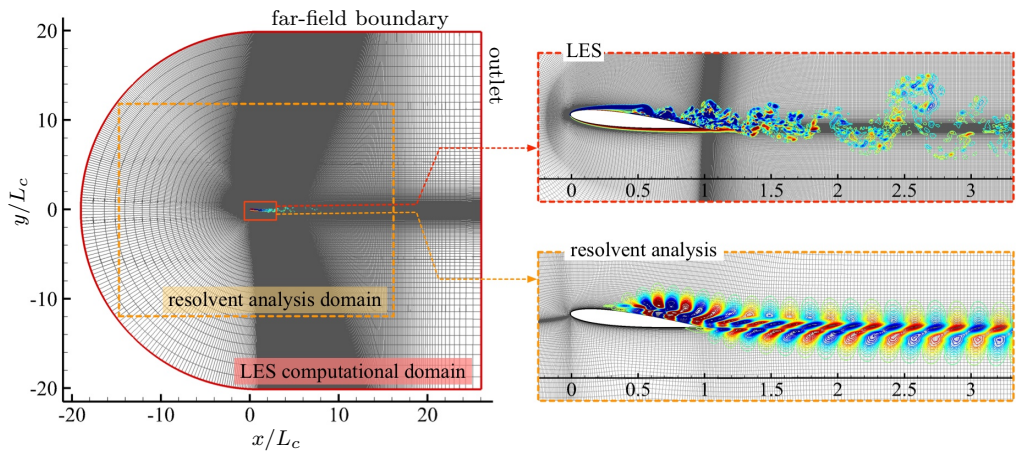
\begin{figure}
	\vspace{0.1in}
	\begin{center}
	\begin{overpic}[scale=0.206]{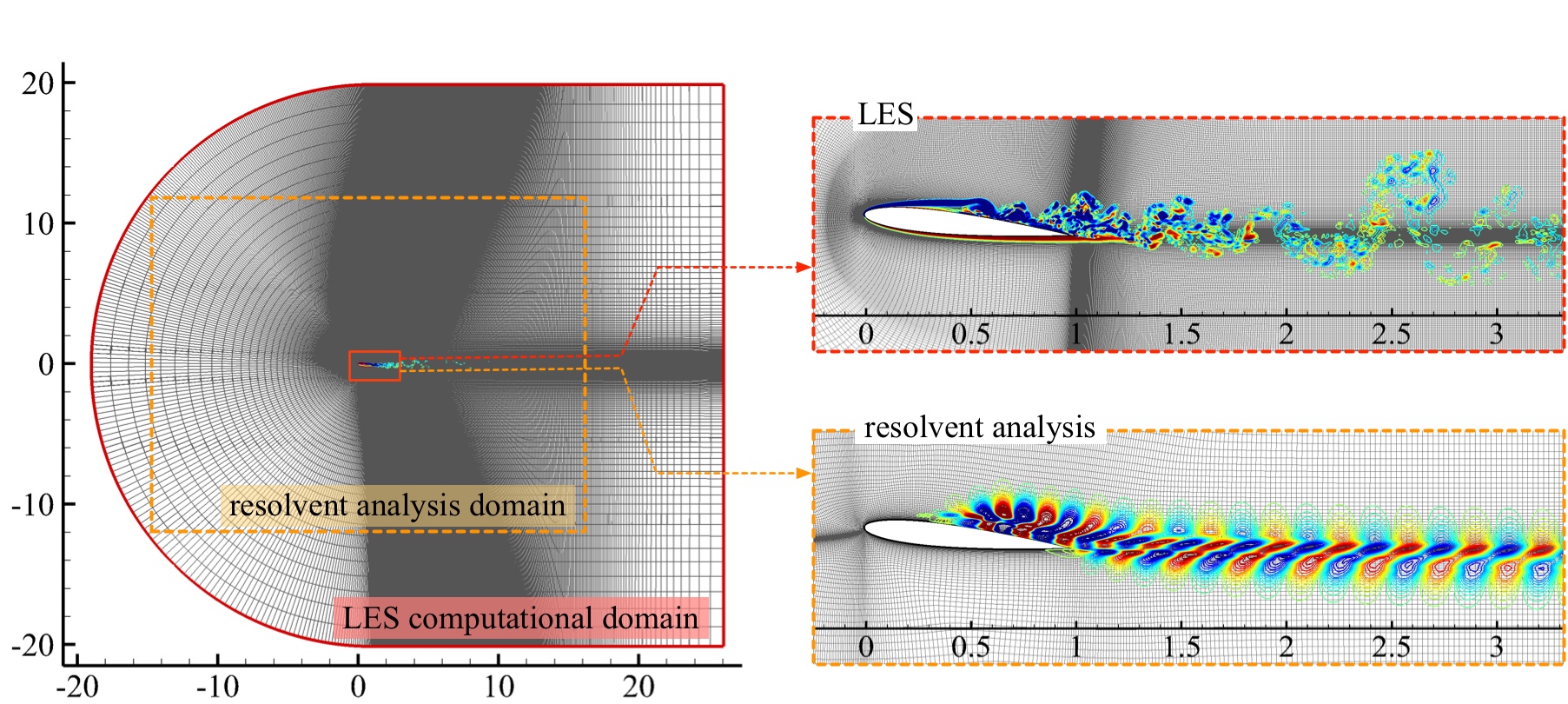}
			\put(23, -2){\indexsize$x/L_c$}
			\put(-1.5, 20){\indexsize\rotatebox{90}{$y/L_c$}}
			\put(25, 40.2){\indexsize far-field boundary}
			\put(47, 38){\rotatebox{270}{\indexsize outlet}}
	\end{overpic}
	\end{center}
	\vspace{0.1in}
	\caption{\label{fig:Mesh} The computational domains ($x$-$y$ plane, shown for $\alpha = 6^\circ$) for LES and resolvent analysis. The near-field mesh (right) is shown along with instantaneous spanwise vorticity from LES and streamwise velocity mode from resolvent analysis. For both meshes, uniform $\Delta x$ is adopted in $x/L_c \in [1.5, 6]$ to resolve the wake structures.}
\end{figure}

We perform LES to simulate spanwise-periodic flows over the airfoil using a finite-volume compressible flow solver {\it{CharLES}} \citep{Khalighi:AIAA11,Bres:AIAAJ2017}, which is second-order accurate in space and third-order accurate in time.  Vremen's sub-grid scale model \citep{Vreman:PoF2004} is utilized in the LES. Figure \ref{fig:Mesh} illustrates the C-shaped computational mesh, with the airfoil positioned with its leading edge at $x/L_c = y/L_c = 0$.  The extent of the computational domain is $x/L_c \in [-19, 26]$, $y/L_c \in [-20, 20]$ and $z/L_c \in [-0.1, 0.1]$ in the streamwise, transverse and spanwise direction, respectively.  This domain is discretized with approximately 35 million grid cells.  We have examined the grid convergence by comparing the flow field and aerodynamics forces from this mesh to other two meshes that are further refined in the near-field with the total of 63 and 82 million grid cells.  From each mesh, the force data is collected for the developed flow over 80 convective time units, and the time-averaged drag and lift were observed to be insensitive to the grid resolution of the three meshes.

For fluid properties, we use the specific heat ratio $\gamma = 1.4$ and the Prandtl number $Pr = 0.7$, which are representative for standard air. The temperature-varying dynamic viscosity, $\mu(T)$, is evaluated with the power law as $\mu = \mu_\infty ( T/T_\infty )^{0.76}$, where $\mu_\infty$ and $T_\infty$ are the free-stream dynamic viscosity and temperature, respectively \citep{Garnier:2009LES}.  The power law models the dynamic viscosity variation for standard air in the range of $T/T_\infty \in [0.5, 1.7]$.  This range is suitable for the current study with local thermal inputs, where we observe that the maximum temperature fluctuation is within $42\%$ of $T_\infty$ for all controlled flows.

The simulations are performed with Dirichlet boundary condition specified at the far-field boundary as $[\rho, v_x, v_y, v_z, T] = [\rho_\infty, v_\infty, 0, 0, T_\infty]$, where $\rho$ is the density, $v_x$, $v_y$ and $v_z$ are respectively the streamwise, transverse and spanwise velocity, and $T$ is the temperature.  Over the airfoil, the no-slip adiabatic boundary condition is prescribed, except for where the actuator is placed for controlled cases.  Along the outlet boundary, a sponge layer \citep{Freund:AIAAJ97} is applied over $x/L_c \in [15,25]$ with the target state set to the running-averaged flow over 10 acoustic time units.  The time integration is performed at a constant time step of $\Delta t v_\infty/L_c = 4.14 \times 10^{-5}$, corresponding to a maximum Courant-Friedrichs-Lewy (CFL) number of $0.84$.  Further details regarding the meshing strategy and computational setup are reported in \citet{Yeh:AIAA2017}.



\subsection{Actuator model}
\label{sec:ActuatorModel}

The thermal actuator is implemented as an oscillatory energy-flux boundary condition to model the fundamental effects of thermoacoustic and plasma-based actuators in the LES.  It is prescribed in the energy equation as an unsteady Neumann boundary condition, along with no-slip boundary condition for the momentum equation in the compressible Navier--Stokes equations.  The actuator model is expressed as
\begin{equation}
	\label{eq:ActuatorModel}
	\phi^+(\omega^+, k_z^+) = \frac{1}{4} \hat{\phi} \sin( \omega^+ t) \left[1 + \cos (k_z^+ z)\right] \left\lbrace1 + \cos\left[\frac{2\pi}{\sigma_a}\left( x - x_a \right)\right]\right\rbrace, 
\end{equation}
where $(x - x_a)/\sigma_a \in [-0.5,0.5]$. This expression provides the boundary heat flux input with a compact spatial support in the form of a Hanning window centered at $x_a/L_c = 0.03$ on the suction surface with width of $\sigma_a/L_c = 0.04$, as illustrated in figure \ref{fig:setup}.  The actuator introduces open-loop control input at the prescribed actuation frequency, $\omega^+$, and spanwise wavenumber, $k_z^+$.  They are parameterized in the LES of controlled cases and will be reported in terms of the actuation Strouhal number $St^+ = \omega^+L_c/(2\pi v_\infty)$ and the normalized wavenumber $k_z^+L_c$ throughout this paper.  Due to the choice of the spanwise extent for the computational domain ($z/L_c \in [-0.1, 0.1]$), the actuation wavenumbers $k_z^+L_c$ are restricted to  integer multiples of $10\pi$.  Hence, we only consider the use of $k_z^+L_c = 0$, $10\pi$, $20\pi$ and $40\pi$ in the LES of controlled flows.  In the actuator model \ref{eq:ActuatorModel}, the actuation amplitude $\hat{\phi}$ is selected such that the normalized total actuation power,
\begin{equation}
\label{eq:NormalizedForcingPower}
	E^+ = \frac{\frac{1}{4} \hat{\phi} \sigma_a}{\frac{1}{2} \rho_\infty v_\infty^3 (L_c \sin \alpha)} = 0.0902
\end{equation}
for all controlled cases throughout this work.  This magnitude is representative of those used in thermally actuated flow control studies \citep{Corke:ARFM10,SinhaSamimy:PoF2012,AkinsLittle:AIAA2015,Yeh:AIAA2015}.  For this thermal actuator, \citet{Yeh:JFM2017} have investigated its control mechanism and flow control capability in free shear layers.  The thermal input from the actuator translates to vortical perturbations in the forms of oscillatory surface vorticity flux and baroclinic torque. The thermal actuation is capable of exciting fundamental and subharmonic instabilities and its capability of modifying shear-layer dynamics is ideal for this study of separation control.

\subsection{Baseline simulations}
\label{sec:Baseline}

\begin{figure}
	\begin{center}
		\vspace{0.2in}
		\begin{overpic}[scale=0.55]{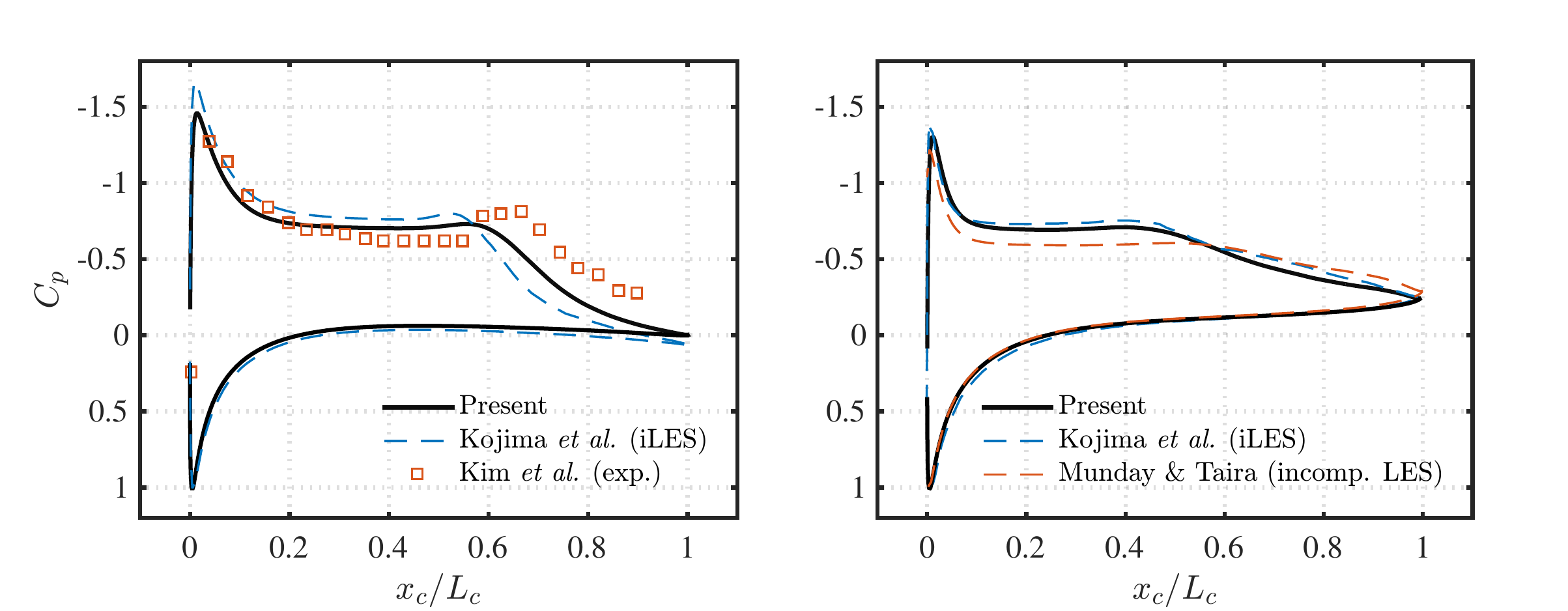}
			\put(4, 36){\indexsize (a)}
			\put(51, 36){\indexsize (b)}
			\put(38, 31){\indexsize $\alpha = 6^\circ$}
			\put(85, 31){\indexsize $\alpha = 9^\circ$}
		\end{overpic}
	\caption{\label{fig:CP_Validation} Surface pressure profiles for $\alpha = 6^\circ$ (a) and $9^\circ$ (b) over the chord-wise coordinate $x_c/L_c = (x\cos\alpha + y\sin\alpha)/L_c$, in comparison with those reported by \citet{Kim:AIAA2009}, \citet{Kojima:JoA2013} and \citet{Munday:AIAAJ2018}.}
	
		\vspace{0.4in}
		\begin{overpic}[scale=0.55]{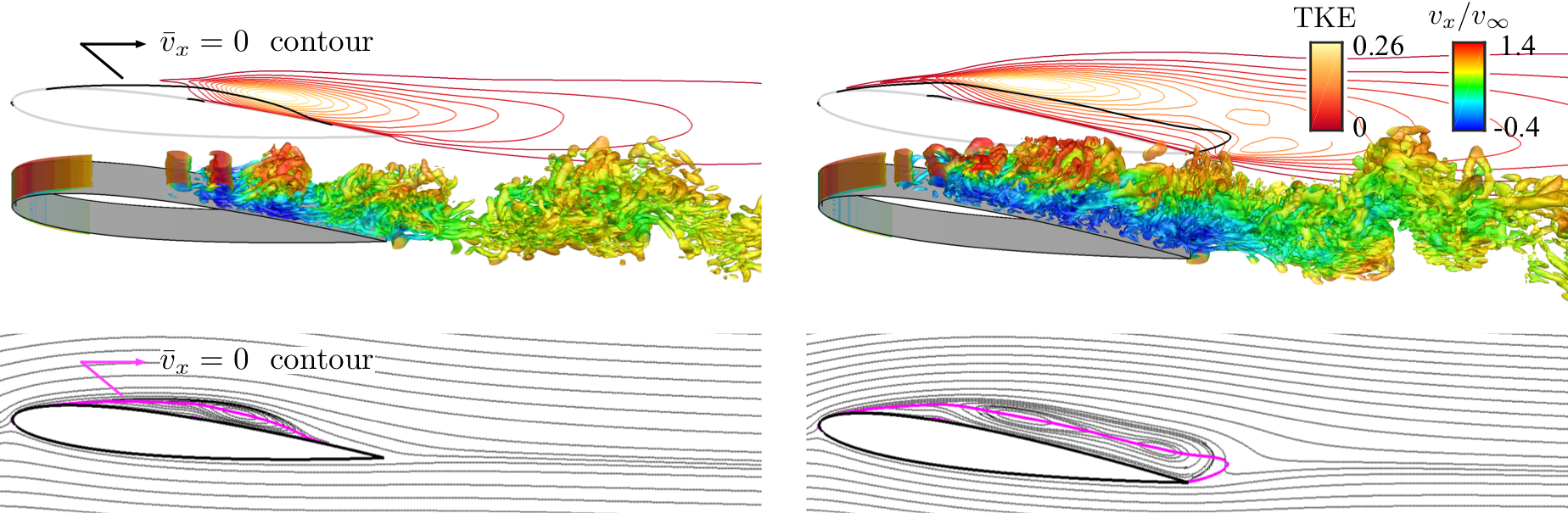}
			\put(0, 32){\indexsize (a)}
			\put(51.2, 32){\indexsize (b)}
			\put(0, 13){\indexsize (c)}
			\put(51.2, 13){\indexsize (d)}
			\put(5, 33){$\alpha = 6^\circ$, $C_L/C_{L,0} = 0.84$}
			\put(56.2, 33){$\alpha = 9^\circ$, $C_L/C_{L,0} = 0.52$}		
		\end{overpic}
	\end{center}
	\caption{\label{fig:Baseline_FlowVis} (a-b) Baseline flow visualization using $Q$-Criterion (iso-surface of $QL_c^2/u_\infty^2 = 50$ colored by streamwise velocity) and spanwise-average turbulent kinetic energy $\text{TKE} = \overline{(v_x'^2 + v_y'^2 + v_z'^2)}/v_\infty^2$ in the background.  (c-d) The time-averaged streamlines. The contour line for $\bar{v}_x = 0$ is shown and will be used for characterizing the extent of the separation region throughout this study.}
\end{figure}

We validate the baseline simulations at angles of attack of $\alpha = 6^\circ$ and $9^\circ$ by comparing the surface pressure distribution and aerodynamic forces to those reported in literatures for $Re_{L_c} = 23,000$.  Throughout this study, the pressure coefficient, $C_p$, lift coefficient, $C_L$, and drag coefficients $C_D$ are defined as
\begin{equation}
		C_p = \frac{p - p_\infty}{\frac{1}{2}\rho_\infty v_\infty^2}
	,~~~C_L = \frac{F_L}{\frac{1}{2}\rho_\infty v_\infty^2 A}
	,~~~C_D = \frac{F_D}{\frac{1}{2}\rho_\infty v_\infty^2 A},
\end{equation}
where $F_L$ and $F_D$ are the total lift and drag forces on the airfoil, respectively, and $A$ is the planform area of the airfoil.  The time-averaged aerodynamic forces and surface pressure profile are respectively presented in table \ref{tab:CD_CL_Validation} and figure \ref{fig:CP_Validation}.  We found reasonable agreements with those reported by \citet{Kim:AIAA2009}, \citet{Kojima:JoA2013} and \citet{Munday:AIAAJ2018}.  We note that the numerical study of \citet{Kojima:JoA2013} was conducted using implicit LES and \citet{Munday:AIAAJ2018} reported the results from incompressible LES.  The discrepancy in the surface pressure with the experimental measurement by \citet{Kim:AIAA2009} can be attributed to the different transverse blockage ratios ($0.26\%$ for the present study). 

\begin{table}
\begin{center}
\begin{tabular}{lll llllllll}
        			&~~&& \multicolumn{3}{c}{$\alpha = 6^\circ$} &~~&& \multicolumn{3}{c}{$\alpha = 9^\circ$} \vspace{0.03in} \\ 
					\cline{4-6} \cline{9-11} \vspace{-0.08in} \\ 
        			&&& \multicolumn{1}{c}{$\bar{C}_D$}	&& \multicolumn{1}{c}{$\bar{C}_L$}	&&& \multicolumn{1}{c}{$\bar{C}_D$}	&& \multicolumn{1}{c}{$\bar{C}_L$} \vspace{-0.06in} \\ 
					\hline \vspace{-0.15in} \\
Present 			&&& $0.066$	&~& $0.609$	&&& $0.113$	&~& $0.570$	\\
\citet{Munday:AIAAJ2018}     &&& $0.062$	&~& $0.637$	&&& $0.117$	&~& $0.565$	\\
\citet{Kojima:JoA2013}        &&& $0.054$	&~& $0.639$	&&& $0.118$	&~& $0.594$          
\end{tabular}
\caption{\label{tab:CD_CL_Validation} The time-averaged drag and lift coefficients on a NACA 0012 airfoil at $\alpha = 6^\circ$ and $9^\circ$ at $Re_{L_c} = 23,000$.  Present study performs compressible LES at a free-stream Mach number $M_\infty = 0.3$, in comparison with the results from the incompressible LES by \citet{Munday:AIAAJ2018} and the implicit LES by \citet{Kojima:JoA2013} at $M_\infty = 0.2$.}
\end{center}
\end{table}

Figure \ref{fig:Baseline_FlowVis} shows the instantaneous flow fields and time-average streamlines for the baseline flows at $\alpha = 6^\circ$ and $9^\circ$. The iso-surface of $Q$-criterion \citep{Hunt:CTR1988} is used to visualize the vortical structures.  The contour line of time- and spanwise-averaged streamwise velocity $\bar{v}_x = 0$ is also shown to identify the flow separation and reattachment.  This contour line is also shown on top of the time-average streamlines, where we see the contour line extends through the separation bubble for each case.  For both angles of attack, laminar separation is observed near the leading edge and forms a shear layer. The shear layer rolls up over the suction surface and evolves into spanwise vortices. This roll-up process leads to the increasing turbulent kinetic energy (TKE) within the shear layer.  Farther downstream, these spanwise vortices break up and lose their spanwise coherence, resulting in the laminar-turbulent transition.  Within this roll-up and transition process, one common feature in the pressure profiles in figure \ref{fig:CP_Validation} is the `plateau' observed for both angles of attack.  Such a plateau in the pressure profile is also observed by \citet{Marxen:JFM2013} and \citet{BentonVisbal:PRF2018} in the transition process that takes place over a laminar separation bubble.  The transition process is accompanied by the maximum TKE over the airfoil at $x/L_c \approx 0.6$ for $\alpha = 6^\circ$ and $x/L_c \approx 0.5$ for $\alpha = 9^\circ$.  The roll-up and break-up processes both result in momentum mixing and entrains the free stream, leading to the flow reattachment for $\alpha = 6^\circ$ at $x/L_c \approx 0.85$.  Over the $\alpha = 9^\circ$ airfoil, the flow is in full stall.  To quantitatively characterize the stall condition,  we calculate the potential-flow lift $C_{L, 0}$ using panel method \citep{Hess:ARFM1990} to mark a theoretical upper bound of the lift for both angle of attacks.  The flow over the $\alpha = 6^\circ$ reattaches and achieves $84\%$ of $C_{L,0}$. Whereas for the $\alpha = 9^\circ$ airfoil, while experiencing deep stall, provides only $52\%$ of the potential flow lift.  This difference in the stall condition will be reflected in the control flows to be discussed in section \ref{sec:Control}.

The excitation of shear-layer instabilities serves as the key to separation control \citep{Greenblatt:PAS2000}.  For the laminar separation bubble that is observed in both baseline flows, \citet{Haggmark:PTRSL2000} have experimentally shown that the Kelvin--Helmholtz instability dominates the laminar-turbulent transition.  In order to leverage the Kelvin--Helmholtz instability for flow control, we place the thermal actuator slightly upstream of the separation point such that the perturbations can be introduced at the onset of the shear layer.

\section{Resolvent analysis of mean baseline flows}
\label{sec:ResolventAnalysis}
Following the baseline LES, we perform resolvent analysis on these turbulent mean flows to provide physical insights into the design of active separation control.

\subsection{Formulation}
Let us consider the compressible Navier--Stokes equations expressed as
\begin{equation}
\label{eq:NSE_OperatorForm}
	\frac{\partial \boldsymbol{q}}{\partial t} = \mathcal{N}(\boldsymbol{q}) + \boldsymbol{f}^+,
\end{equation}
where $\mathcal{N}$ is the nonlinear Navier--Stokes operator that acts on the flow state variable $\boldsymbol{q} = [\rho, v_x, v_y, v_z, T]^T$, and $\boldsymbol{f}^+$ represents the external actuation input from active flow control.  Note that the external forcing $\boldsymbol{f}^+$ can be absent.  We perform the Reynolds decomposition of $\boldsymbol{q} = \bar{\boldsymbol{q}} + \check{\boldsymbol{q}}$ so that the flow state variable $\boldsymbol{q}$ is decomposed into a statistically stationery long-time mean component $\bar{\boldsymbol{q}}$ and a fluctuating component $\check{\boldsymbol{q}}$.  Substituting $\boldsymbol{q}$ with its Reynolds decomposition into the Navier--Stokes equations \ref{eq:NSE_OperatorForm} yields
\begin{equation}
\label{eq:DeriveLinearNSE}
	\frac{\partial \bar{\boldsymbol{q}}}{\partial t} + \frac{\partial \check{\boldsymbol{q}}}{\partial t} = \underbrace{\mathcal{L}(\check{\boldsymbol{q}}, \bar{\boldsymbol{q}} \bnabla \check{\boldsymbol{q}}, \check{\boldsymbol{q}} \bnabla \bar{\boldsymbol{q}}, \nabla^2 \check{\boldsymbol{q}}, \cdots)}_{\mathcal{L}_{\bar{\boldsymbol{q}}}(\check{\boldsymbol{q}})} + \underbrace{\mathcal{N}(\bar{\boldsymbol{q}}) + \bar{f}(\check{\boldsymbol{q}}^n) + \boldsymbol{f}^+}_{\check{\boldsymbol{u}}}.
\end{equation}
With the Reynolds decomposition, the linear operations for $\check{\boldsymbol{q}}$ are extracted from the nonlinear operation of $\mathcal{N}(\bar{\boldsymbol{q}} + \check{\boldsymbol{q}})$.  We collect these terms that are linear with respect to $\check{\boldsymbol{q}}$ and denote them as $\mathcal{L}_{\bar{\boldsymbol{q}}}(\check{\boldsymbol{q}})$.  The term $\mathcal{N}(\bar{\boldsymbol{q}})$ accounts for the Navier--Stokes operation taking place only on $\bar{\boldsymbol{q}}$, and $\bar{f}(\check{\boldsymbol{q}}^n)$ collects the nonlinear higher-order terms for $\check{\boldsymbol{q}}$ in $\mathcal{O}(\boldsymbol{q}^n)$, where $n > 1$.  In particular, we note that $\mathcal{N}(\bar{\boldsymbol{q}}) + \bar{f}(\check{\boldsymbol{q}}^n)$ can be interpreted as the internal forcing in the turbulent flow due to the nonlinear interaction \citep{FarrellIoannou:PRL1994,MckeonSharma:JFM2010}.  This internal forcing together with the external forcing $\boldsymbol{f}^+$ is further denoted as $\check{\boldsymbol{u}}$.  Noting that $\partial_t \bar{\boldsymbol{q}} = 0$, equation \ref{eq:DeriveLinearNSE} can be simplified as
\begin{equation}
\label{eq:LinearNSE}
	\frac{\partial \check{\boldsymbol{q}}}{\partial t} = \mathcal{L}_{\bar{\boldsymbol{q}}}(\check{\boldsymbol{q}}) + \check{\boldsymbol{u}}.
\end{equation}
Thus far, no assumptions have been made in the formulation except for the statistical stationarity of the mean flow about which the Navier--Stokes equations are rewritten in the above form.

Now, we cast the linearized Navier--Stokes equations \ref{eq:LinearNSE} for the spanwise-periodic flow over the airfoil.  Considering the two-dimensional airfoil geometry in this study, the time- and spanwise-average flow obtained from the baseline flow simulation is used as the mean component so that the Reynolds decomposition can be written as
\begin{equation}
\label{eq:ReyDecomp}
	\boldsymbol{q}(x, y, z, t) = \bar{\boldsymbol{q}}(x, y) + \check{\boldsymbol{q}}(x, y, z, t).
\end{equation}
The spanwise-periodic setup in the present study allows for the biglobal-mode representation for $\check{\boldsymbol{q}}$ and $\check{\boldsymbol{u}}$ as the sum of temporal and spanwise Fourier modes \citep{Theofilis:PAS2003} respectively as
\begin{equation}
\label{eq:ansatzQ}
	\check{\boldsymbol{q}}(x, y, z, t) = \int_{-\infty}^{\infty}\int_{-\infty}^{\infty} \hat{\boldsymbol{q}}_{k_z,\omega}(x, y) e^{i(k_z z - \omega t)} {\rm d}\omega{\rm d}k_z
\end{equation}
and
\begin{equation}
\label{eq:ansatzU}
	\check{\boldsymbol{u}}(x, y, z, t) = \int_{-\infty}^{\infty}\int_{-\infty}^{\infty} \hat{\boldsymbol{u}}_{k_z,\omega}(x, y) e^{i(k_z z - \omega t)} {\rm d}\omega{\rm d}k_z.
\end{equation}
Here, $i = \sqrt{-1}$, $\omega$ is the complex radian frequency, $k_z$ is the real spanwise wavenumber, and $\hat{\boldsymbol{q}}_{k_z, \omega}$ and $\hat{\boldsymbol{u}}_{k_z, \omega}$ are the biglobal modes for spanwise wavenumber $k_z$ and temporal frequency $\omega$.  Substituting the modal expressions \ref{eq:ansatzQ} and \ref{eq:ansatzU} for $\check{\boldsymbol{q}}$ and $\check{\boldsymbol{u}}$ into equation \ref{eq:LinearNSE}, we arrive at the linearized Navier--Stokes equations in Fourier space as
\begin{equation}
\label{eq:Modal_LinearNSE}
	-i\omega \hat{\boldsymbol{q}}_{k_z,\omega} = \mathcal{L}_{\bar{\boldsymbol{q}}}(\hat{\boldsymbol{q}}_{k_z,\omega};k_z) + \hat{\boldsymbol{u}}_{k_z,\omega}.
\end{equation}
By treating $\hat{\boldsymbol{u}}_{k_z,\omega}$ as a known forcing, equation \ref{eq:Modal_LinearNSE} (or \ref{eq:LinearNSE} equivalently) represents an inhomogeneous linear differential equation that governs the time evolution of perturbation $\hat{\boldsymbol{q}}_{k_z,\omega}$, with $\hat{\boldsymbol{u}}_{k_z,\omega}$ being the inhomogeneous forcing term on the right hand side.  Its general solution comprises of a homogeneous solution and a particular solution.  The homogeneous solution can be found by solving equation \ref{eq:Modal_LinearNSE} without the forcing term. That is, 
\begin{equation}
\label{eq:homo}
	- i\omega \hat{\boldsymbol{q}}_{k_z,\omega} = \mathcal{L}_{\bar{\boldsymbol{q}}}(\hat{\boldsymbol{q}}_{k_z,\omega};k_z),
\end{equation}
which forms an eigenvalue problem so that the homogeneous solution associates with the {\it spectrum} of $\mathcal{L}_{\bar{\boldsymbol{q}}}$.  On the other hand, the particular solution of \ref{eq:Modal_LinearNSE} can be expressed as
\begin{equation}
\label{eq:inhomo}
	\hat{\boldsymbol{q}}_{k_z,\omega} = \left[ - i\omega - \mathcal{L}_{\bar{\boldsymbol{q}}}(k_z)\right]^{-1} \hat{\boldsymbol{u}}_{k_z,\omega},
\end{equation}
where the operator
\begin{equation}
\label{eq:ResolventOperator}
	\mathcal{H}_{\bar{\boldsymbol{q}}}(k_z,\omega) = \left[ - i\omega - \mathcal{L}_{\bar{\boldsymbol{q}}}(k_z)\right]^{-1}
\end{equation}
is referred to as the {\it resolvent} and is associated with the {\it pseudospectrum} of $\mathcal{L}_{\bar{\boldsymbol{q}}}$  \citep{TrefethenEmbree:2005}.  

Our objective is not to solve the differential equation \ref{eq:Modal_LinearNSE}, which requires knowledge of the initial condition and the explicit forcing $\hat{\boldsymbol{u}}_{k_z,\omega}$.  However, we characterize its general solution by analyzing the spectrum and pseudospectrum of the linear operator $\mathcal{L}_{\bar{\boldsymbol{q}}}$.  Moreover, we note that the particular solution \ref{eq:inhomo} describes a linear operation that takes place between a sustained input $\hat{\boldsymbol{u}}_{k_z,\omega}$ and the harmonic output $\hat{\boldsymbol{q}}_{k_z,\omega}$ through the resolvent operator $\mathcal{H}_{\bar{\boldsymbol{q}}}(k_z,\omega)$.  For this reason, the pseudospectrum of $\mathcal{L}_{\bar{\boldsymbol{q}}}$, which captures energy amplification through the input-output process, is the main focus of this study on active flow control.

With the knowledge of $\bar{\boldsymbol{q}}$ and appropriate boundary conditions, the linear operator $\mathcal{L}_{\bar{\boldsymbol{q}}}$ can be explicitly constructed in its discretized form $\mathsfbi{L}_{\bar{\boldsymbol{q}}}$ for a prescribed spanwise wavenumber $k_z$.  Equation \ref{eq:Modal_LinearNSE} can be rewritten in discrete form as
\begin{equation}
\label{eq:Modal_DLinearNSE}
	- i\omega \hat{\boldsymbol{q}}_{k_z,\omega} = \mathsfbi{L}_{\bar{\boldsymbol{q}}}(k_z)\hat{\boldsymbol{q}}_{k_z,\omega} + \hat{\boldsymbol{u}}_{k_z,\omega},
\end{equation}
where the operation of $\mathcal{L}_{\bar{\boldsymbol{q}}}$ on $\hat{\boldsymbol{q}}_{k_z,\omega}$ is represented by a matrix-vector multiplication of $\mathsfbi{L}_{\bar{\boldsymbol{q}}}(k_z)\hat{\boldsymbol{q}}_{k_z,\omega}$. The modal wavenumber $k_z$ is embedded in $\mathsfbi{L}_{\bar{\boldsymbol{q}}}$ since it emerges from the spatial differentiation in the construction of $\mathsfbi{L}_{\bar{\boldsymbol{q}}}$.  With the discrete linear operator $\mathsfbi{L}_{\bar{\boldsymbol{q}}}$ constructed, its spectrum and pseudospectrum can be found numerically.  Below, we document the domain discretization and boundary conditions for constructing the discrete linear operator $\mathsfbi{L}_{\bar{\boldsymbol{q}}}$.  The numerical approach for computing its spectrum and pseudospectrum is also offered.

\subsection{Numerical setup}
The discretization for equation \ref{eq:Modal_LinearNSE} is performed on the computational mesh as shown in figure \ref{fig:Mesh} highlighted in orange on top of the LES domain.  This 2D domain has an extent of $x/L_c \in [-15, 16]$, $y/L_c \in [-12, 12]$ and is composed of approximately $0.14$ million grid points.  For the far-field boundary and over the airfoil, the Dirichlet boundary condition is set for $[\rho', u', v', w'] = [0, 0, 0, 0]$ and the Neumann boundary condition is set for $T'$ such that $\boldsymbol{e}_n \cdot \bnabla T' = 0$, where $\boldsymbol{e}_n$ is the unit normal boundary vector.  At the outlet boundary, the same Neumann boundary condition is set for all flow variables.  With these boundary conditions and the turbulent mean flow $\bar{\boldsymbol{q}} = [\bar{\rho}, \bar{u}, \bar{v}, \bar{w}, \bar{T}]^T$ obtained from the baseline LES, we construct the linear operator in its discrete form $\mathsfbi{L}_{\bar{\boldsymbol{q}}}(k_z)$ for a chosen spanwise wavenumber $k_z$.  

In the current study, the size of $\mathsfbi{L}_{\bar{\boldsymbol{q}}}$ is approximately $0.7$-million $\times$ $0.7$-million.  Considering the large size of $\mathsfbi{L}_{\bar{\boldsymbol{q}}}$, the implicitly restarted Arnoldi method \citep{Lehoucq:ARPACK} is used to handle the large-scale eigenvalue problems to solve for its spectrum and pseudospectrum.  The eigenvalues and the resolvent norm (for pseudospectrum) are computed with a Krylov space of $128$ vectors and a residual tolerance of $10^{-10}$.  The domain size and mesh resolution were examined to ensure that the results converge to at least 7 significant digits.

\subsection{Spectrum and pseudospectrum of $\mathsfbi{L}_{\bar{\boldsymbol{q}}}$}
The mean-flow-based linear operator $\mathsfbi{L}_{\bar{\boldsymbol{q}}}$ can be characterized by its spectrum (eigenvalues) and pseudospectrum.  Arising from the general solution of the Navier--Stokes equations \ref{eq:Modal_LinearNSE}, they describes the dynamical response of the fluid-flow system.

\subsubsection{Spectrum}
The eigenvalue problem arising from the homogeneous problem \ref{eq:homo} can be expressed in its discretized form
\begin{equation}
\label{eq:Eigen}
	\mathsfbi{L}_{\bar{\boldsymbol{q}}}(k_z)\hat{\boldsymbol{q}}_{k_z,\omega} = - i\omega \hat{\boldsymbol{q}}_{k_z,\omega},
\end{equation}
where  $-i\omega$ and $\hat{\boldsymbol{q}}$ are the eigenvalue and eigenmode, respectively.  The eigenvalue $-i\omega = -i\omega_r + \omega_i$ determines the temporal stability with modal frequency $\omega_r$ and growth or decay rate $\omega_i$.  An instability is identified if the complex modal frequency $\omega = \omega_r + i\omega_i$ resides on the positive imaginary plane with $\omega_i > 0$.  Upon prescribing a modal wavenumber $k_z$ for $\mathsfbi{L}_{\bar{\boldsymbol{q}}}(k_z)$, the eigenvalue problem \ref{eq:Eigen} can be referred to as the biglobal linear stability analysis \citep{Theofilis:ARFM2011} at $k_z$ with the turbulent mean flow $\bar{\boldsymbol{q}}$ as the base state.  
 
\begin{figure}
	\begin{center}
		\vspace{0.2in}
		\begin{overpic}[scale=0.55]{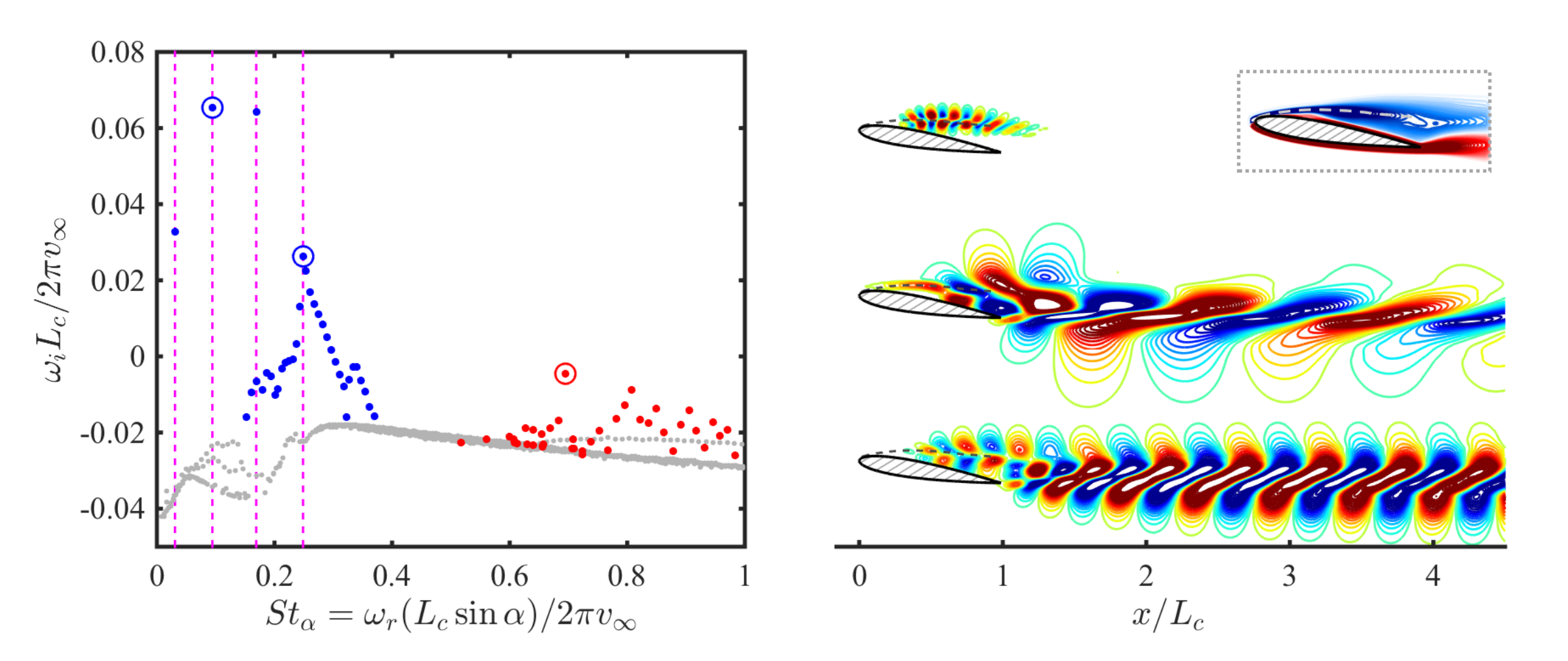}
			\put(27, 41){(a)}
			\put(55, 41){(b)}
			\put(78, 41){(c)}
			
			\put(10.4, 36.5){\color{blue}\scriptsize (2)}
			\put(20, 27){\color{blue}\scriptsize (3)}
			\put(36.7, 19.5){\color{red}\scriptsize (1)}
			\put(25, 32.5){\color{blue}{\scriptsize$\bullet$}\indexsize~wake modes}
			\put(25, 35.5){\color{red}{\scriptsize$\bullet$}\indexsize~shear-layer modes}
			\put(82.5, 39){$\bar{\zeta}_z L_c/v_\infty$}
			\put(53, 36.5){\indexsize\color{red}mode (1)}
			\put(53, 26){\indexsize\color{blue}mode (2)}
			\put(53, 15.5){\indexsize\color{blue}mode (3)}			
		\end{overpic}
	\end{center}
	\caption{\label{fig:A93D_BiG} Spectrum (a) and dominant eigenmodes (b) of $\mathsfbi{L}_{\bar{\boldsymbol{q}}}(k_z = 0)$ for $\alpha = 9^\circ$ mean flow. Spurious eigenvalues that associate with unphysical structures are colored in gray in the spectrum (a).  The magenta dashed lines in (a) highlight the frequencies of the dominant wake modes and are also shown in the frequency spectra in figure \ref{fig:Pseudospectra} (a-b). The shear-layer over the separation bubble is identified by the time-averaged spanwise vorticity $\bar{\zeta}_z$ as shown in (c) and is marked with gray dashed line to highlight the shear-layer structures in the eigenmodes.}
\end{figure}
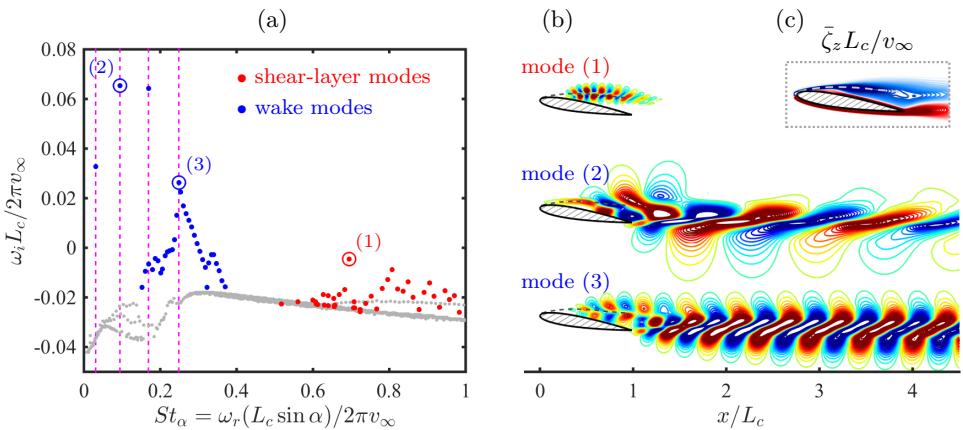

In figure \ref{fig:A93D_BiG}, we show the results of the spectrum of $\mathsfbi{L}_{\bar{\boldsymbol{q}}}(k_z = 0)$ and three representative eigenmodes for $\alpha = 9^\circ$.  We note that the spectrum is symmetric about the $\omega_i$ axis, since the modal phase velocity does not exhibit preferential spanwise direction due to the two-dimensional geometry of the airfoil.  Thus, in figure \ref{fig:A93D_BiG}, we only show the spectrum on the positive frequency plane ($\omega_r \ge 0$).  In the spectrum, two branches can be identified: the wake-mode branch and the shear-layer-mode branch. These two branches can be characterized by the frequency bandwidth of the eigenvalues or through the examination of their modal structures.  Three eigenmodes are chosen in the spectrum with $\boldsymbol{\circ}$ and their modal structures are visualized in figure \ref{fig:A93D_BiG} (b) with the streamwise velocity profile $\hat{u}$: the dominant shear-layer mode {\color{red}(1)}, the dominant wake mode  {\color{blue}(2)}, and a coupling mode {\color{blue}(3)} of shear-layer and wake.  On top of each modal structure, a dashed line is shown to mark the location of the time-average shear layer. This line is determined by examining the time-averaged spanwise vorticity $\bar{\zeta}_z$ for its local maximum magnitude over the separation bubble, as shown in figure \ref{fig:A93D_BiG} (c).  The shear-layer mode presents distinctively strong structure along the shear layer.  While the shear-layer mode gradually vanishes in the wake, the wake-mode structure extends farther downstream and resembles the pattern of von K\'arm\'an vortex street behind a bluff body.    On the wake branch, the frequencies of the two dominant modes are highlighted with magenta lines. These frequencies, marked again in the frequency spectrum of lift $\hat{C}_L$ in figure \ref{fig:Pseudospectra} (b), are found to be in agreement with the peaks obtained from LES.  Similar agreement holds for $\alpha = 6^\circ$ results in figure \ref{fig:Pseudospectra} (a).  The agreement between the $\mathsfbi{L}_{\bar{\boldsymbol{q}}}$ spectrum and the dominant frequency identified from the baseline flow shows that the nonlinear vortex-shedding physics can be revealed by the linear analysis.  Comparing the lift spectra for $\alpha = 6^\circ$ and $9^\circ$ in figure \ref{fig:Pseudospectra} (a-b),  we find that the frequency content of $\hat{C}_L$ scales well with the frontal-height-based Strouhal number $St_\alpha = \omega(L_c\sin\alpha)/2\pi v_\infty$.  The $St_\alpha$ scaling for the lift spectra has been studied by \citet{FageJohansen:PRSLA1927}, reporting the appearance of the $\hat{C}_L$ peaks near $St_\alpha \approx 0.2$.

The linear operator $\mathsfbi{L}_{\bar{\boldsymbol{q}}}(k_z)$ is observed to be unstable for $k_zL_c = 0$ as it possesses eigenvalues with positive growth rates.  In fact, $\mathsfbi{L}_{\bar{\boldsymbol{q}}}(k_z)$ is found to be unstable for $k_zL_c \lesssim 8\pi$.  The identification of the critical $k_z$ that yields instability in $\mathsfbi{L}_{\bar{\boldsymbol{q}}}(k_z)$ is out of scope of the present study. However, we make a cautious note here that its unstable nature for low $k_z$ necessitates further care when performing the resolvent analysis of $\mathsfbi{L}_{\bar{\boldsymbol{q}}}$, which will be discussed in detail in section \ref{sec:Discouted_InputOutput}. 

\subsubsection{Pseudospectrum}

\begin{figure}
	\begin{center}
		\begin{overpic}[scale=0.55]{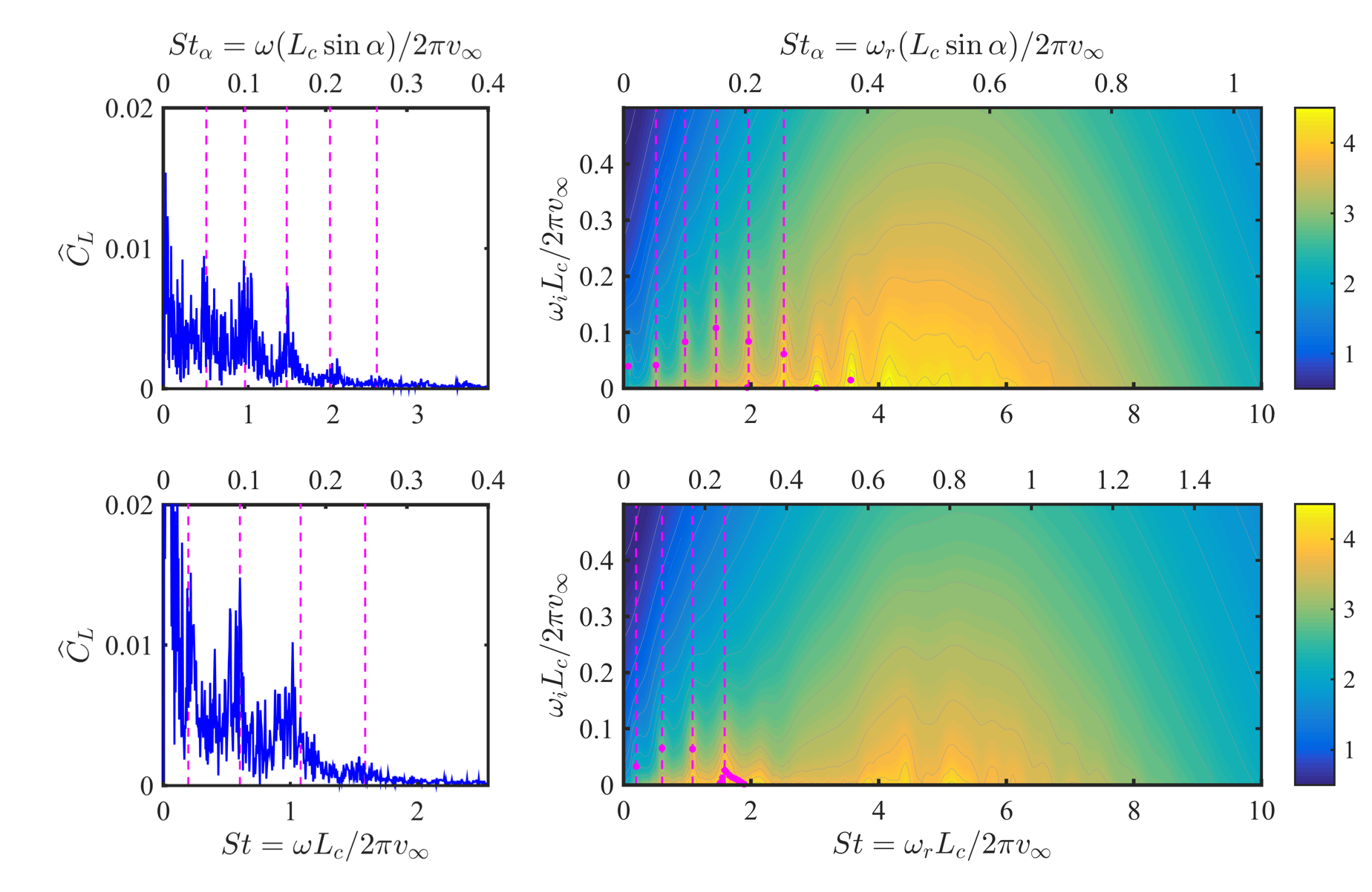}
			\put(1, 43){\rotatebox{90}{$\alpha = 6^\circ$}}
			\put(1, 14){\rotatebox{90}{$\alpha = 9^\circ$}}
			\put(31, 54){\indexsize(a)}
			\put(31, 25){\indexsize(b)}
			\put(86, 54){\indexsize\color{white}(c)}
			\put(86, 25){\indexsize\color{white}(d)}
			\put(92, 59){\indexsize$\log_{10}(\sigma_1)$}
			\put(92, 30){\indexsize$\log_{10}(\sigma_1)$}
		\end{overpic}
	\end{center}
	\caption{\label{fig:Pseudospectra} Frequency spectra of lift $\hat{C}_L$ from baseline LES (a-b) and pseudospectra of $\mathsfbi{L}_{\bar{\boldsymbol{q}}}$ with $k_zL_c = 0$ (c-d) for both $\alpha = 6^\circ$ (top) and $9^\circ$ (bottom).  Magenta dots in (c-d) depict the eigenvalues of the corresponding $\mathsfbi{L}_{\bar{\boldsymbol{q}}}$. Over the horizontal axis, two frequency scales are provided: the Fage--Johansen Strouhal number $St_\alpha = \omega(L_c\sin\alpha)/2\pi v_\infty$ on the upper axis and the chord-based Strouhal number $St = \omega L_c/2\pi v_\infty$ on the lower axis.  In each panel, the magenta dashed lines mark the frequencies of dominant wake modes from the spectra of $\mathsfbi{L}_{\bar{\boldsymbol{q}}}$.}
\end{figure}
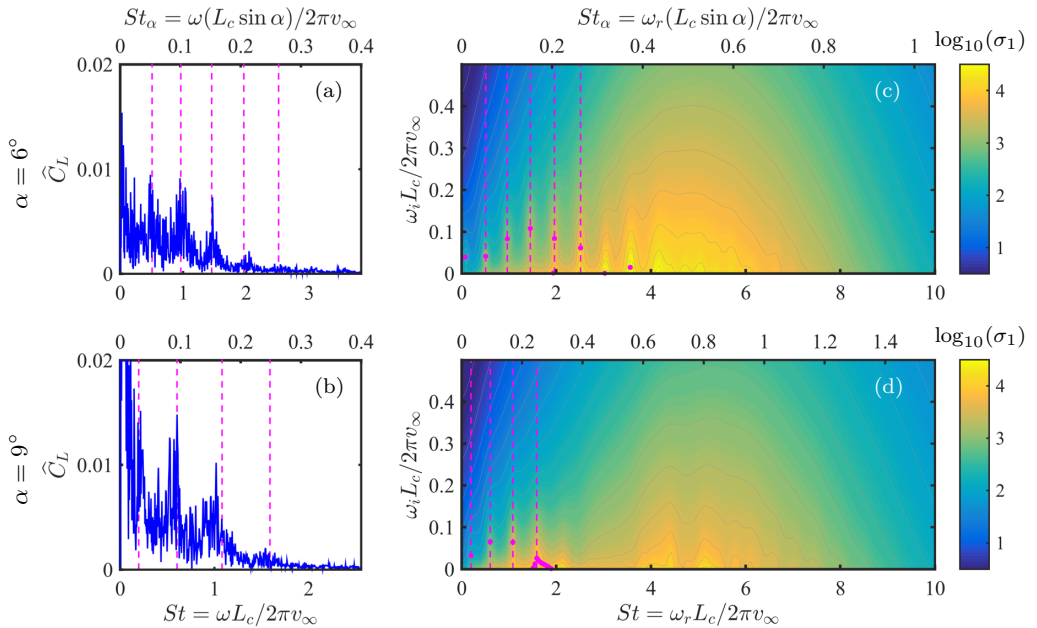

A normal operator satisfies $\mathsfbi{L}\mathsfbi{L}^* = \mathsfbi{L}^*\mathsfbi{L}$, where the superscript $^*$ denotes the Hermitian transpose. It has orthonormal eigenmodes with corresponding eigenvalues that govern the dynamical behavior.  For a nonnormal operator (\ie $\mathsfbi{L}\mathsfbi{L}^* \ne \mathsfbi{L}^*\mathsfbi{L}$),  its transient behavior is not described simply by the eigenvalues and eigenvectors.  Instead of just the spectrum, the pseudospectrum is needed to analyze the dynamics resulted from a nonnormal operator.  \citet{TrefethenEmbree:2005} examined pseudospectra of nonnormal operators and explained how they align with the dynamical behaviors governed by these operators.  In fluid-flow systems, shear is a source of nonnormality \citep{Trefethen:1993Science,SchmidHenningson:2001,MckeonSharma:JFM2010}.  From the baseline flows, we readily identify the presence of strong shear particularly over the separation bubble. They can be recognized in the mean flow profile for which $\mathsfbi{L}_{\bar{\boldsymbol{q}}}$ is constructed.

We have mentioned that the pseudospectrum of $\mathsfbi{L}_{\bar{\boldsymbol{q}}}(k_z)$ arises from the resolvent operator $\mathcal{H}_{\bar{\boldsymbol{q}}}(k_z, \omega)$ in the particular solution \ref{eq:inhomo}.  Here, we work with the discrete resolvent operator
\begin{equation}
\label{eq:ResolventMatrix}
	\mathsfbi{H}_{\bar{\boldsymbol{q}}}(k_z, \omega) = \left[ - i\omega \mathsfbi{I} - \mathsfbi{L}_{\bar{\boldsymbol{q}}}(k_z) \right]^{-1},
\end{equation}
where $\mathsfbi{I}$ is the identity matrix. The pseudospectrum of $\mathsfbi{L}_{\bar{\boldsymbol{q}}}(k_z)$ is to be mapped out over the complex $\omega$ plane by seeking a 2-norm measure through the singular value decomposition (SVD) of its resolvent matrix $\mathsfbi{H}_{\bar{\boldsymbol{q}}}$. An appropriate 2-norm for this fluid-flow study can be introduced as the weighted inner product between two state vectors
\begin{equation}
\label{eq:EnergyNorm}
	\langle \boldsymbol{q}_1, \boldsymbol{q}_2 \rangle_E = \int_\Omega \boldsymbol{q}_1^*\textsf{diag}\left(\frac{R\bar{T}}{\bar{\rho}}, \bar{\rho}, \bar{\rho}, \bar{\rho}, \frac{R\bar{\rho}}{(\gamma-1)\bar{T}}\right) \boldsymbol{q}_2 {\rm d}\boldsymbol{x},
\end{equation}
where $\Omega$ is the domain of interest and $R$ is the ideal gas constant. The inner product $\langle \boldsymbol{q}_1, \boldsymbol{q}_2 \rangle_E$ is referred to as the energy norm \citep{SchmidHenningson:2001}. We adopt the compressible disturbance energy proposed by \citet{Chu:Acta1965} and use this 2-norm for our computation of pseudospectra.  For the discrete flow fields, the energy norm is evaluated as
\begin{equation}
\label{eq:EnergyNorm_Matrix}
	\langle \boldsymbol{q}_1, \boldsymbol{q}_2 \rangle_E = \boldsymbol{q}_1^* \mathsfbi{W} \boldsymbol{q}_2,
\end{equation}
where the weight matrix $\mathsfbi{W}$ is the numerical quadrature that accounts for both the energy weight and domain integration. By introducing the similarity transformation of $\mathsfbi{H}_{\bar{\boldsymbol{q}}} \mapsto \mathsfbi{H}_{\bar{\boldsymbol{q}}, \mathsfbi{W}} = \mathsfbi{W}^{\frac{1}{2}} \mathsfbi{H}_{\bar{\boldsymbol{q}}} \mathsfbi{W}^{-\frac{1}{2}}$, the energy norm for $\mathsfbi{H}_{\bar{\boldsymbol{q}}}$ can be handled within the 2-norm framework for $\mathsfbi{H}_{\bar{\boldsymbol{q}}, \mathsfbi{W}}$ \citep{TrefethenEmbree:2005}.  Also, the similarity transformation performed for $\mathsfbi{H}_{\bar{\boldsymbol{q}}}$ translates to $\mathsfbi{L}_{\bar{\boldsymbol{q}}}$ and preserves its eigenvalues. The pseudospectrum of $\mathsfbi{L}_{\bar{\boldsymbol{q}}}$ with respect to the energy norm \ref{eq:EnergyNorm} can be evaluated through the SVD of $\mathsfbi{H}_{\bar{\boldsymbol{q}},\mathsfbi{W}}$ as
\begin{equation}
\label{eq:SVD}
	\mathsfbi{H}_{\bar{\boldsymbol{q}}, \mathsfbi{W}}(k_z,\omega) = \mathsfbi{Q}_\mathsfbi{W}\boldsymbol{\Sigma}\mathsfbi{U}_\mathsfbi{W}^*.
\end{equation}
By seeking the leading singular value $\sigma_1$ in $\boldsymbol{\Sigma}$, the pseudospectrum of $\mathsfbi{L}_{\bar{\boldsymbol{q}}}(k_z)$ is obtained at the complex $\omega$. 

Following the approach, in figure \ref{fig:Pseudospectra} (c-d), we present the pseudospectra of $\mathsfbi{L}_{\bar{\boldsymbol{q}}}(k_z)$ with respect to the energy norm for both $\alpha = 6^\circ$ and $9^\circ$ with $k_z = 0$, along with the frequency spectra of the lift coefficients from LES (a-b).  For all the four panels, we provide two different frequency scalings over the horizontal axes: the Fage--Johansen Strouhal number $St_\alpha = \omega(L_c\sin\alpha)/2\pi v_\infty$ on the top, and the chord-based Strouhal number $St = \omega L_c/2\pi v_\infty$ on the bottom.  Comparing the results from two angles of attack, we observe that, while the lift spectra scale well with  $St_\alpha$, the general behavior of the pseudospectra agrees better with $St$, especially in the high $\omega_i$ region.  The pseudospectra levels spread out from the region where most of the shear-layer eigenmodes reside for both angles of attack.  This observation can be explained by the high nonnormal nature of these shear-layer modes, whose structures are supported by the separation bubble above the airfoil that exhibits the strongest shear in the mean flow.  The high nonnormality in these shear-layer modes expands the {\it pseudospectral radius} about them such that they are centered by the roll-off in the pseudospectra levels.  Therefore, instead of the $St_\alpha$ scaling which emphasizes the wake physics, the shear-layer dominated behavior is better supported by the $St$ scaling for the pseudospectra. 


\subsection{Resolvent analysis for active flow control}
\label{sec:ResolventAFC}
To provide physical interpretation for the right- and left-singular vectors of the SVD \ref{eq:SVD},  let us recall the resolvent operator as part of the particular solution,
\begin{equation}
\label{eq:ResolventOperation}
	\hat{\boldsymbol{q}} = \mathsfbi{H}_{\bar{\boldsymbol{q}}} \hat{\boldsymbol{u}}.
\end{equation}
Here, we have left out the subscript $k_z$ and $\omega$ for simplicity.  The similarity transformation for $\mathsfbi{H}_{\bar{\boldsymbol{q}}}$ can be brought into the particular solution as $\mathsfbi{W}^{\frac{1}{2}}\hat{\boldsymbol{q}} = (\mathsfbi{W}^{\frac{1}{2}}\mathsfbi{H}_{\bar{\boldsymbol{q}}}\mathsfbi{W}^{-\frac{1}{2}})\mathsfbi{W}^{\frac{1}{2}} \hat{\boldsymbol{u}}$.  With the SVD for $\mathsfbi{H}_{\bar{\boldsymbol{q}}, \mathsfbi{W}}$ in \ref{eq:SVD}, the particular solution can be rewritten considering the energy norm as
\begin{equation}
\label{eq:inhomoSVD}
	\mathsfbi{W}^{\frac{1}{2}}\hat{\boldsymbol{q}} = \left( \mathsfbi{Q}_\mathsfbi{W}\boldsymbol{\Sigma}\mathsfbi{U}_\mathsfbi{W}^* \right) \mathsfbi{W}^{\frac{1}{2}} \hat{\boldsymbol{u}}.
\end{equation}
Starting from the right side of this equation, we see the projection of the weighted forcing $\mathsfbi{W}^{\frac{1}{2}} \hat{\boldsymbol{u}}$ onto the vector space spanned by the right-singular vectors $\mathsfbi{U}_\mathsfbi{W}$.  Such a projection takes the inner product with respect to the energy norm and decomposes $\mathsfbi{W}^{\frac{1}{2}} \hat{\boldsymbol{u}}$ into the vector components in $\mathsfbi{U}_\mathsfbi{W}$ with a series of projection coefficients.  Each forcing component is amplified by the corresponding singular value in $\boldsymbol{\Sigma}$, producing a set of scaled coefficients for the corresponding left-singular vectors.  The output $\mathsfbi{W}^{\frac{1}{2}} \hat{\boldsymbol{q}}$ is generated through the linear combination of the left-singular vectors using this set of scaled coefficients.  Thus, in the SVD of $\mathsfbi{H}_{\bar{\boldsymbol{q}}, \mathsfbi{W}}$, the left-singular vectors $\mathsfbi{Q}_\mathsfbi{W} = \mathsfbi{W}^{\frac{1}{2}}[\hat{\boldsymbol{q}}_1, \hat{\boldsymbol{q}}_2, \dots, \hat{\boldsymbol{q}}_n]$ can be interpreted as response modes, whereas the right-singular vector $\mathsfbi{U}_\mathsfbi{W} = \mathsfbi{W}^{\frac{1}{2}}[\hat{\boldsymbol{u}}_1, \hat{\boldsymbol{u}}_2, \dots, \hat{\boldsymbol{u}}_n]$ can be interpreted as forcing modes.  Each forcing-response pair is subjected to the corresponding amplification in $\boldsymbol{\Sigma} = \mathsf{diag}(\sigma_1, \sigma_2, \dots, \sigma_n)$, where $\sigma_k$ can be arranged in a descending order.  If $\sigma_1 \gg \sigma_2$, the rank-1 assumption \citep{MckeonSharma:JFM2010,Luhar:JFM2014_1,Gomez:JFMR2016,Beneddine:JFM2016} can be appropriately made, expecting that the input-output process is dominated by the leading forcing-response pair, \ie $\hat{\boldsymbol{q}} \approx \hat{\boldsymbol{q}}_1 \sigma_1 \langle \hat{\boldsymbol{u}}_1, \hat{\boldsymbol{u}} \rangle_E$, as long as $\langle \hat{\boldsymbol{u}}_1, \hat{\boldsymbol{u}} \rangle_E$ has reasonable magnitude.  This assumption will be shortly justified with the results presented in the next section.

\begin{figure}
	\vspace{0.1in}
	\begin{center}
		\includegraphics[scale=0.55]{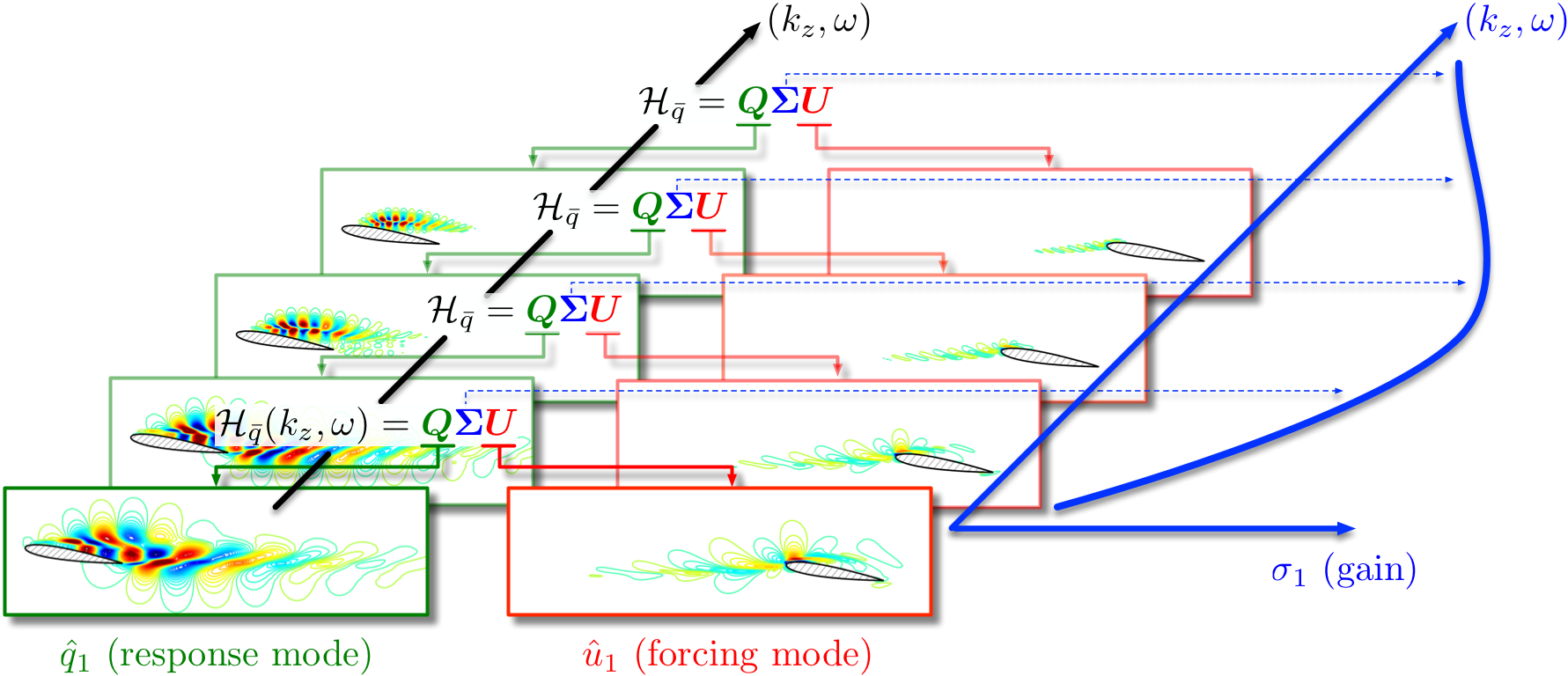}
	\end{center}
	\caption{\label{fig:Resolvent_Demo} Schematic demonstration of resolvent analysis: each SVD provides an optimal forcing-response pair with the associated amplification (gain) while sweeping through frequency $\omega$ and wavenumber $k_z$. }
\end{figure}

Recognizing that the SVD is performed for $\mathsfbi{H}_{\bar{\boldsymbol{q}}}(k_z, \omega)$ for prescribed $k_z$ and $\omega$, a concept of `Bode plot' can be realized by sweeping through the frequency $\omega$ for each $k_z$, seeking for the leading amplification (as the `gain') from each SVD \citep{Jovanovic:JFM2005}.  Such an approach is illustrated in figure \ref{fig:Resolvent_Demo}, where each SVD gives a leading forcing-response pair along with the associated gain.   With the Bode plot constructed based on the pseudospectral analysis of $\mathsfbi{L}_{\bar{\boldsymbol{q}}}$, efficient ways of forcing may be predicted by looking for the $k_z$ and $\omega$ that produce high gain.  Such a forcing input will be highly amplified by $\mathsfbi{H}_{\bar{\boldsymbol{q}}}$ to produce perturbation $\hat{\boldsymbol{q}}$ about $\bar{\boldsymbol{q}}$.  The amplitude of perturbation may grow beyond the validity of linear regime governed by $\mathsfbi{L}_{\bar{\boldsymbol{q}}}$. Through nonlinearity, the highly amplified perturbation can modify the mean flow ${\bar{\boldsymbol{q}}}$, which is the objective of flow control.  For this reason, resolvent analysis, arising from the input-output process in the particular solution \ref{eq:ResolventOperation}, provides insightful information for the design of flow control.  While following this approach, we provide a couple of cautionary comments on the  resolvent analysis in the context of designing flow control techniques:
\begin{enumerate}
	\item Even though the effective forcing predicted by the resolvent analysis may have a good chance to modify ${\bar{\boldsymbol{q}}}$, the direction of the change (\eg increase or decrease in lift) may be beyond the insights that can be provided by the amplification. The achievement of an aerodynamically favorable change may require further knowledge, such as the structure of the harmonic response rather than just the knowledge on amplifications;
	\item Once the base flow ${\bar{\boldsymbol{q}}}$ is modified with control, the results from the analysis performed with respect to the operator for uncontrolled base state $\mathsfbi{L}_{\bar{\boldsymbol{q}}}$ may no longer be valid.  However, resolvent analysis shall still provide valuable insights for the effective forcing before the system departs from the linear regime about the uncontrolled ${\bar{\boldsymbol{q}}}$. 
\end{enumerate}

We have presented a control-oriented interpretation of the results from resolvent analysis.  Traditionally, resolvent analysis used in fluid mechanics deals with {\it asymptotically stable} base flows (the Lyapunov stability).  With asymptotic stability, the gain obtained from the {\it sustained} forcing is bounded over the {\it infinite-time horizon}.  However, the linear operators $\mathsfbi{L}_{\bar{\boldsymbol{q}}}$ for the present flows are unstable, as pointed out in figure \ref{fig:A93D_BiG}.  To address this matter for the present flow control effort, we discuss an extension to the standard resolvent analysis in the following section.

\subsection{Finite-time horizon resolvent analysis}
\label{sec:Discouted_InputOutput}
While the analysis of asymptotic stability requires an infinite-time horizon, the dynamical behavior of a nonnormal system within a finite-time horizon is also relevant.  For an asymptotically stable system, the perturbation energy can undergo transient growth due to nonnormality of the operator.  Such dynamics is not described by the asymptotic behavior of the operator, but can be characterized through an initial-value problem by specifying a finite-time horizon \citep{Schmid:ASME2014}.  Even if the system is characterized as unstable (unbounded) asymptotically, a bounded amplification can be found when a finite-time horizon is specified.  For the present fluid-flow problem, some nonlinear dynamic processes, such as the shear-layer roll-up, the break-up of spanwise vortical structures, and the vortex merging process can all take place within a short time window.  Therefore, we do not concern ourselves with the concept of asymptotic stability, but rather focus on the short-term dynamics by considering a finite-time horizon for the input-output analysis, following the approach proposed by \citet{Jovanovic:Thesis2004}. 


\citet{Jovanovic:Thesis2004} introduced an input-output analysis on an unstable system with an exponential discount. This analysis starts with the introduction of a temporal filter performed on both response and forcing such that $\check{\boldsymbol{q}}_{\beta} = \check{\boldsymbol{q}}e^{-t/t_\beta}$ and $\check{\boldsymbol{u}}_{\beta} = \check{\boldsymbol{u}}e^{-t/t_\beta}$. The time constant $t_\beta > 0$ is chosen such that the decay rate $\beta = 1/t_\beta$ in the temporal filter $e^{-\beta t}$ overtakes the growth rate of the dominant unstable eigenvalue of $\mathsfbi{L}_{\bar{\boldsymbol{q}}}$.  That is, $\beta > \max (\omega_i)$.  The use of such temporal filter ensures that we examine the dominant transient growth that takes place over a time window characterized by $t_\beta$.  Therefore, the value of $t_\beta$ can be chosen according to physical interests.  Upon substituting these growth-discounted modes of $\check{\boldsymbol{q}}_{\beta}$ and $\check{\boldsymbol{u}}_{\beta}$ into the Navier--Stokes equation \ref{eq:LinearNSE}, we have
\begin{equation}
\label{eq:DiscountedNSE}
	\left( \beta - i\omega \right) \hat{\boldsymbol{q}}_{\beta} = \mathsfbi{L}_{\bar{\boldsymbol{q}}} \hat{\boldsymbol{q}}_{\beta} + \hat{\boldsymbol{u}}_{\beta}.
\end{equation}
Thus, we can express the discounted resolvent analysis as
\begin{equation}
\label{eq:DiscountedResolvent}
	\hat{\boldsymbol{q}}_{\beta} = \left[ - i\omega \mathsfbi{I} - \left( \mathsfbi{L}_{\bar{\boldsymbol{q}}} - \beta\mathsfbi{I}\right) \right]^{-1} \hat{\boldsymbol{u}}_{\beta},
\end{equation}
with the discounted resolvent operator $\mathsfbi{H}_{\bar{\boldsymbol{q}},\beta}$
\begin{equation}
	\mathsfbi{H}_{\bar{\boldsymbol{q}},\beta} 
		= \left[ - i\omega \mathsfbi{I} - \left( \mathsfbi{L}_{\bar{\boldsymbol{q}}} - \beta\mathsfbi{I}\right) \right]^{-1}. 
		\label{eq:DiscountedResolventMatrix1}
\end{equation}
This expression constructs the discounted resolvent operator $\mathsfbi{H}_{\bar{\boldsymbol{q}},\beta}$ using the shifted linear operator $\left( \mathsfbi{L}_{\bar{\boldsymbol{q}}} - \beta\mathsfbi{I}\right)$.  The eigenvalues of $\mathsfbi{L}_{\bar{\boldsymbol{q}}}$ are now shifted by $-\beta$ and all reside on the stable complex plane so that the standard resolvent analysis can be performed with $\mathsfbi{H}_{\bar{\boldsymbol{q}},\beta}$ along the real axis of $\omega = \omega_r$.  Note that $\mathsfbi{H}_{\bar{\boldsymbol{q}},\beta}$ can also be expressed as $\mathsfbi{H}_{\bar{\boldsymbol{q}},\beta} = \left[ - i\left( \omega +i\beta\right) \mathsfbi{I} - \mathsfbi{L}_{\bar{\boldsymbol{q}}} \right]^{-1}$, suggesting that an equivalent exercise can be performed by directly evaluating the pseudospectrum of $\mathsfbi{L}_{\bar{\boldsymbol{q}}}$ on a raised frequency axis of $( \omega_r +i\beta)$.  The traditional approach is recovered by setting $\beta = 0$ (\ie $t_\beta \rightarrow \infty$ for infinite-time horizon). 

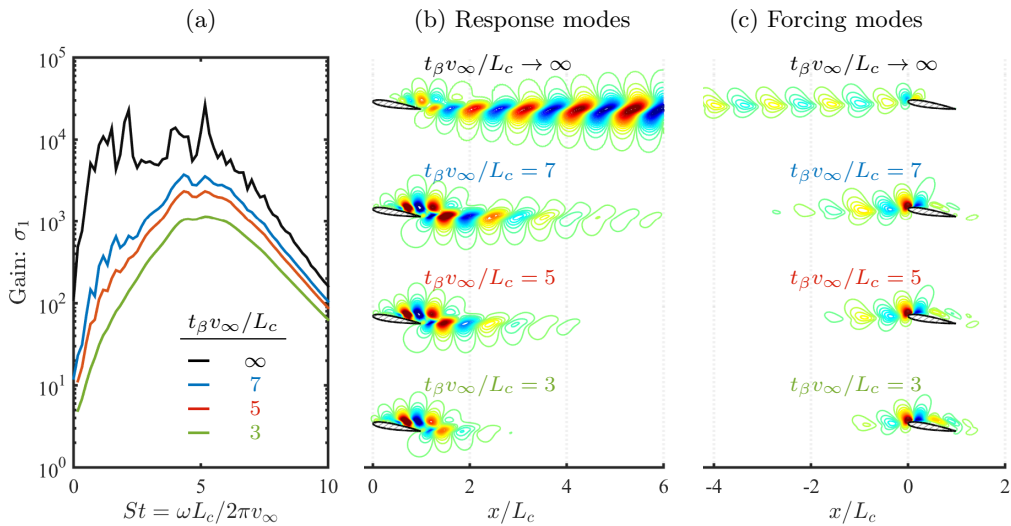
\begin{figure}
	\begin{center}
	\vspace{0.2in}
		\begin{overpic}[scale=0.55]{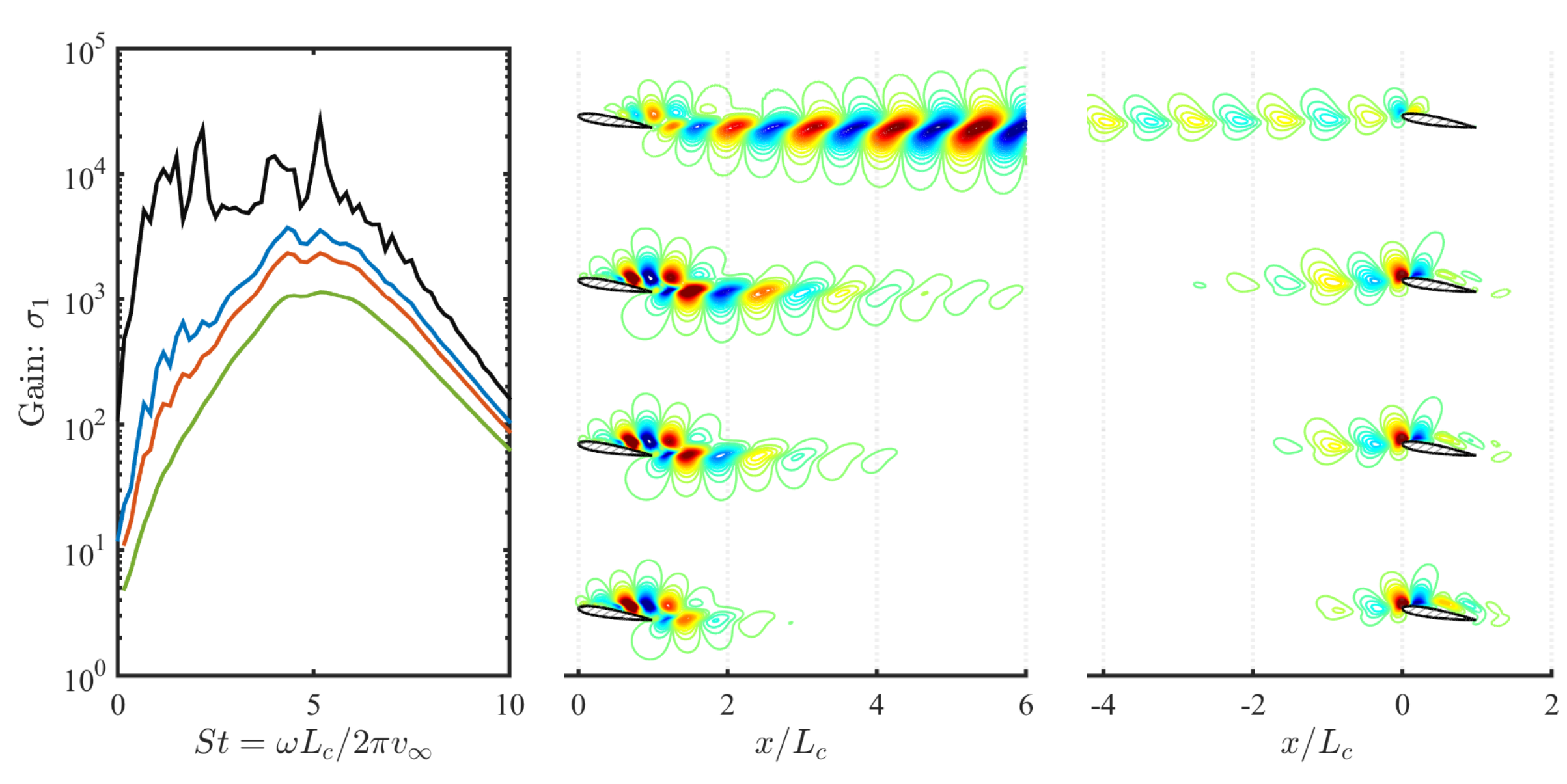}
			\put(18.5, 50){(a)}
			\put(41, 50){(b) Response modes}
			\put(72, 50){(c) Forcing modes}
			\put(18, 15){\indexsize
							\begin{tabular}{r c} 
								\multicolumn{2}{c}{$t_\beta v_\infty/L_c$} \vspace{-0.07in}\\ \hline\vspace{-0.150in}\\
								{\color{black}\solid}&{\color{black}$\infty$} \\
								{\color{blue2}\solid}&{\color{blue2}$7$} \\
								{\color{red2}\solid}&{\color{red2}$5$} \\ 
								{\color{green2}\solid}&{\color{green2}$3$}
							\end{tabular}}
			\put(42, 45.75){\indexsize$t_\beta v_\infty/L_c \rightarrow \infty$}
			\put(42, 35.3){\indexsize\color{blue2} $t_\beta v_\infty/L_c = 7$}
			\put(42, 24.85){\indexsize\color{red2} $t_\beta v_\infty/L_c = 5$}
			\put(42, 14.4){\indexsize\color{green2} $t_\beta v_\infty/L_c = 3$}

			\put(78, 45.75){\indexsize$t_\beta v_\infty/L_c \rightarrow \infty$}
			\put(78, 35.3){\indexsize\color{blue2} $t_\beta v_\infty/L_c = 7$}
			\put(78, 24.85){\indexsize\color{red2} $t_\beta v_\infty/L_c = 5$}
			\put(78, 14.4){\indexsize\color{green2} $t_\beta v_\infty/L_c = 3$}
		\end{overpic}
	\end{center}
	\caption{\label{fig:Discount_Demo} Finite-time-horizon resolvent analysis with different choices of $t_\beta$, considering $\alpha = 9^\circ$ mean flow with $k_zL_c = 0$. (a) Gain over frequency in $St$; (b) resolvent response modes; (c) forcing modes. The lowest magnitude marked by contour lines is $1\%$ of the modal maximum. The streamwise extent of the modal structures shortens with decreasing $t_\beta$.}
\end{figure}

We demonstrate this finite-time horizon resolvent analysis in figure \ref{fig:Discount_Demo} by showing representative results over varied choices of $t_\beta$.  Here, we use the operator $\mathsfbi{L}_{\bar{\boldsymbol{q}}}$ constructed with $k_z = 0$ about the $\alpha = 9^\circ$ mean baseline flow and choose $t_\beta$ such that $t_\beta v_\infty/L_c = 3$, $5$, and $7$. The results from these choices of $t_\beta$ will be compared with those from the infinite-time horizon analysis ($t_\beta \rightarrow \infty$). 

Let us analyze the gain distribution over frequency shown in figure \ref{fig:Discount_Demo} (a).  By decreasing $t_\beta$ from $7$ to $3$, we observe that the gain over $St$ decreases with $t_\beta$.  The decrease in gain can be explained by the shorter time horizon over which the growth in perturbation energy is evaluated.  It can also be understood as the decreasing pseudospectral level with increasing $\omega_i$ (moving away from the neutral stability axis) as we can observe in figure \ref{fig:Pseudospectra}.  The finite-time horizon analysis removes the spikes appearing in the gain distribution evaluated with the infinite-time horizon.  The spikiness is attributed to the response of pseudospectral level to subdominant and spurious eigenmodes populating densely near the frequency axis, which can be seen in the spectrum in figure \ref{fig:A93D_BiG} (a). 

In figures \ref{fig:Discount_Demo} (b) and (c), the leading response modes and forcing modes are respectively shown for the corresponding $t_\beta$.  From the response modes in figure \ref{fig:Discount_Demo} (b), we observe that all choices of $t_\beta$ reveal the flow responses in the shear-layer over the airfoil and in the wake.  In figure \ref{fig:Discount_Demo} (c), the forcing modes exhibit advective structure near the airfoil and the upstream.  Note that the time scale, $t_\beta v_\infty /L_c$, can also be interpreted as the advective length scale over the finite-time window.  The streamwise coverage of the structures in both response and forcing modes is well characterized by each time constant $t_\beta$ used in the temporal filter.  

The advective feature of the forcing mode motivates the use of local actuation, since the locally introduced perturbation that advects with the flow can leverage this feature as long as the forcing mode structures extend farther downstream of the actuator.  Moreover, we observe that the forcing modes exhibit high level of fluctuation near the leading edge in all values of $t_\beta$ examined.  The forcing mode shape suggests that the amplification from the input-output process can be efficiently leveraged if actuation is introduced near the leading edge.  Our choice of the actuator location ($x_a/L_c = 0.03$) is hence supported by the observation on the forcing mode structure.

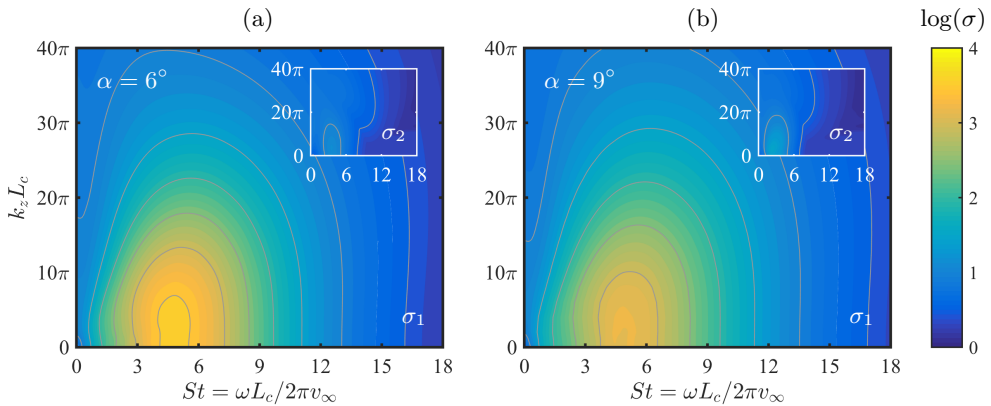
\begin{figure}
	\begin{center}
		\vspace{0.15in}
		\begin{overpic}[scale=0.55]{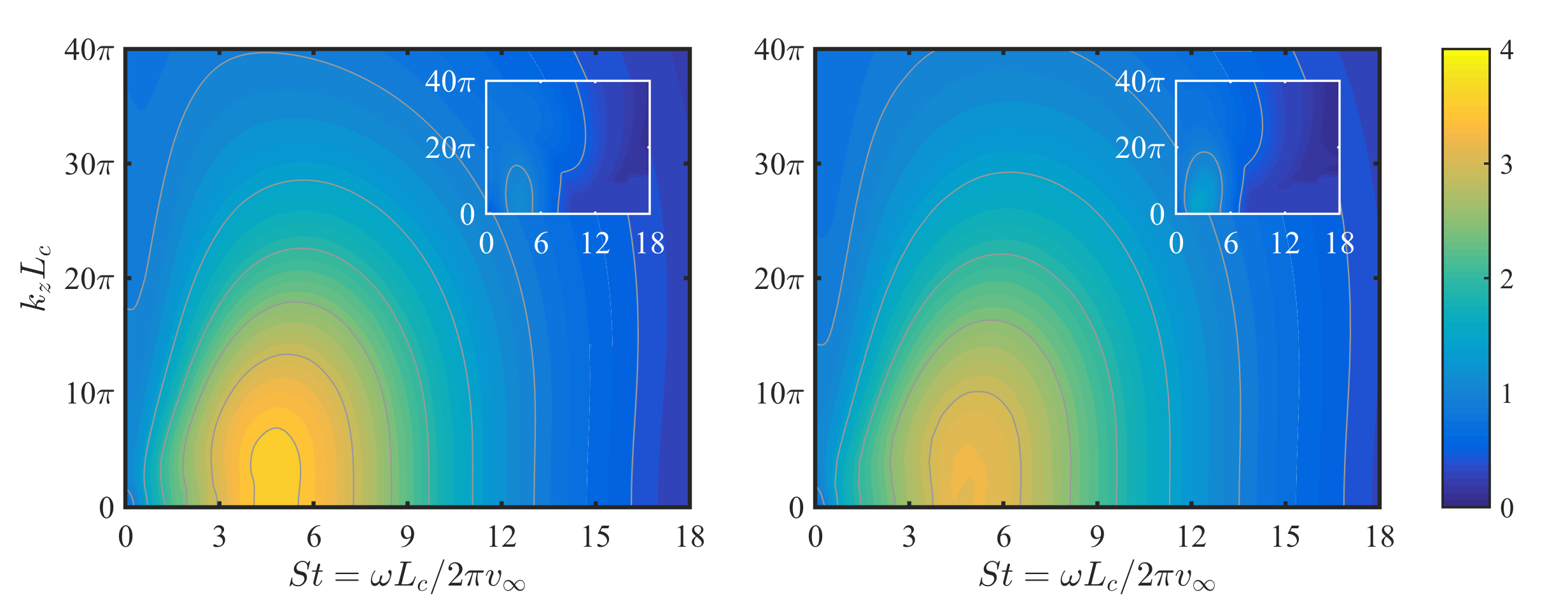}
			\put(24.3, 38){(a)}
			\put(68.0, 38){(b)}
			\put(10, 32){\color{white}$\alpha = 6^\circ$}
			\put(54, 32){\color{white}$\alpha = 9^\circ$}
			\put(91, 38){$\log(\sigma)$}
			\put(40, 9){\color{white}$\sigma_1$}
			\put(84, 9){\color{white}$\sigma_1$}
			\put(38, 27){\color{white}$\sigma_2$}
			\put(82, 27){\color{white}$\sigma_2$}
		\end{overpic}
	\end{center}
	\caption{\label{fig:Bode}Gain distribution over the $\omega$-$k_z$ space for $\alpha = 6^\circ$ (a) and $\alpha = 9^\circ$ (b). Approximately $20$ dB difference from the leading to second singular value is observed.}
\end{figure}

In figure \ref{fig:Bode}, we present the gain distribution over the $\omega$-$k_z$ plane with $t_\beta v_\infty/L_c = 5$.  In the rest of this work, we will focus on this choice of $t_\beta$.  For each $\alpha$, the gain constructed from the second singular value $\sigma_2$ is also presented in comparison with that from $\sigma_1$ over the same frequency-wavenumber plane.  The difference between $\sigma_1$ and $\sigma_2$ is typically greater than an order of magnitude.  This gap between the leading and second singular value justifies the rank-1 assumption discussed in the previous section.   Comparing the results from both angles of attack, we find that leading gain over the entire $\omega$-$k_z$ plane is well-scaled in the chord-based Strouhal number $St = \omega L_c/2\pi v_\infty$ and wavenumber $k_zL_c$.  The resemblance stems from the highly nonnormal shear-layer modes residing near $St \approx 5$ for both angles of attack, which are observed from their pseudospectra in figure \ref{fig:Pseudospectra}.  Also, the gain exhibits a general decreasing trend with increasing $k_zL_c$.  This behavior can be attributed to the attenuation of 3D instability, which has been studied by \citet{Pierrehumbert:JFM1982} and \citet{Hwang:JFM2013} for free shear layer and wake, respectively.  

\begin{figure}
\begin{center}
\indexsize
 	\begin{tabular}{	>{\centering\arraybackslash} m{0.25in}  
						>{\centering\arraybackslash} m{0.25in}  
						>{\centering\arraybackslash} m{1.45in}
						>{\centering\arraybackslash} m{1.45in}
						>{\centering\arraybackslash} m{1.45in}}
	$k_zL_c$\vspace{-0.12in} & $St$\vspace{-0.12in}
			& $\hat{v}_x$ mode\vspace{-0.12in} 
			& $\hat{v}_y$ mode\vspace{-0.12in} 
			& Reynolds stress $\hat{R}_z$\vspace{-0.12in}\\
    \hline
    \vspace{-0.05in}$0$&\vspace{-0.05in}$1.5$
    		&	\vspace{-0.05in}\includegraphics[width=1.45in]{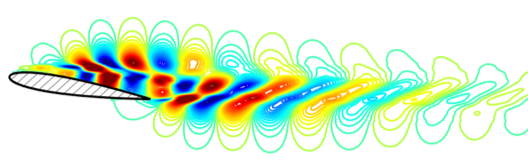}
			&	\vspace{-0.05in}\includegraphics[width=1.45in]{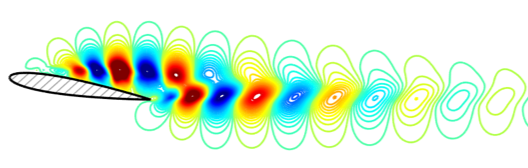}
			&	\vspace{-0.05in}\includegraphics[width=1.45in]{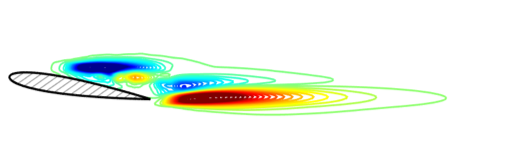}\\
	\vspace{-0.06in}$0$&\vspace{-0.06in}$2.5$
    		&	\vspace{-0.06in}\includegraphics[width=1.45in]{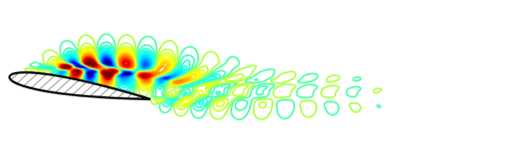}
			&	\vspace{-0.06in}\includegraphics[width=1.45in]{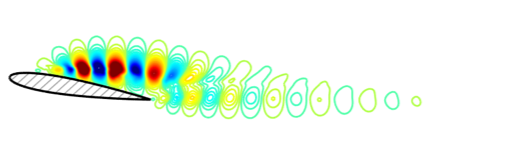}
			&	\vspace{-0.06in}\includegraphics[width=1.45in]{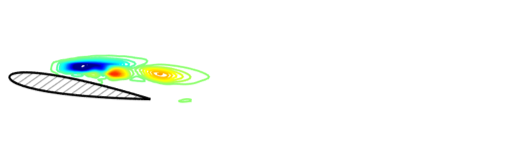}\\
	\vspace{-0.06in}$0$&\vspace{-0.06in}$5$
    		&	\vspace{-0.06in}\includegraphics[width=1.45in]{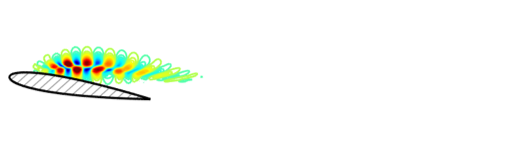}
			&	\vspace{-0.06in}\includegraphics[width=1.45in]{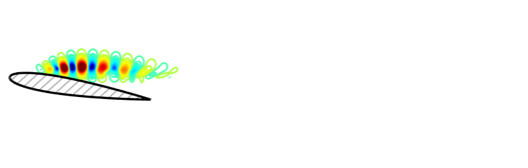}
			&	\vspace{-0.06in}\includegraphics[width=1.45in]{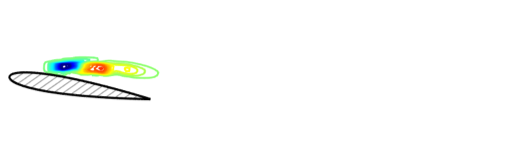}\\	
	\vspace{-0.06in}$0$\vspace{-0.12in}&\vspace{-0.06in}$10$\vspace{-0.12in}
    		&	\vspace{-0.06in}\includegraphics[width=1.45in]{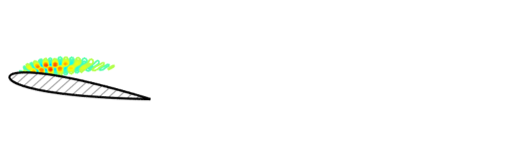}\vspace{-0.12in}
			&	\vspace{-0.06in}\includegraphics[width=1.45in]{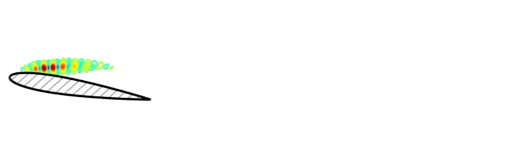}\vspace{-0.12in}
			&	\vspace{-0.06in}\includegraphics[width=1.45in]{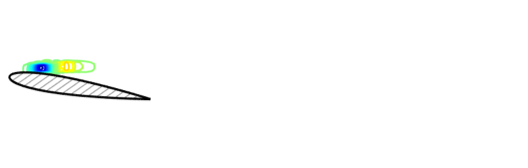}\vspace{-0.12in}\\
	\hline
    \vspace{-0.05in}$4\pi$&\vspace{-0.05in}$1.5$
    		&	\vspace{-0.05in}\includegraphics[width=1.45in]{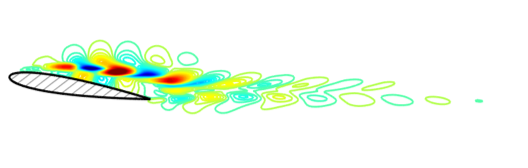}
			&	\vspace{-0.05in}\includegraphics[width=1.45in]{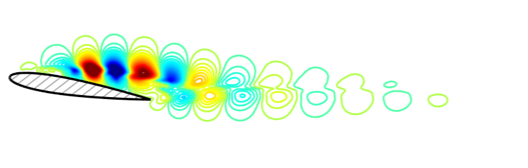}
			&	\vspace{-0.05in}\includegraphics[width=1.45in]{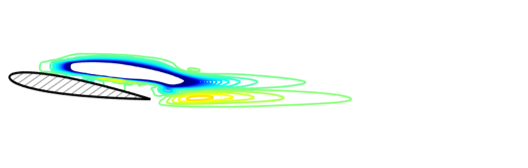}\\
	\vspace{-0.06in}$4\pi$\vspace{-0.12in}&\vspace{-0.06in}$5$\vspace{-0.12in}
    		&	\vspace{-0.06in}\includegraphics[width=1.45in]{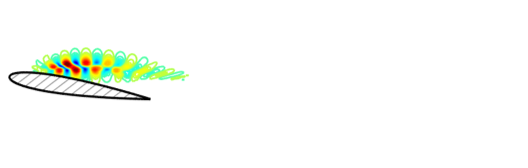}\vspace{-0.12in}
			&	\vspace{-0.06in}\includegraphics[width=1.45in]{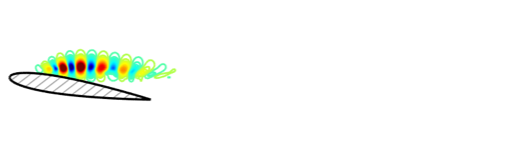}\vspace{-0.12in}
			&	\vspace{-0.06in}\includegraphics[width=1.45in]{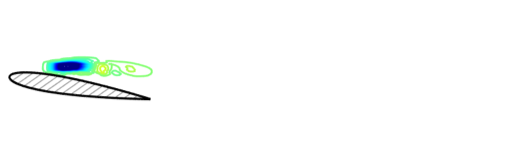}\vspace{-0.12in}\\
	\hline
    \vspace{-0.05in}$10\pi$&\vspace{-0.05in}$1.5$
    		&	\vspace{-0.05in}\includegraphics[width=1.45in]{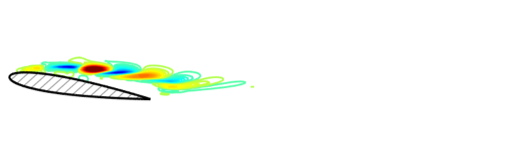}
			&	\vspace{-0.05in}\includegraphics[width=1.45in]{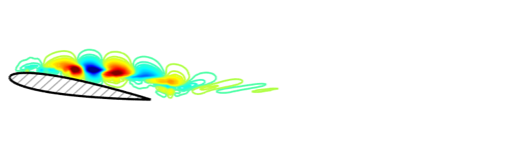}
			&	\vspace{-0.05in}\includegraphics[width=1.45in]{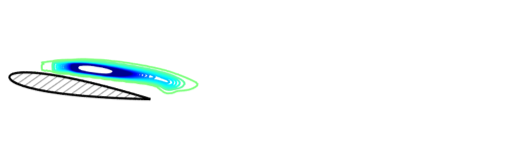}\\
	\vspace{-0.06in}$10\pi$&\vspace{-0.06in}$5$
    		&	\vspace{-0.06in}\includegraphics[width=1.45in]{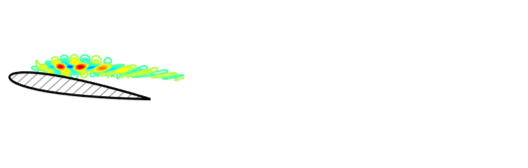}
			&	\vspace{-0.06in}\includegraphics[width=1.45in]{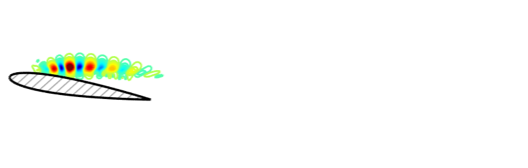}
			&	\vspace{-0.06in}\begin{overpic}[width=1.45in]{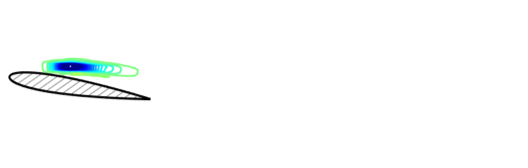}
									\put (70, 3) {\includegraphics[scale=0.55]{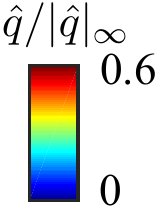}} 
								\end{overpic}
			\\
	\end{tabular}
	\caption{\label{fig:ResolventModes} Streamwise velocity mode $\hat{v}_x$, transverse velocity mode $\hat{v}_y$ and spanwise modal Reynolds stress $\hat{R}_z$ of representative $k_z$-$St$ combinations for $\alpha = 9^\circ$ mean baseline flow. The response modes are obtained with $t_\beta v_\infty/L_c = 5$ and are visualized by the contour lines of $\hat{q}/|\hat{q}|_\infty \in \pm [0.01, 0.9]$.}
\end{center}
\end{figure}

The structure of the response mode can also provide knowledge for identifying the actuation $k_z^+$ and $\omega^+$ that result in aerodynamically favorable control.  Given a response mode $\hat{\boldsymbol{q}} \equiv [\hat{\rho}, \hat{v}_x, \hat{v}_y, \hat{v}_z, \hat{T}]^T$ at specified $k_z$ and $\omega$,  we also evaluate the associated streamwise, transverse, and spanwise Reynolds stress respectively by
\begin{equation}
		\hat{R}_x(k_z, \omega) = \Re(\hat{v}_y^* \hat{v}_z)
	,~~~\hat{R}_y(k_z, \omega) = \Re(\hat{v}_z^* \hat{v}_x)
	,~~~\hat{R}_z(k_z, \omega) = \Re(\hat{v}_x^* \hat{v}_y),
\end{equation}
where $\Re(\cdot)$ denotes the real component of the argument.  In figure \ref{fig:ResolventModes}, we visualize the response modes using their streamwise velocity $\hat{v}_x$, transverse velocity $\hat{v}_y$ and the associated spanwise Reynolds stress $\hat{R}_z$ with representative $k_zL_c$-$St$ combinations for the mean baseline flow at $\alpha = 9^\circ$.  For modes of $St = 1.5$ and $2.5$, response structure develops from the shear layer above the suction surface and extends farther into the wake.  Particularly for $(k_zL_c, St) = (0, 1.5)$,   we observe an extended wake structure in the velocity modes as well as the resolvent Reynolds stress.  The Reynolds stress exhibits a pattern of von K\'arm\'an vortex shedding with negative correlation developing in the shear layer above the airfoil and positive correlation extending from the trailing edge over the bottom.  By either increasing $St$ or $k_zL_c$, the streamwise extent of the modal structure reduces to the shear layer.  Further increase of frequency moves the response structure towards the leading edge where the shear layer remains thin and is capable of supporting small-scale structures from high-frequency perturbations.  In section \ref{sec:Resovent_vs_CtrlLES}, we will further leverage these results on response mode structures to provide quantitative guidance to suppress stall.

We have performed resolvent analysis for the mean baseline flows of $\alpha = 6^\circ$ and $9^\circ$ and discussed an extension to the standard approach for the two unstable linear operators.  From the gain distribution over frequency and wavenumber, we have seen the shear-layer dominated feature for the baseline flows at both angles of attack.  In section \ref{sec:Resovent_vs_CtrlLES}, we will leverage the insights from resolvent analysis and provide guidelines for the design of active separation control.  


\section{Large-eddy simulations of controlled flows}
\label{sec:Control}


In this section, we examine the open-loop separation control using the thermal actuator modeled by equation \ref{eq:ActuatorModel}.  To assess the effectiveness of flow control and to develop a data base to relate flow control to resolvent analysis, we conduct a parametric study with LES over the open-loop actuation frequency $St^+$ and wavenumber $k_z^+$.  We will start our discussion by giving an overall picture of how aerodynamic forces (lift and drag) respond to the chosen $St^+$ and $k_z^+$.  We then analyze the controlled flow fields to correlate the flow physics to the change in the aerodynamic forces and their fluctuation magnitudes. The near-field velocity profiles and surface pressure distributions are also investigated to reveal the mechanism of aerodynamic force modification.  With the results obtained from LES, the control effects will be compared to the results of resolvent analysis in the next section.

\begin{figure}
	\begin{center}
		\vspace{0.10in}
		\begin{overpic}[scale=0.55]{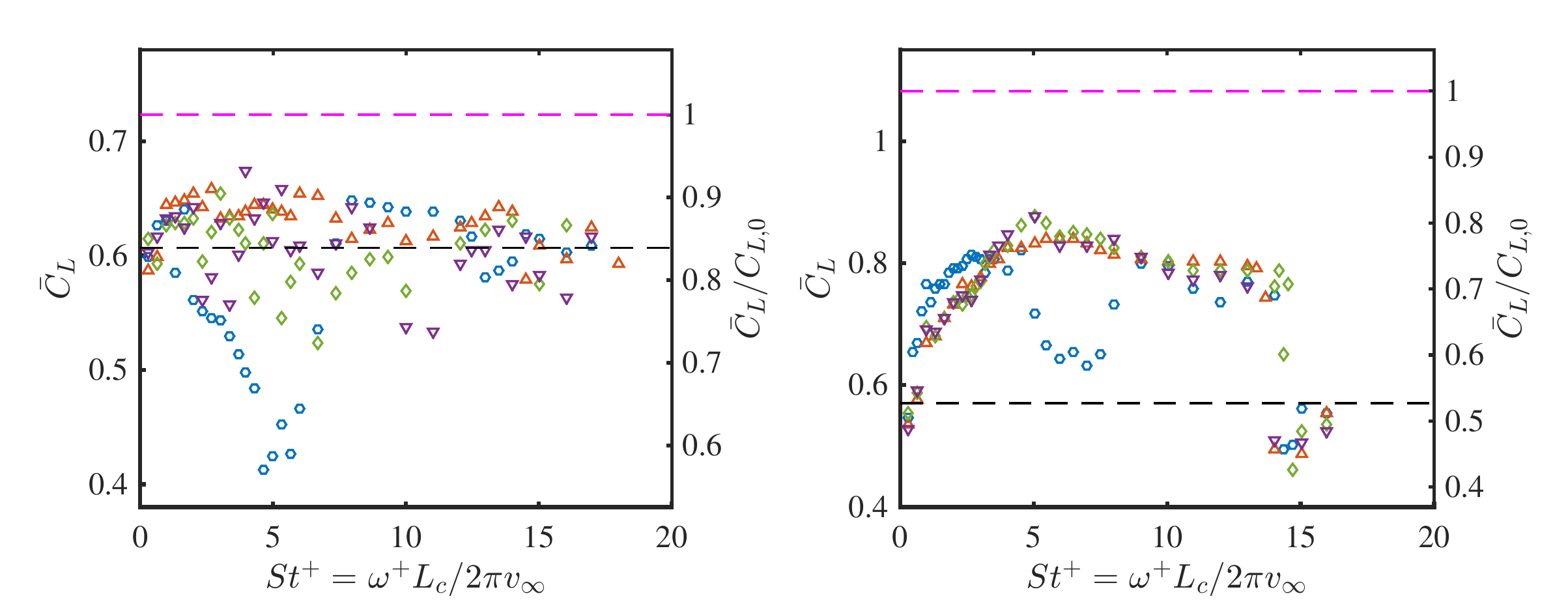}
			\put(20.8, 36.9){(a): $\alpha = 6^\circ$}
			\put(69, 36.9){(b): $\alpha = 9^\circ$}
			\put(32.5, 12.2){\scriptsize	
							\begin{tabular}{c c} 
								\multicolumn{2}{c}{$k_z^+L_c$} \vspace{-0.07in}\\ \hline\vspace{-0.19in}\\
								{${\color{blue2}\boldsymbol{\circ}}$} & $0$ \\ 
								{\tiny${\color{red2}\boldsymbol{\triangle}}$} & $10\pi$ \\ 
								{${\color{green2}\boldsymbol{\diamond}}$} & $20\pi$ \\ 
								{\tiny\color{purple2}$\nabla$} & $40\pi$ \\ 
							\end{tabular}
						}
			\put(66, 11.2){\indexsize Baseline}
			\put(16, 32.6){\indexsize\color{magenta} Potential flow ($C_{L, 0}$)}
			\put(65, 30.85){\indexsize\color{magenta} Potential flow ($C_{L, 0}$)}
		\end{overpic}
	\end{center}
	\caption{\label{fig:Ctrl_Lift} The time-averaged lift coefficients $\bar{C}_L$ of controlled flows for angles of attack of $\alpha = 6^\circ$ (a) and $\alpha = 9^\circ$ (b). In each figure, the black dashed line marks the baseline value for the corresponding angle of attack. The magenta dashed line marks the potential flow lift coefficient computed using panel method.}

	\vspace{0.15in}
	\begin{center}
		\begin{overpic}[scale=0.55]{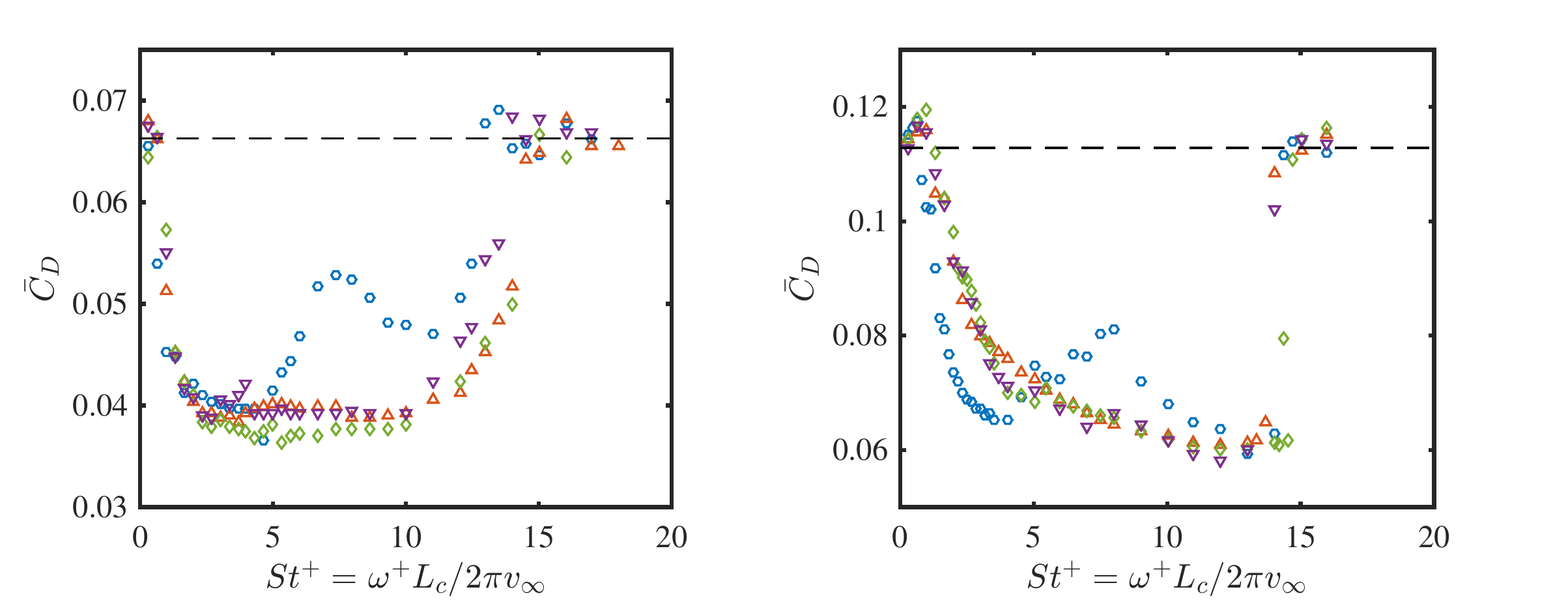}
			\put(20.8, 36.9){(a): $\alpha = 6^\circ$}
			\put(69, 36.9){(b): $\alpha = 9^\circ$}
			\put(17, 30.5){\indexsize Baseline}
			\put(67, 30){\indexsize Baseline}
		\end{overpic}
	\end{center}
	\caption{\label{fig:Ctrl_Drag} The time-averaged drag coefficients $\bar{C}_D$ of controlled flows for $\alpha = 6^\circ$ (a) and $\alpha = 9^\circ$ (b).  The black dashed line marks the baseline value for the corresponding angle of attack. Symbols share the same legend in figure \ref{fig:Ctrl_Lift} (a). }
	
	\vspace{0.15in}
	\begin{center}
		\begin{overpic}[scale=0.55]{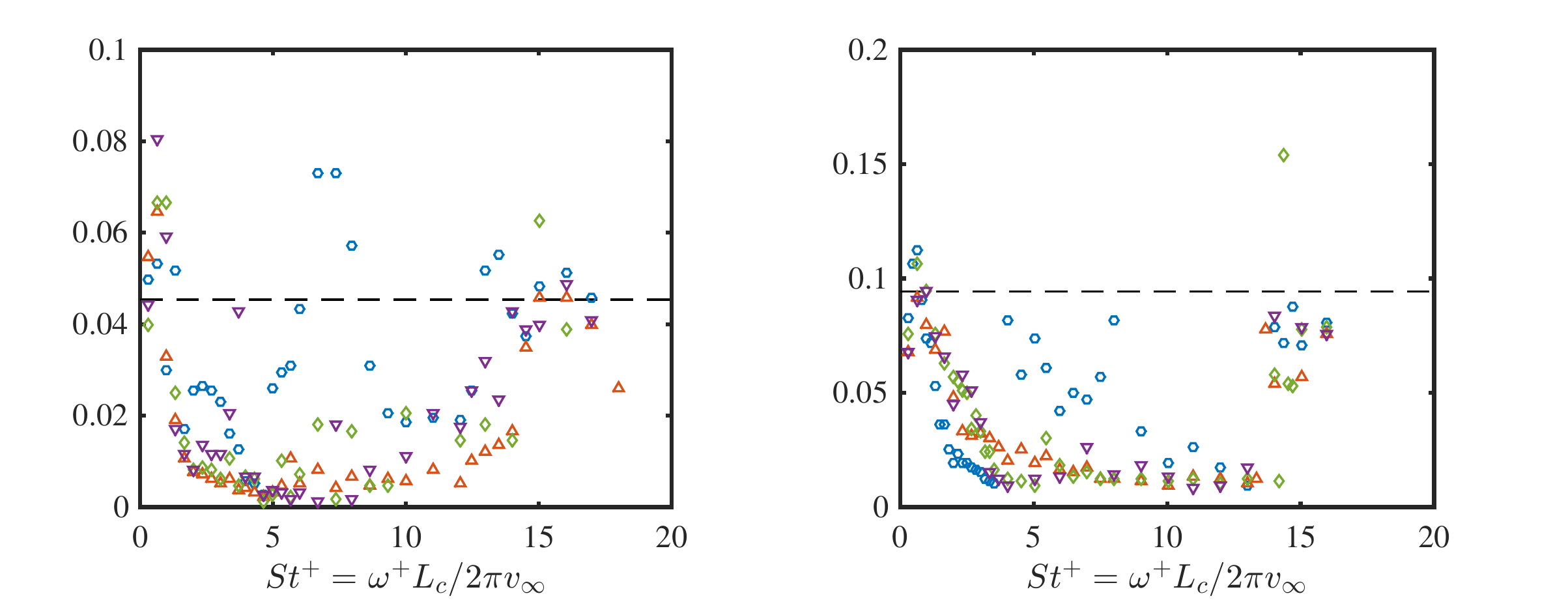}
			\put(2.0, 18){\rotatebox{90}{$C_{L, \text{rms}}$}}
			\put(50.3, 18){\rotatebox{90}{$C_{L, \text{rms}}$}}
			\put(20.8, 36.9){(a): $\alpha = 6^\circ$}
			\put(69, 36.9){(b): $\alpha = 9^\circ$}
		\end{overpic}
	\end{center}
	\caption{\label{fig:A6A9_CL_RMS} Root-mean-square of the lift coefficients $C_{L, \text{rms}}$ of controlled flow for $\alpha = 6^\circ$ (a) and $\alpha = 9^\circ$ (b). The black dashed line marks the baseline value for the corresponding angle of attack.  Symbols share the same legend in figure \ref{fig:Ctrl_Lift} (a).}
\end{figure}

For both angle of attacks, we present the drag and lift coefficients respectively in figures \ref{fig:Ctrl_Lift} and \ref{fig:Ctrl_Drag} for the controlled flows by sweeping through actuation frequencies and wavenumbers.  Let us now direct our attention to the change in lift in figure \ref{fig:Ctrl_Lift}.  While the controlled lift data appears scattered for $\alpha = 6^\circ$, the flow control for $\alpha = 9^\circ$ achieves enhancement in lift by up to $54\%$ with the thermal-based actuation. On the right of both lift plots, we provide an additional scale of $\bar{C}_L/C_{L,0}$ with $C_{L,0}$ being the potential-flow lift for the baseline.   We recall that, while the $\alpha = 9^\circ$ airfoil is in deep stall, the mildly separated baseline flow at $\alpha = 6^\circ$ reattaches and achieves $84\%$ of $C_{L,0}$,  leaving a smaller room for lift enhancement with active flow control.  The lift enhancement at $\alpha = 6^\circ$ does not exhibit a clean trend as at $\alpha = 9^\circ$, which is likely due to difference in the baseline $\bar{C}_L/C_{L,0}$.  However, for both angles of attack, the fluctuation in lift is generally reduced by over $85\%$ with active flow control, as shown in figure \ref{fig:A6A9_CL_RMS}.


\begin{figure}
 	\begin{tabular}{	>{\centering\arraybackslash} m{1.2in}  
						>{\centering\arraybackslash} m{2.1in}
						c
						>{\centering\arraybackslash} m{1.6in}
				   }
	~& Baseline flow ($\alpha = 6^\circ$)\vspace{-0.12in} && \vspace{-0.12in}\\
	\hline	
	{\indexsize\vspace{-0.1in}
			\begin{tabular}{l r} 
				\multicolumn{2}{c}{Forces} \vspace{-0.07in}\\ \hline\vspace{-0.15in}\\
				$\bar{C}_D$ & $0.066$ \\ $\bar{C}_L$ & $0.609$ \\ $\bar{C}_L/\bar{C}_D$ & $9.23$ 
			\end{tabular}}
			& 	\begin{overpic}[height=0.688889in]{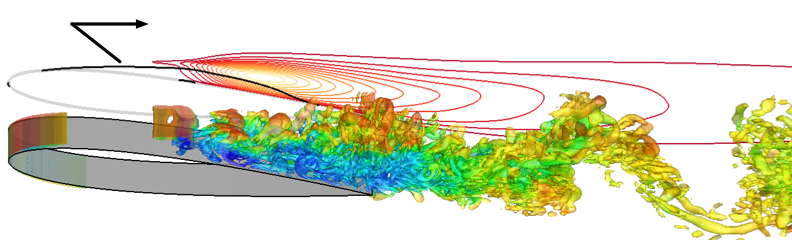}
					\put (20, 26) {\indexsize $\bar{v} = 0$ contour}
					\put (102, 1) {\includegraphics[scale=0.55]{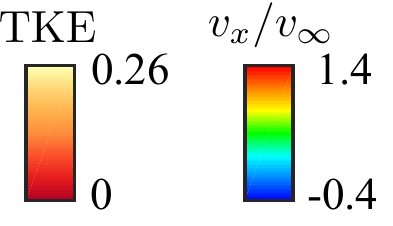}} 
			  	\end{overpic}
			&~~~
    		& \\
			\vspace{0.2in}\\
	Case ($k_z^+L_c = 0$)\vspace{-0.12in} & Controlled flows\vspace{-0.12in} && Resovent mode ($\hat{v}_x$)\vspace{-0.12in}\\
	\hline
	{\indexsize\vspace{-0.1in}
			\begin{tabular}{l r} 
				\multicolumn{2}{l}{~~6-0A: $St^+ = 1.67$} \vspace{-0.07in}\\ \hline\vspace{-0.15in}\\
				$\Delta \bar{C}_D$ & $-38\%$ \\ $\Delta \bar{C}_L$ & $+5.5\%$ \\ $\Delta (\bar{C}_L/\bar{C}_D)$ & $+69\%$ 
			\end{tabular}}
			& 	\begin{overpic}[height=0.62in]{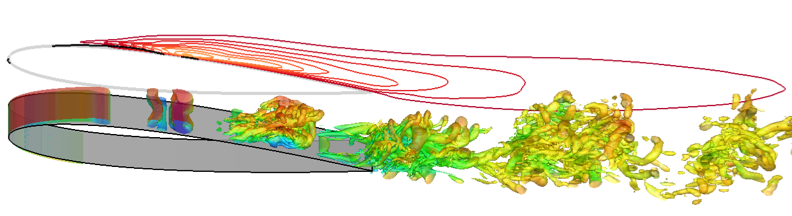}
			  	\end{overpic}
			&~~~
    		& \includegraphics[height=0.62in]{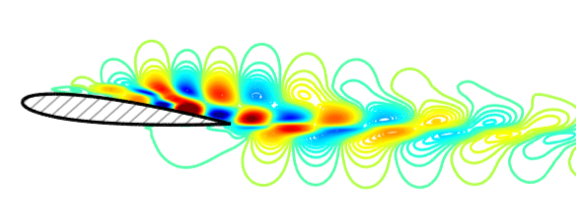}\\ 
	\vspace{-0.06in}&~\vspace{-0.06in}~\vspace{-0.06in}&~\vspace{-0.06in}\\				
	{\indexsize\vspace{-0.1in}
			\begin{tabular}{l r} 
				\multicolumn{2}{l}{~~6-0B: $St^+ = 3$} \vspace{-0.07in}\\ \hline\vspace{-0.15in}\\
				$\Delta \bar{C}_D$ & $-39\%$ \\ $\Delta \bar{C}_L$ & $-11\%$ \\ $\Delta (\bar{C}_L/\bar{C}_D)$ & $+48\%$ 
			\end{tabular}}
			& 	\begin{overpic}[height=0.62in]{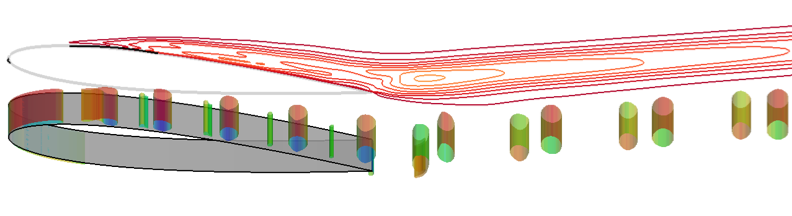}
			  	\end{overpic}
			&~~~
    		& \includegraphics[height=0.62in]{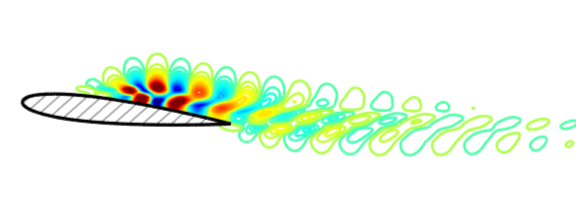}\\
	\vspace{-0.06in}&~\vspace{-0.06in}~\vspace{-0.06in}&~\vspace{-0.06in}\\			
	{\indexsize\vspace{-0.1in}
			\begin{tabular}{l r} 
				\multicolumn{2}{l}{~~6-0C: $St^+ = 7.33$} \vspace{-0.07in}\\ \hline\vspace{-0.15in}\\
				$\Delta \bar{C}_D$ & $-20\%$ \\ $\Delta \bar{C}_L$ & $+0.6\%$ \\ $\Delta (\bar{C}_L/\bar{C}_D)$ & $+25\%$ 
			\end{tabular}}
			& 	\begin{overpic}[height=0.62in]{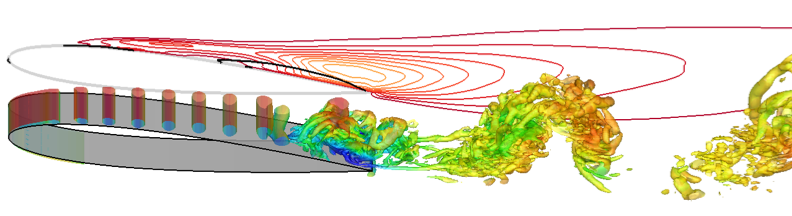}
			  	\end{overpic}
			&~~~
    		& \includegraphics[height=0.62in]{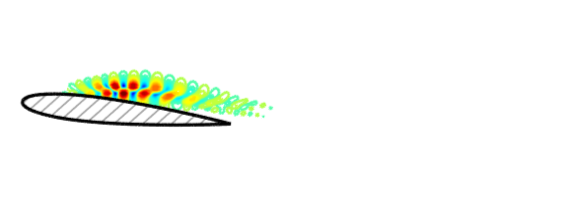}\\
	\vspace{-0.06in}&~\vspace{-0.06in}~\vspace{-0.06in}&~\vspace{-0.06in}\\			
	{\indexsize\vspace{-0.1in}
			\begin{tabular}{l r} 
				\multicolumn{2}{l}{~~6-0D: $St^+ = 11$} \vspace{-0.07in}\\ \hline\vspace{-0.15in}\\
				$\Delta \bar{C}_D$ & $-29\%$ \\ $\Delta \bar{C}_L$ & $+3.8\%$ \\ $\Delta (\bar{C}_L/\bar{C}_D)$ & $+46\%$ 
			\end{tabular}}
			& 	\begin{overpic}[height=0.62in]{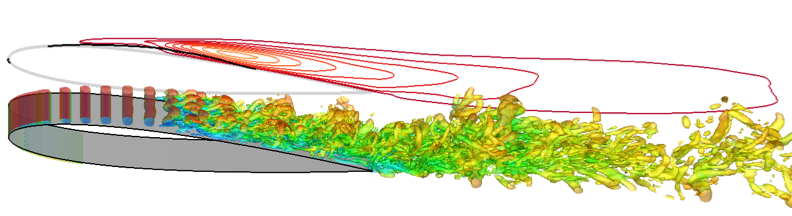}
			  	\end{overpic}
			&~~~
    		& \includegraphics[height=0.62in]{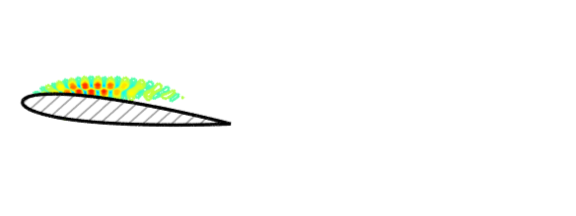}\\
	\vspace{-0.06in}&~\vspace{-0.06in}~\vspace{-0.06in}&~\vspace{-0.06in}\\			
	{\indexsize\vspace{-0.1in}
			\begin{tabular}{l r} 
				\multicolumn{2}{l}{~~6-0E: $St^+ = 15$} \vspace{-0.07in}\\ \hline\vspace{-0.15in}\\
				$\Delta \bar{C}_D$ & $+2.3\%$ \\ $\Delta \bar{C}_L$ & $-3.1\%$ \\ $\Delta (\bar{C}_L/\bar{C}_D)$ & $-5.2\%$ 
			\end{tabular}}
			& 	\begin{overpic}[height=0.62in]{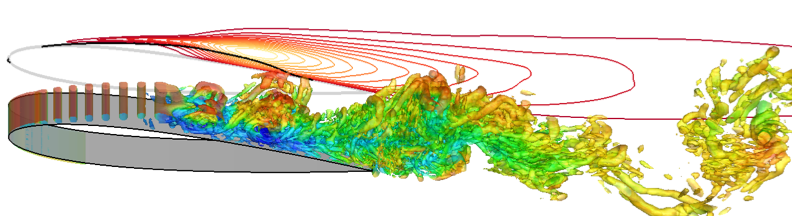}
			  	\end{overpic}
			&~~~
    		& \includegraphics[height=0.62in]{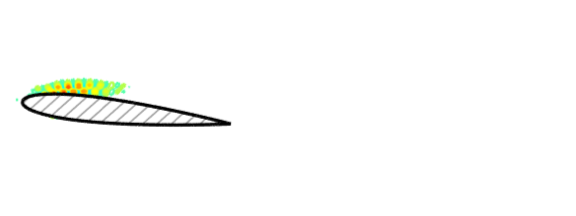}		
	\end{tabular}
	
	\caption{\label{fig:A6_Ctrl_Kz0} Controlled flows for $\alpha = 6^\circ$ with $k_z^+L_c = 0$ and the resolvent response modes (streamwise velocity $\hat{v}_x$) at the corresponding $k_z$-$St$.   The percentage change in the drag coefficient is computed using $\Delta \bar{C}_D = (\bar{C}_{D,\text{control}} - \bar{C}_{D,\text{baseline}})/\bar{C}_{D,\text{baseline}}$ and similarly for lift and lift-to-drag ratio.  Note that the resolvent response modes are computed based on mean baseline flow.  Iso-surface of $QL_c^2/u_\infty^2 = 50$ colored by streamwise velocity is used in the flow visualization. The response modes are obtained with $t_\beta v_\infty/L_c = 5$ and are shown by the contour lines of $\hat{v}_x/|\hat{v}_x|_\infty \in \pm [0.01, 0.9]$.}
\end{figure}

Drag for both angles of attack exhibits significant reduction with active flow control, as shown in figure \ref{fig:Ctrl_Drag}.  The thermal actuation achieves drag reduction of up to $45\%$ for $\alpha = 6^\circ$ and $49\%$ for $\alpha = 9^\circ$.  More importantly, by comparing the drag reduction for both angles of attack, we observe that the effective range of the actuation frequency scales well with the chord-based actuation Strouhal number $St^+ = \omega^+L_c/2\pi v_\infty$.  Significant drag reduction is achieved over $3 \lesssim St^+ \lesssim 15$ but a sharp loss in the drag reduction is observed at $St^+ \approx 15$  for both $\alpha = 6^\circ$ and $9^\circ$.  Beyond $St^+ \gtrsim 15$, control effect diminishes and no control case exhibits apparent change in the aerodynamics forces.  Similar to effective frequency range for drag reduction, the lift fluctuation shown in figure \ref{fig:A6A9_CL_RMS} is also observed to decrease significantly over $3 \lesssim St^+ \lesssim 15$ for both angles of attack.  The frequency scaling with $St^+$ rather than the wake-based Fage--Johansen $St_\alpha^+$ once again implies a shear-layer dominated nature for separation control.

Another interesting feature in the change of aerodynamic forces is the distinct trend exhibited by the $k_z^+L_c = 0$ (\ie 2D actuation) cases.  We observe that drag, while still below the baseline value, increases near $St^+ \approx 7.5$ for both angles of attack when using $k_z^+L_c = 0$.  When a spanwise variation ($k_z^+L_c > 0$) is introduced to the actuation profile, such increase in drag is absent from the intermediate range of actuation frequency.  In fact, little difference can be observed in the change of aerodynamics forces with $k_z^+L_c = 10\pi$, $20\pi$ and $40\pi$ using the actuation power $E^+ = 0.0902$ in \ref{eq:NormalizedForcingPower} for the present study.

To reveal the cause for the distinctive trend in drag with $k_z^+L_c = 0$, we visualize the instantaneous flows for representative cases of $\alpha = 6^\circ$ in figure \ref{fig:A6_Ctrl_Kz0}.  Behind the $Q$-criterion visualization, we also show the TKE contour as well as a black curve that marks $\bar{v}_x = 0$ to indicate the separation region for each case.  Along with the flow visualization, the percentage change of aerodynamic forces is tabulated on the left.  In all cases, we find that the thermal actuation is able to excite the roll-up of the shear layer.  The periodic thermal input chops the shear layer at the actuation frequency.   Each chopping forms a compact 2D spanwise vortex, advecting along the suction side of the airfoil.  These vortical structures enhance momentum mixing and entrain the free-stream.  Similar to the discussion in \citet{Glezer:AIAAJ2005}, the entrainment results in the Coand\u{a}-like effect and suppresses flow separation, which can be seen in cases 6-0A to 6-0D by comparing the $\bar{v}_x = 0$ contours to that of the baseline.  In what follows, we split the discussion into four ranges of frequencies according to the distinctive change in drag as well as similar flow responses to the actuation.

{\it Frequency range $0.6 \lesssim St^+ \lesssim 4.33$ (represented by cases 6-0A and 6-0B)}\\
In this frequency range, the flow response is characterized by the coupling between the roll-up of the shear layer over the airfoil and the vortex shedding in the wake.  Particularly for case 6-0B, we observe that the formation of strong spanwise vortices advect farther downstream into the wake, diminishing the development of 3D structures and fully laminarizing the flow.   Such a global laminarization is observed over $2 \lesssim St^+ \lesssim 4.33$ with 2D actuation for $\alpha = 6^\circ$.   Although such flow laminarization is not observed in $0.6 \lesssim St^+ \lesssim 1.67$, the coupling between the excited shear-layer roll-up and the wake shedding holds for this frequency range.  In this frequency range of $0.6 \lesssim St^+ \lesssim 4.33$, the drag generally decreases with increasing actuation frequency with the coupling of instabilities.

{\it Frequency range: $4.67 \lesssim St^+ \lesssim 7.33$ (represented by case 6-0C)}\\
In this range, the pairing between the spanwise vortices takes place near the trailing edge.  Though the flow is reattached before mid-chord due to actuation, the vortex pairing process results in trailing-edge separation and causes the drag to increase.  The pairing process also stimulates the laminar-turbulent transition and increase TKE near the trailing edge. The wake also becomes turbulent.  The drag reaches the local maximum with $St^+ \approx 7.33$ over the varied actuation frequency in this range.

{\it Frequency range: $8 \lesssim St^+ \lesssim 11$ (represented by case 6-0D)}\\
The flow response in this frequency range is characterized by the break-up of the spanwise vortices over the suction surface, accompanied by the laminar-turbulent transition before the pairing process takes place.  It is also marked by the removal of von K\'arm\'an shedding structures that are prominent in other regimes as well as the baseline.  The break-up of the spanwise vortices occurs near the mid-chord with increased TKE, after which turbulent structures covers the rest of the suction surface.  Compared to the baseline flow, these turbulent structures in case 6-0D possess higher streamwise momentum and advect close to the suction surface.  The break-up process allows for 3D mixing and keeps high-momentum turbulent structures staying adjacent to the suction surface,  suppressing the trailing-edge separation.  As a result, the drag further decreases and reaches the local minimum at case 6-0D with $St^+ = 11$.  

{\it Frequency range: $St^+ \gtrsim 11$ (represented by case 6-0E)}\\
The drag increases beyond $St^+ \gtrsim 11$. In this range, the spanwise vortices are not sufficiently large and strong to induce enough momentum mixing for free-stream entrainment.  By comparing the flow fields of 6-0E to that of the baseline, the appearance of the actuation induced spanwise vortices are still visibly clear.  However, while these smaller spanwise structures advecting downstream, they also move away from the suction surface, as oppose to their trajectories in cases 6-0A to 6-0D.  Even though the actuation still excites the shear-layer roll-up, it does not effectively entrain the free-stream momentum and leads to the drag to remain at the baseline level near $St^+ \approx 15$.

Along with the above observations made from the controlled flows, we also examine the response modes from resolvent analysis in figure \ref{fig:A6_Ctrl_Kz0}.  We remind that these response modes are obtained from the resolvent analysis on the mean baseline flow.  The response mode is provided at the frequency used for the unsteady actuation in each corresponding control cases in the middle column.   For case 6-0A and 6-0B, the corresponding response structure develops from the shear layer above the suction surface and extends farther into the wake.  For higher frequencies, the streamwise extent of the modal structure reduces to the shear layer, starting from the mode at $St = 7.33$ (case 6-0C) and for higher frequency cases.  According to these observations, we see that the response mode structure is capable of providing insights on the global flow receptivity to perturbation of specified frequency.  When the modal structures cover both the shear layer and the wake, in corresponding controlled flows we observe that the perturbation amplified through the shear layer also advects into the wake and stimulates the shedding instability.  Similarly, when the modal structures appear only within the shear layer, the corresponding controlled flow shows that the actuation-induced spanwise vortices either merge near the trailing edge or break up over the airfoil, never able to advect into the wake while remaining compact.  Such a qualitative agreement between resolvent analysis and controlled flows has made it promising for resolvent analysis to provide quantitative design guidelines.  We will further elaborate on this point in the next section.


\begin{figure}
 	\begin{tabular}{	>{\centering\arraybackslash} m{0.2in}  
						>{\centering\arraybackslash} m{1.6in}
						>{\centering\arraybackslash} m{1.6in}
						>{\centering\arraybackslash} m{1.6in}}
	$St^+$\vspace{-0.12in}& $k_z^+L_c = 10\pi$\vspace{-0.12in} & $k_z^+L_c = 20\pi$\vspace{-0.12in} & $k_z^+L_c = 40\pi$\vspace{-0.12in}\\
    \hline\\~\vspace{+0.07in}\\
    $4$
    		&	\begin{overpic}[width=1.6in]{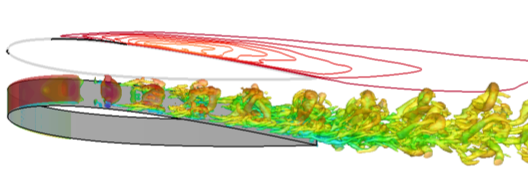}
					\put (1, 42) {\scriptsize	\begin{tabular}{c} 
													Case 6-1A \vspace{-0.07in}\\ \hline\vspace{-0.15in}\\
													$\Delta \bar{C}_D = -41\%$, $\Delta \bar{C}_L = +5.1\%$ 
												\end{tabular}}
				\end{overpic}
			&	\begin{overpic}[width=1.6in]{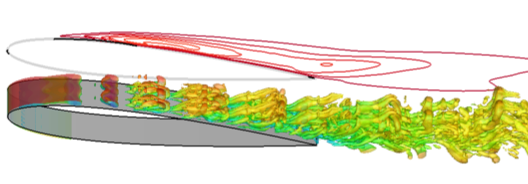}
					\put (1, 42) {\scriptsize	\begin{tabular}{c} 
													Case 6-2A \vspace{-0.07in}\\ \hline\vspace{-0.15in}\\
													$\Delta \bar{C}_D = -43\%$, $\Delta \bar{C}_L = +1.2\%$ 
												\end{tabular}}
				\end{overpic}
			&	\begin{overpic}[width=1.6in]{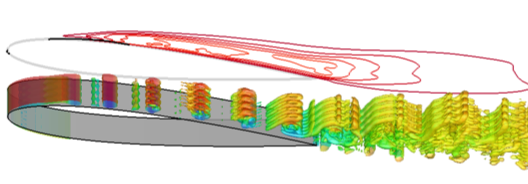}
					\put (1, 42) {\scriptsize	\begin{tabular}{c} 
													Case 6-4A \vspace{-0.07in}\\ \hline\vspace{-0.15in}\\
													$\Delta \bar{C}_D = -36\%$, $\Delta \bar{C}_L = +11\%$ 
												\end{tabular}}
				\end{overpic}\\~\vspace{+0.3in}\\
    $6$
    		&	\begin{overpic}[width=1.6in]{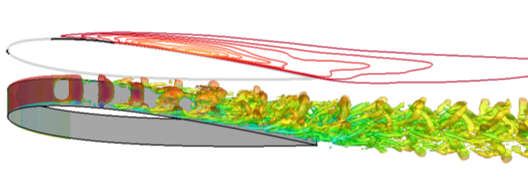}
					\put (1, 42) {\scriptsize	\begin{tabular}{c} 
													Case 6-1B \vspace{-0.07in}\\ \hline\vspace{-0.15in}\\
													$\Delta \bar{C}_D = -40\%$, $\Delta \bar{C}_L = +7.8\%$
												\end{tabular}}
				\end{overpic}
			&	\begin{overpic}[width=1.6in]{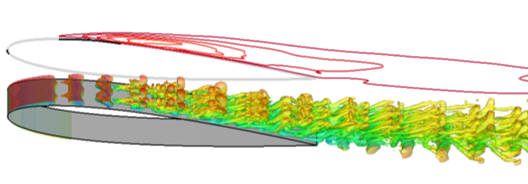}
					\put (1, 42) {\scriptsize	\begin{tabular}{c} 
													Case 6-2B \vspace{-0.07in}\\ \hline\vspace{-0.15in}\\
													$\Delta \bar{C}_D = -44\%$, $\Delta \bar{C}_L = -2.2\%$
												\end{tabular}}
				\end{overpic}
			&	\begin{overpic}[width=1.6in]{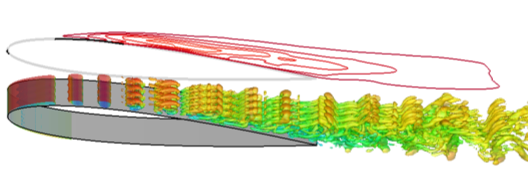}
					\put (1, 42) {\scriptsize	\begin{tabular}{c} 
													Case 6-4B \vspace{-0.07in}\\ \hline\vspace{-0.15in}\\
													$\Delta \bar{C}_D = -41\%$, $\Delta \bar{C}_L = +0.3\%$  
												\end{tabular}}
				\end{overpic}\\~\vspace{+0.3in}\\
    $12$
    		&	\begin{overpic}[width=1.6in]{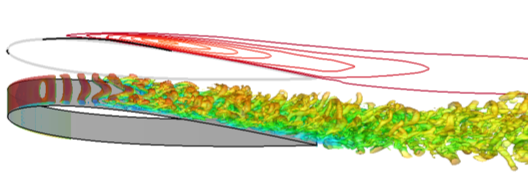}
					\put (1, 42) {\scriptsize	\begin{tabular}{c} 
													Case 6-1C \vspace{-0.07in}\\ \hline\vspace{-0.15in}\\
													$\Delta \bar{C}_D = -38\%$, $\Delta \bar{C}_L = +2.8\%$
												\end{tabular}}
				\end{overpic}
			&	\begin{overpic}[width=1.6in]{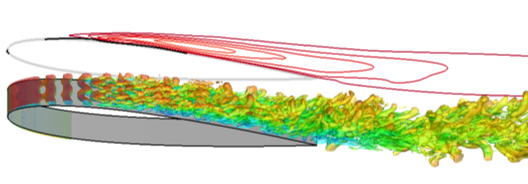}
					\put (1, 42) {\scriptsize	\begin{tabular}{c} 
													Case 6-2C \vspace{-0.07in}\\ \hline\vspace{-0.15in}\\
													$\Delta \bar{C}_D = -37\%$, $\Delta \bar{C}_L = +0.3\%$ 
												\end{tabular}}
				\end{overpic}
			&	\begin{overpic}[width=1.6in]{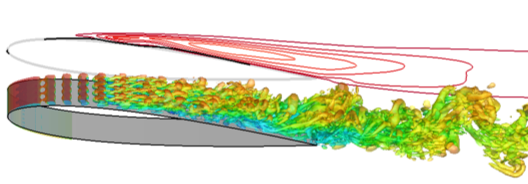}
					\put (1, 42) {\scriptsize	\begin{tabular}{c} 
													Case 6-4C \vspace{-0.07in}\\ \hline\vspace{-0.15in}\\
													$\Delta \bar{C}_D = -30\%$, $\Delta \bar{C}_L = -2.4\%$ 
												\end{tabular}}
				\end{overpic}\\~\vspace{+0.3in}\\ 	
    $15$
    		&	\begin{overpic}[width=1.6in]{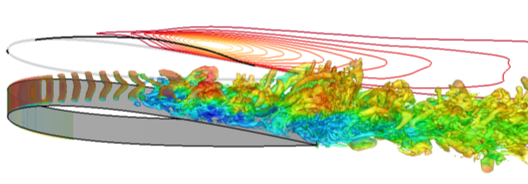}
					\put (0.5, 42) {\scriptsize	\begin{tabular}{c} 
													Case 6-1D \vspace{-0.07in}\\ \hline\vspace{-0.15in}\\
													$\Delta \bar{C}_D = -2.2\%$, $\Delta \bar{C}_L = +0.4\%$ 
												\end{tabular}}
				\end{overpic}
			&	\begin{overpic}[width=1.6in]{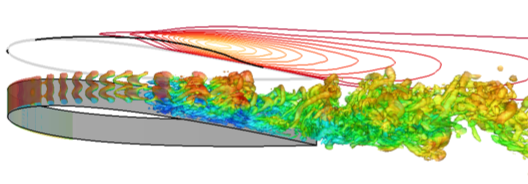}
					\put (0.5, 42) {\scriptsize	\begin{tabular}{c} 
													Case 6-2D \vspace{-0.07in}\\ \hline\vspace{-0.15in}\\
													$\Delta \bar{C}_D = -2.8\%$, $\Delta \bar{C}_L = -0.4\%$ 
												\end{tabular}}
				\end{overpic}
			&	\begin{overpic}[width=1.6in]{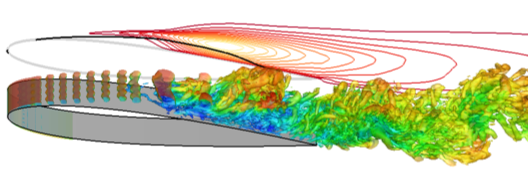}
					\put (0.5, 42) {\scriptsize	\begin{tabular}{c} 
													Case 6-4D \vspace{-0.07in}\\ \hline\vspace{-0.15in}\\
													$\Delta \bar{C}_D = +2.8\%$, $\Delta \bar{C}_L = -4.1\%$ 
												\end{tabular}}
				\end{overpic}\\
				~&\multicolumn{3}{c}{\includegraphics[scale=0.56]{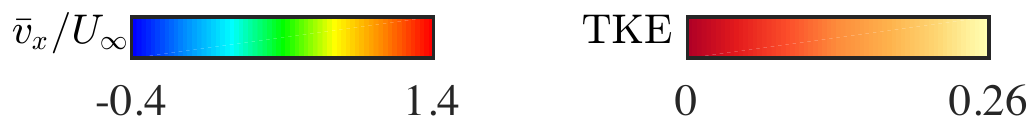}}\\
	\end{tabular}
	\caption{\label{fig:A6_Ctrl_Kz} Instantaneous flow fields and TKE (in the background) for controlled cases with $k_z^+L_c > 0$ of $\alpha = 6^\circ$. Iso-surface of $QL_c^2/u_\infty^2 = 50$ colored by streamwise velocity is utilized in the flow visualization.}
\end{figure} 

Continuing the discussion for control cases at $\alpha = 6^\circ$, we present the flow visualization for cases where a spanwise variation is introduced into the actuation with $k_z^+L_c > 0$ in figure \ref{fig:A6_Ctrl_Kz}.  We also refer to the drag value reported in figure \ref{fig:Ctrl_Drag} (a) for the controlled cases.  For all $k_z^+L_c > 0$ examined, the drag decrease reaches $\bar{C}_D \approx 0.04$ at $St^+ \approx 3$ and continues to maintain this level of approximately $40\%$ drag reduction from the baseline.  The control effect degrades at $St^+ \approx 10$ and returns to the baseline drag level by $St^+ \approx 15$. Similar to the $k_z^+L_c = 0$ cases, the thermal actuation generates spanwise vortices near the leading edge, which can be seen in the flow visualization. These vortices carry the spanwise variation introduced by the actuation input for the actuation wavenumbers of $k_z^+L_c = 10\pi$, $20\pi$ and $40\pi$ (respectively corresponding to one, two and four waves across the spanwise extent in the current LES).   These spanwise vortices advect along the suction surface and evolve into turbulent structures near mid-chord.  Similar to the comments we made previously for case 6-0D on the effect of mid-chord transition, the same mechanism holds here for drag reduction in all effective cases with $k_z^+L_c > 0$.  Therefore, as opposed to the controlled cases with $k_z^+L_c = 0$, drag reduction achieved from $k_z^+L_c > 0$ remains at a comparable level over the intermediate actuation frequencies. 

\begin{figure}
	\vspace{0.1in}
 	\begin{tabular}{	>{\centering\arraybackslash} m{0.2in}  
						>{\centering\arraybackslash} m{1.6in}
						>{\centering\arraybackslash} m{1.6in}
						>{\centering\arraybackslash} m{1.6in}}
	~& \multicolumn{3}{c}{Baseline LES ($\alpha = 9^\circ$)\vspace{-0.05in}}\\
	\hline	
	~& \multicolumn{3}{c}{\hspace{-0.5in}\hspace{0.5in}
			\begin{overpic}[height=0.688889in]{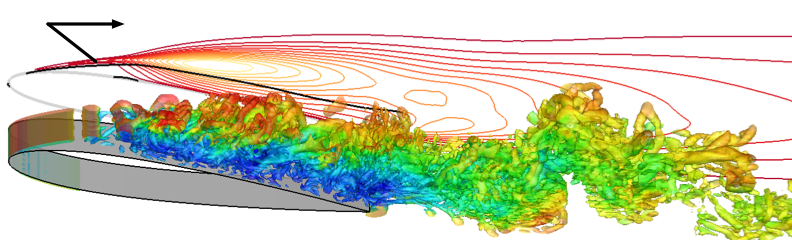}
				\put (18, 27) {\indexsize $\bar{v} = 0$ contour}
				\put (-45, 14) {\scriptsize\vspace{-0.1in}
								\begin{tabular}{l r} 
									\multicolumn{2}{c}{Forces} \vspace{-0.07in}\\ \hline\vspace{-0.15in}\\
									$\bar{C}_D$ & $0.113$ \\ $\bar{C}_L$ & $0.570$ \\ $\bar{C}_L/\bar{C}_D$ & $5.04$ 
								\end{tabular}}
				\put (102, 1) {\includegraphics[scale=0.55]{Figs/FlowVis_ColorBar}} 
			\end{overpic}}
			\vspace{0.05in}\\
	$St^+$\vspace{-0.12in}& $k_z^+L_c = 0$\vspace{-0.12in} & $k_z^+L_c = 10\pi$\vspace{-0.12in} & $k_z^+L_c = 20\pi$\vspace{-0.12in}\\
    \hline\\~\vspace{+0.07in}\\
    $2$
    		&	\begin{overpic}[width=1.6in]{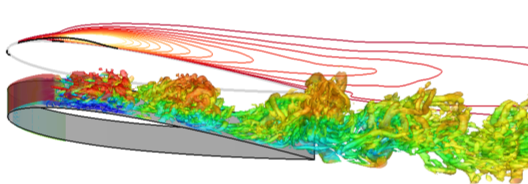}
					\put (1, 42) {\scriptsize	\begin{tabular}{c} 
													Case 9-0A \vspace{-0.07in}\\ \hline\vspace{-0.15in}\\
													$\Delta \bar{C}_D = -35\%$, $\Delta \bar{C}_L = +39\%$  
												\end{tabular}}
				\end{overpic}
			&	\begin{overpic}[width=1.6in]{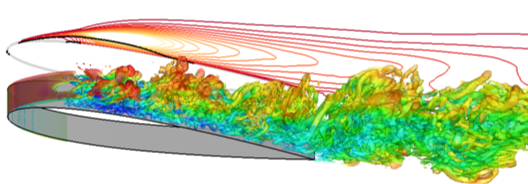}
					\put (1, 42) {\scriptsize	\begin{tabular}{c} 
													Case 9-1A \vspace{-0.07in}\\ \hline\vspace{-0.15in}\\
													$\Delta \bar{C}_D = -19\%$, $\Delta \bar{C}_L = +28\%$ 
												\end{tabular}}
				\end{overpic}
			&	\begin{overpic}[width=1.6in]{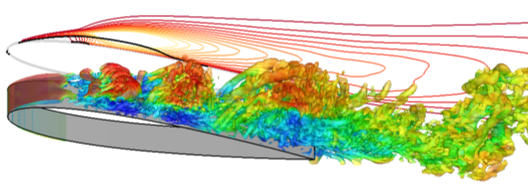}
					\put (1, 42) {\scriptsize	\begin{tabular}{c} 
													Case 9-2A \vspace{-0.07in}\\ \hline\vspace{-0.15in}\\
													$\Delta \bar{C}_D = -17\%$, $\Delta \bar{C}_L = +29\%$ 
												\end{tabular}}
				\end{overpic}\\~\vspace{+0.3in}\\
    $5.5$
    		&	\begin{overpic}[width=1.6in]{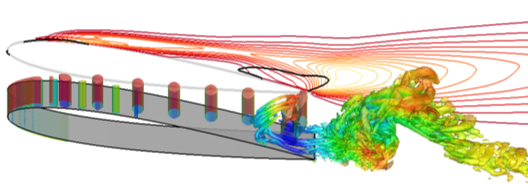}
					\put (1, 42) {\scriptsize	\begin{tabular}{c} 
													Case 9-0B \vspace{-0.07in}\\ \hline\vspace{-0.15in}\\
													$\Delta \bar{C}_D = -35\%$, $\Delta \bar{C}_L = +16\%$ 
												\end{tabular}}
				\end{overpic}
			&	\begin{overpic}[width=1.6in]{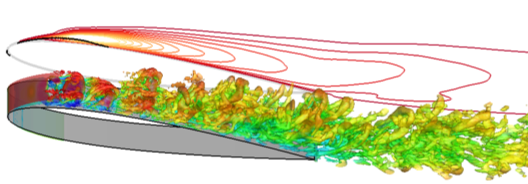}
					\put (1, 42) {\scriptsize	\begin{tabular}{c} 
													Case 9-1B \vspace{-0.07in}\\ \hline\vspace{-0.15in}\\
													$\Delta \bar{C}_D = -38\%$, $\Delta \bar{C}_L = +47\%$ 
												\end{tabular}}
				\end{overpic}
			&	\begin{overpic}[width=1.6in]{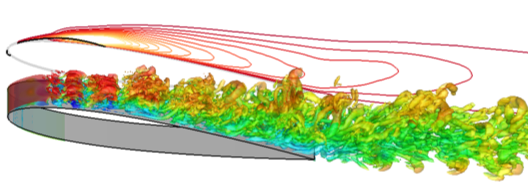}
					\put (1, 42) {\scriptsize	\begin{tabular}{c} 
													Case 9-2B \vspace{-0.07in}\\ \hline\vspace{-0.15in}\\
													$\Delta \bar{C}_D = -37\%$, $\Delta \bar{C}_L = +53\%$  
												\end{tabular}}
				\end{overpic}\\~\vspace{+0.3in}\\
    $12$
    		&	\begin{overpic}[width=1.6in]{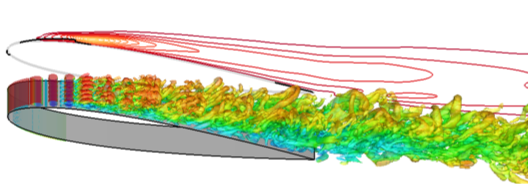}
					\put (1, 42) {\scriptsize	\begin{tabular}{c} 
													Case 9-0C \vspace{-0.07in}\\ \hline\vspace{-0.15in}\\
													$\Delta \bar{C}_D = -43\%$, $\Delta \bar{C}_L = +28\%$ 
												\end{tabular}}
				\end{overpic}
			&	\begin{overpic}[width=1.6in]{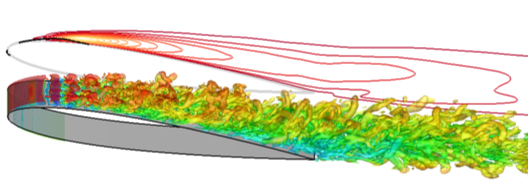}
					\put (1, 42) {\scriptsize	\begin{tabular}{c} 
													Case 9-1C \vspace{-0.07in}\\ \hline\vspace{-0.15in}\\
													$\Delta \bar{C}_D = -46\%$, $\Delta \bar{C}_L = +41\%$ 
												\end{tabular}}
				\end{overpic}
			&	\begin{overpic}[width=1.6in]{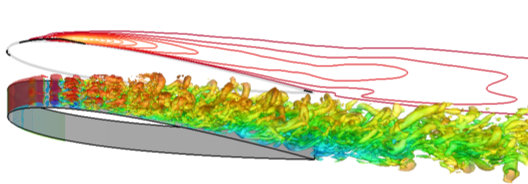}
					\put (1, 42) {\scriptsize	\begin{tabular}{c} 
													Case 9-2C \vspace{-0.07in}\\ \hline\vspace{-0.15in}\\
													$\Delta \bar{C}_D = -49\%$, $\Delta \bar{C}_L = +37\%$ 
												\end{tabular}}
				\end{overpic}\\~\vspace{+0.3in}\\ 	
    $16$
    		&	\begin{overpic}[width=1.6in]{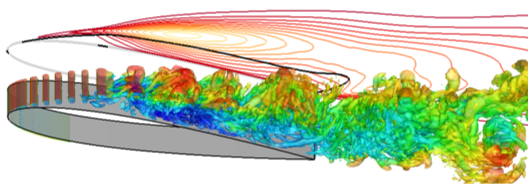}
					\put (0.5, 42) {\scriptsize	\begin{tabular}{c} 
													Case 9-0D \vspace{-0.07in}\\ \hline\vspace{-0.15in}\\
													$\Delta \bar{C}_D = -1.7\%$, $\Delta \bar{C}_L = -3.5\%$ 
												\end{tabular}}
				\end{overpic}
			&	\begin{overpic}[width=1.6in]{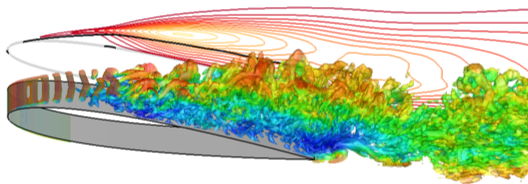}
					\put (0.5, 42) {\scriptsize	\begin{tabular}{c} 
													Case 9-1D \vspace{-0.07in}\\ \hline\vspace{-0.15in}\\
													$\Delta \bar{C}_D = +1.9\%$, $\Delta \bar{C}_L = -3.0\%$ 
												\end{tabular}}
				\end{overpic}
			&	\begin{overpic}[width=1.6in]{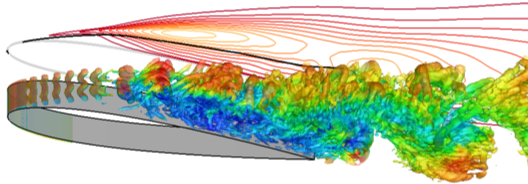}
					\put (0.5, 42) {\scriptsize	\begin{tabular}{c} 
													Case 9-2D \vspace{-0.07in}\\ \hline\vspace{-0.15in}\\
													$\Delta \bar{C}_D = +0.7\%$, $\Delta \bar{C}_L = -7.4\%$ 
												\end{tabular}}
				\end{overpic}\\~\vspace{+0.1in}\\ 	
	\end{tabular}
	\caption{\label{fig:A9Ctrl} Instantaneous flow fields and TKE (in the background) for baseline and controlled cases of $\alpha = 9^\circ$. Iso-surface of $QL_c^2/u_\infty^2 = 50$ colored by streamwise velocity is utilized in the flow visualization.}
\end{figure}

Analogous to the discussions on $\alpha = 6^\circ$ cases, we show representative control cases at $9^\circ$ with their flow visualizations in figure \ref{fig:A9Ctrl}.  A qualitative difference between the controlled flows of $\alpha = 9^\circ$ and those of $6^\circ$ is that the global laminarization by the thermal actuation is not observed in any examined controlled cases with $k_z^+L_c = 0$ for $\alpha = 9^\circ$.  Apart from these two differences, similar flow physics associated with the change in drag for $\alpha = 6^\circ$ also holds for the $\alpha = 9^\circ$ controlled cases.  Cases 9-0A, 9-0B, 9-0C and 9-0D are respectively associated with four frequency ranges as discussed for $\alpha = 6^\circ$ with $k_z^+L_c = 0$ in figure \ref{fig:A6_Ctrl_Kz0}.  In each frequency range, similar trend in the drag reduction is observed with the use of 2D actuation in both $\alpha = 6^\circ$ and $9^\circ$ controlled cases.  For $\alpha = 9^\circ$, the partial laminarization of the flow by 2D actuation is only observed over the suction surface in $5 \lesssim St^+ \lesssim 7.5$.  Along with drag reduction, significant lift enhancement from baseline flow of $\alpha = 9^\circ$ is also observed in cases where separation is effectively suppressed by the thermal actuation.  Suppression of separation can be attributed to the accelerated laminar-turbulent transition over separation bubble that occurs immediately after the shear-layer roll-up.  In the case of $St^+ = 16$, we observe that the small spanwise vortices depart from the suction surface and fails to suppress flow separation. As a consequence, the lift and drag returns to the baseline level at $St^+ \approx 15$.  Qualitative agreement between the controlled flows and the resolvent response modes are also found for $\alpha = 9^\circ$ cases, similar to the the discussions for $\alpha = 6^\circ$.

\begin{figure}
	\vspace{0.6in}
	\begin{center}
		\begin{overpic}[scale=0.55]{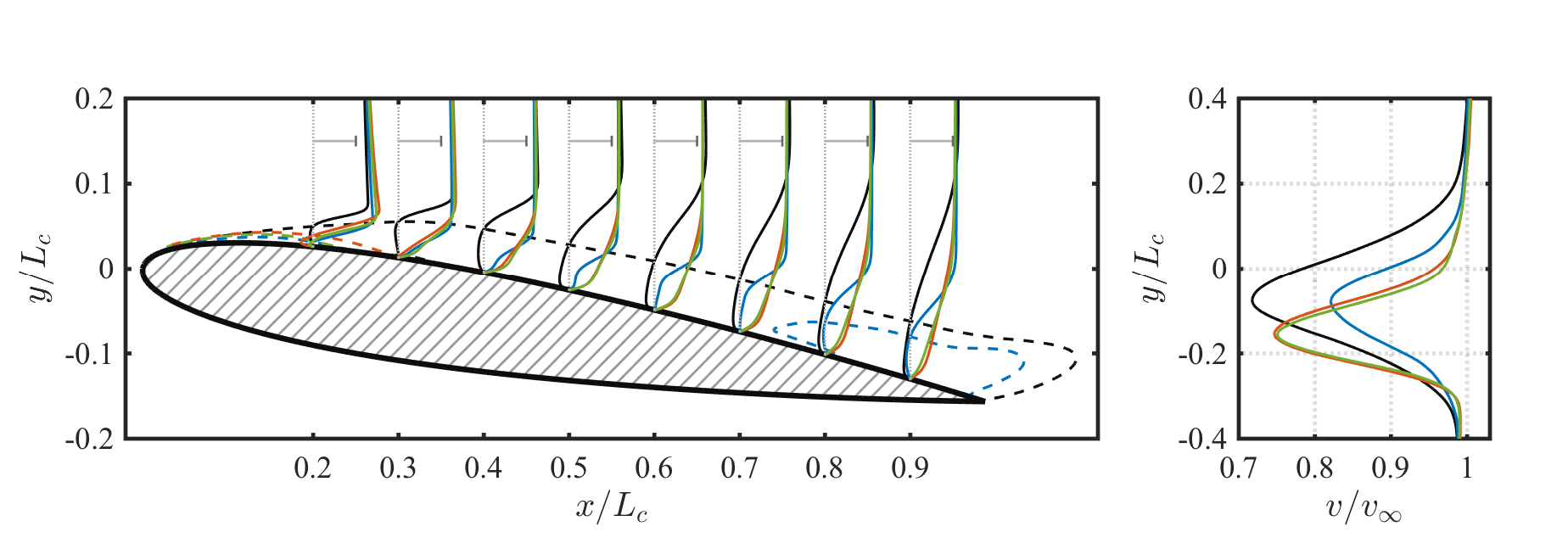}
			\put (11, 25) {(a)}
			\put (81, 25) {(b)}
			\put (20.2, 27) {\scriptsize $v_\infty$}
			
			\put (7, 37) {\indexsize	\begin{tabular}{l r}
											\multicolumn{2}{c}{\color{black}\solid Baseline}\\
											\multicolumn{2}{c}{\color{white}$St^+$, $k_z^+$}
											\vspace{-0.07in}\\ \hline\vspace{-0.15in}\\
											$\bar{C}_L$ & $0.113$ \\ $\bar{C}_D$ & $0.570$ \\ $C_{L,\text{rms}}$ & $0.094$
										\end{tabular}}
			\put (25, 37) {\indexsize	\begin{tabular}{l r} 
											\multicolumn{2}{c}{\color{blue2}\solid 9-0B}\\
											\multicolumn{2}{c}{\color{blue2}$St^+: 5.5$, $k_z^+L_c: 0$}
											\vspace{-0.07in}\\ \hline\vspace{-0.15in}\\
											$\Delta \bar{C}_L$ & $+16\%$ \\ $\Delta \bar{C}_D$ & $-35\%$\\ $\Delta C_{L,\text{rms}}$ & $-31\%$
										\end{tabular}}
			\put (48, 37) {\indexsize	\begin{tabular}{l r} 
											\multicolumn{2}{c}{\color{red2}\solid 9-1B}\\
											\multicolumn{2}{c}{\color{red2}$St^+: 5.5$, $k_z^+L_c: 10\pi$}
											\vspace{-0.07in}\\ \hline\vspace{-0.15in}\\
											$\Delta \bar{C}_L$ & $+47\%$ \\ $\Delta \bar{C}_D$ & $-38\%$\\ $\Delta C_{L,\text{rms}}$ & $-77\%$
										\end{tabular}}
			\put (73.5, 37) {\indexsize	\begin{tabular}{l r} 
											\multicolumn{2}{c}{\color{green2}\solid 9-1C}\\
											\multicolumn{2}{c}{\color{green2}$St^+: 12$, $k_z^+L_c: 10\pi$}
											\vspace{-0.07in}\\ \hline\vspace{-0.15in}\\
											$\Delta \bar{C}_L$ & $+41\%$ \\ $\Delta \bar{C}_D$ & $-46\%$\\ $\Delta C_{L,\text{rms}}$ & $-69\%$
										\end{tabular}}
		\end{overpic}
	\end{center}
	\caption{\label{fig:A9_Ctrl_Cp_UProf}Time- and spanwise-averaged streamwise velocity profiles over the airfoil (a) and in near-wake at $x/L_c = 2.0$ (b). Dashed curves mark the contours of $\bar{v}_x = 0$ for the four cases.} 
	\vspace{0.3in}
	\begin{center}
		\begin{overpic}[scale=0.55]{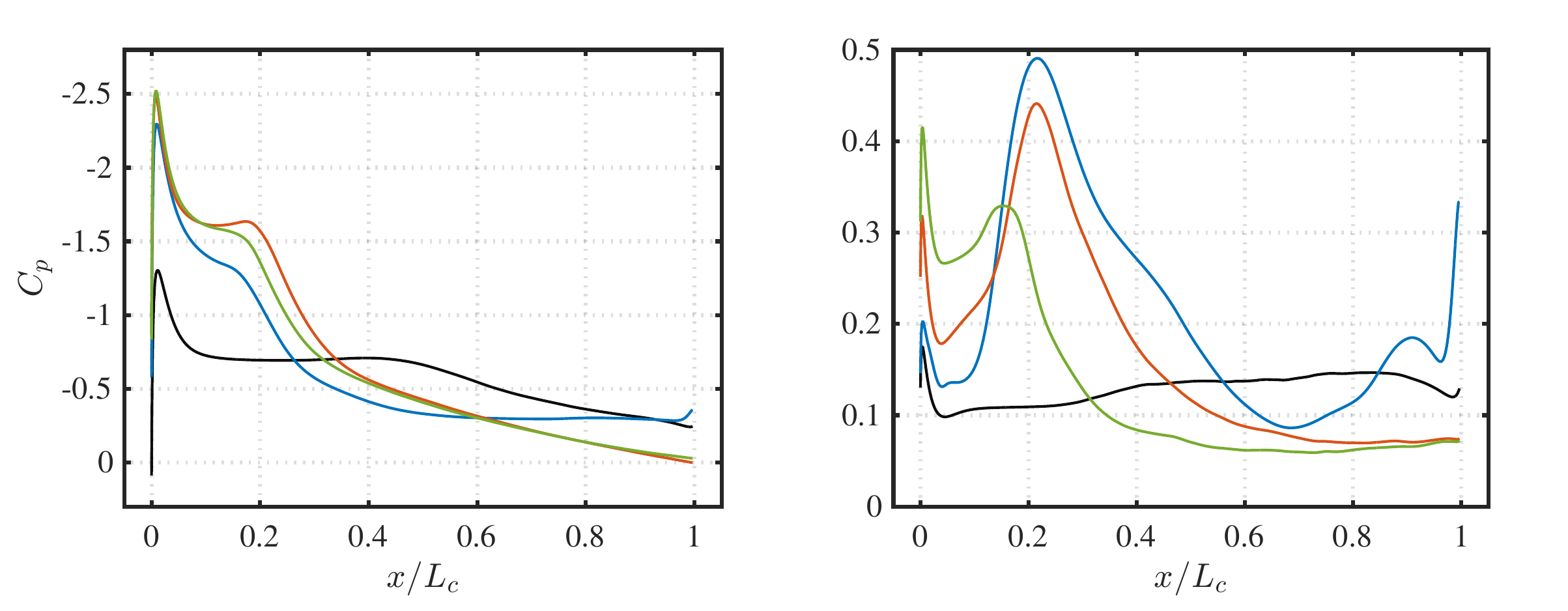}
			\put(0, 35){(a)}
			\put(49, 35){(b)}
			\put(50, 18){\rotatebox{90}{$C_{p, \text{rms}}$}}
			\put (27, 28) {\indexsize	\begin{tabular}{c l}
											{\color{black}\solid} & {\color{black}Baseline}\\
											{\color{blue2}\solid} & {\color{blue2}Case 9-0B}\\
											{\color{red2}\solid} & {\color{red2}Case 9-1B}\\
											{\color{green2}\solid} & {\color{green2}Case 9-1C}\\
										\end{tabular}}
					
		\end{overpic}
	\end{center}
	\caption{\label{fig:A9_Cp_CpRMS}Suction-surface pressure profiles (a) and their root-mean-square (b) of controlled flows and baseline for $\alpha = 9^\circ$.  Legends follow figure \ref{fig:A9_Ctrl_Cp_UProf}.}
\end{figure}

To provide further insights into the mechanism for suppressing flow separation, we examine three selective control cases from figure \ref{fig:A9Ctrl} along with the $\alpha = 9^\circ$ baseline in their near-field mean flows.  The change in the aerodynamics forces of these three control cases, 9-0B, 9-1B and 9-1C, are listed on the top of figure \ref{fig:A9_Ctrl_Cp_UProf} with the baseline values for quick reference.  Cases 9-0B and 9-1B employ the same actuation frequency ($St^+ = 5.5$) but with different wavenumbers.  While the levels of drag reduction are comparable for these two control cases, the introduction of spanwise-varying actuation in case 9-1B achieves further enhancement in lift compared to case 9-0B.  Cases 9-1B and 9-1C both use $k_z^+L_c = 10\pi$ but different $St^+$.  These two cases achieve comparable levels in lift enhancement and drag reduction across all drag data presented in figure \ref{fig:Ctrl_Drag}.  

For these three control cases, the time- and spanwise-averaged velocity profiles are provided over the airfoil and one chord downstream in the near-wake ($x/L_c = 2$) in figure \ref{fig:A9_Ctrl_Cp_UProf}.  Dashed curves are also shown in figure \ref{fig:A9_Ctrl_Cp_UProf} (a) to mark the contour of $\bar{v}_x = 0$ for the comparison of separation region for the four cases.  While the separation region covers the entire chord in the baseline flow, the periodically excited flow immediately reattaches after separation.  In case 9-0B, the flow over the airfoil is laminarized with formation of compact spanwise vortices.  These vortices merge near the trailing edge and the flow separates again near $x/L_c \approx 0.75$.   The occurrence of the trailing-edge separation can be envisioned from the increasing deficit in the streamwise velocity profiles observed farther upstream.  In figure \ref{fig:A9Ctrl}, we also observe that the spanwise vortices gradually depart from the suction surface as they advect downstream in case 9-0B.  As opposed to case 9-0B, the accelerated transition by spanwise actuation in cases 9-1B and 9-1C provides further 3D mixing and effectively entrains free-stream momentum, resulting in fully attached boundary layer that extends to trailing edge with $\boldsymbol{e}_n \cdot \bnabla \bar{v}_x > 0$.  Similar observations on the modification of velocity profiles have been made by \citet{Amitay:AIAAJ2002} using actuation frequencies of $St^+ \sim \mathcal{O}(10)$ for separation control with synthetic jets.  In cases 9-1B and 9-1C, the effective entrainment due to 3D mixing further enhances the lift performance from that of case 9-0B.  The wake profiles in figure \ref{fig:A9_Ctrl_Cp_UProf} (b) also provide insight on the drag reduction.  All control cases exhibit reduced momentum deficit in the streamwise velocity profiles in their near wakes.  In particular, we observe that the transverse locations where the wake profiles exhibit the maximum deficit move downwards in cases 9-1B and 9-1C. Such a transverse displacement suggests a stronger downwash and is reflecting the enhanced lift for cases 9-1B and 9-1C as well.

In figure \ref{fig:A9_Ctrl_Cp_UProf} (a), case 9-0B exhibits the smallest separated region.  The 2D actuation used in case 9-0B appears to reattach the flow more effectively than 3D actuation.  In spite of the earlier reattachment, 9-0B provides the least suction over this separation region in $0 \lesssim x/L_c \lesssim 2$ compared to cases 9-1B and 9-1C, as shown in figure \ref{fig:A9_Cp_CpRMS} (a).  While all control cases provide higher suction than the baseline flow over this region, the use of 3D actuation further enhances suction compared to 2D actuation.  As the discussed for baseline flows, the laminar-turbulent transition occurs with a plateau in the pressure profile for the controlled cases also.  Such a pressure plateau is clear in cases 9-1B and 9-1C. However, in case 9-0B where only laminar spanwise vortices are presented, the airfoil does not benefit from the additional suction provided by the pressure plateau associated with laminar-turbulent transition.  

The shear-layer roll-up and transition processes can be identified from the the pressure fluctuation profiles in figure \ref{fig:A9_Cp_CpRMS} (b).  These two processes take place with the pressure fluctuation reaching the local maximum near $x/L_c \approx 0.21$ for cases 9-0B and 9-1B under the same actuation frequency of $St^+ = 5.5$.  With the higher $St^+$ in case 9-1C, the local maximum shifts upstream and suggests the accelerated roll-up and transition processes.  Through the discussion on the velocity and pressure profiles, we have noted that both the excited roll-up and laminar-turbulent transition processes are crucial for the suppression of separation.  Both processes can encourage momentum mixing and entrain free-stream momentum to achieve flow reattachment, which provides enhanced aerodynamic performance.

Let us recapitulate our findings on the important flow physics for suppressing flow separation and the connection between those and the results from resolvent analysis.  The mechanism for suppression of separation relies on enhanced momentum mixing.  The mixing entrains free-stream momentum and can be provided by the excited roll-up of the shear layer over the suction surface as well as the laminar-turbulent transition process that follows the roll-up.  As an observation from the study of controlled flows, the shear-layer dominated physics for separation control aligns with the discussions in \citet{Greenblatt:PAS2000}.  Recalling that resolvent analysis also reveals the shear-layer dominated energy amplification,  capitalizing upon the shear-layer instability becomes critical for developing effective and efficient separation control techniques. In what follows, we incorporate the knowledge from LES with resolvent analysis and leverage its insights for the design of active separation control.

\section{Assessment of control effect via resolvent analysis}
\label{sec:Resovent_vs_CtrlLES}

We have performed resolvent analysis to reveal its insights on energy amplification over a range of frequencies and wavenumbers in section \ref{sec:ResolventAnalysis}.  The amplification can be leveraged for flow control, since highly-amplified perturbations may change the mean flow through nonlinear effects.   By comparing the controlled flows to the resolvent response modes, we found that the modal structures provide insights on the global receptivity to a specified perturbation.  We have also learned from controlled flows that momentum mixing over the airfoil plays an important role in suppressing separation in section \ref{sec:Control}.  The enhancement of aerodynamic performance can be quantified by the momentum mixing taking place over the airfoil.  This section
takes the insights from the resolvent analysis and the LES of controlled flows to provide quantitative guidelines for the design of unsteady separation control.

\begin{figure}
\indexsize
\begin{center}
 	\begin{tabular}{	>{\centering\arraybackslash} m{0.5in}  
						>{\centering\arraybackslash} m{0.5in}  
						>{\centering\arraybackslash} m{0.5in}
						>{\centering\arraybackslash} m{0.5in}
						>{\centering\arraybackslash} m{1.8in}}
	$k_z^+L_c$\vspace{-0.04in} & $St^+$\vspace{-0.04in}
			& $\Delta \bar{C}_L$ \vspace{-0.04in} 
			& $\Delta \bar{C}_D$ \vspace{-0.04in} 
			& Resolvent $\hat{R}_z(k_zL_c, St)$\vspace{-0.04in}\\
    \hline
    \vspace{-0.04in}$0$	&
    \vspace{-0.04in}$1$	&
    \vspace{-0.04in}$+34\%$	&
    \vspace{-0.04in}$-10\%$	&
    \vspace{-0.04in}\includegraphics[width=1.8in]{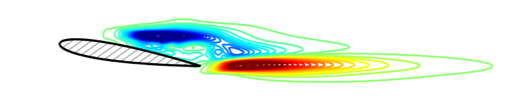}\\
	$0$		&	$3$	&	$+41\%$	&	$-40\%$	&	\includegraphics[width=1.8in]{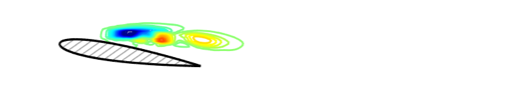}\\
	$10\pi$	&	$4$	&	$+45\%$	&	$-33\%$	&	\includegraphics[width=1.8in]{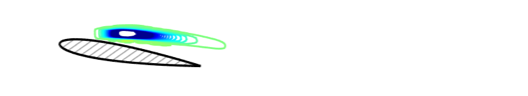}\\
	$10\pi$	&	$8$	&	$+43\%$	&	$-43\%$	&	\begin{overpic}[width=1.8in]{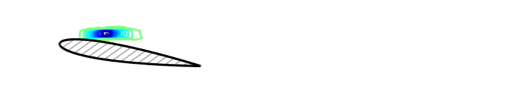}
													\put (80, 0) {\includegraphics[scale=0.55]{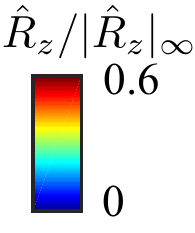}} 
												\end{overpic}\\
	\end{tabular}
	\caption{\label{fig:Ctrl_vs_ReyStress} Comparison of the enhancement in $\bar{C}_L$ and $\bar{C}_D$ and the spanwise Reynolds stress of resolvent response mode for the corresponding $k_zL_c$ and $St$. Note that the resolvent response modes are computed based on mean baseline flow.  The response modes are obtained with $t_\beta v_\infty/L_c = 5$ and the associated $\hat{R}_z$ are visualized by the contour lines of $\hat{R}_z/|\hat{R}_z|_\infty \in \pm [0.01, 0.9]$.}
\end{center}
\end{figure}

While resolvent response mode can capture coherent structures, mixing provided by these coherent structures can be examined through the Reynolds stress associated with the mode \citep{Luhar:JFM2015}.  We have also noted that the location of momentum mixing is crucial to modify the base state and alter the aerodynamic performance.  Over the airfoil, the roll-up and transition processes enhance mixing and suppress flow separation.  On the other hand, momentum mixing induced by large-scale von K\'arm\'an structure in the wake widens the wake and results in increased streamwise momentum deficit and higher drag.  Such mixing is thus unfavorable to aerodynamic stall control.  To address the different effects of these two kinds of mixing, we discuss four representative controlled cases along with the resolvent Reynolds stress obtained from the mean baseline flow for the corresponding $k_zL_c$-$St$ in figure \ref{fig:Ctrl_vs_ReyStress}.  For the case with $(k_z^+L_c, St^+) = (0, 1)$, we observe an extended wake structure in the Reynolds stress with a strong vortex-shedding pattern, causing an unfavorable mixing for drag reduction.  Such mixing in the wake is absent in the other three wavenumber-frequency combinations.  Correspondingly, the use of $(k_z^+L_c, St^+) = (0, 1.5)$ results in less performance enhancement compared to the other three controlled cases, particularly in drag. 
Therefore, for aerodynamically favorable control, we should leverage mixing that takes place over the airfoil by considering the resolvent Reynolds stress as a possible metric for guidance.

\begin{figure}
	\vspace{0.1in}
	\begin{center}
		\begin{overpic}[scale=0.55]{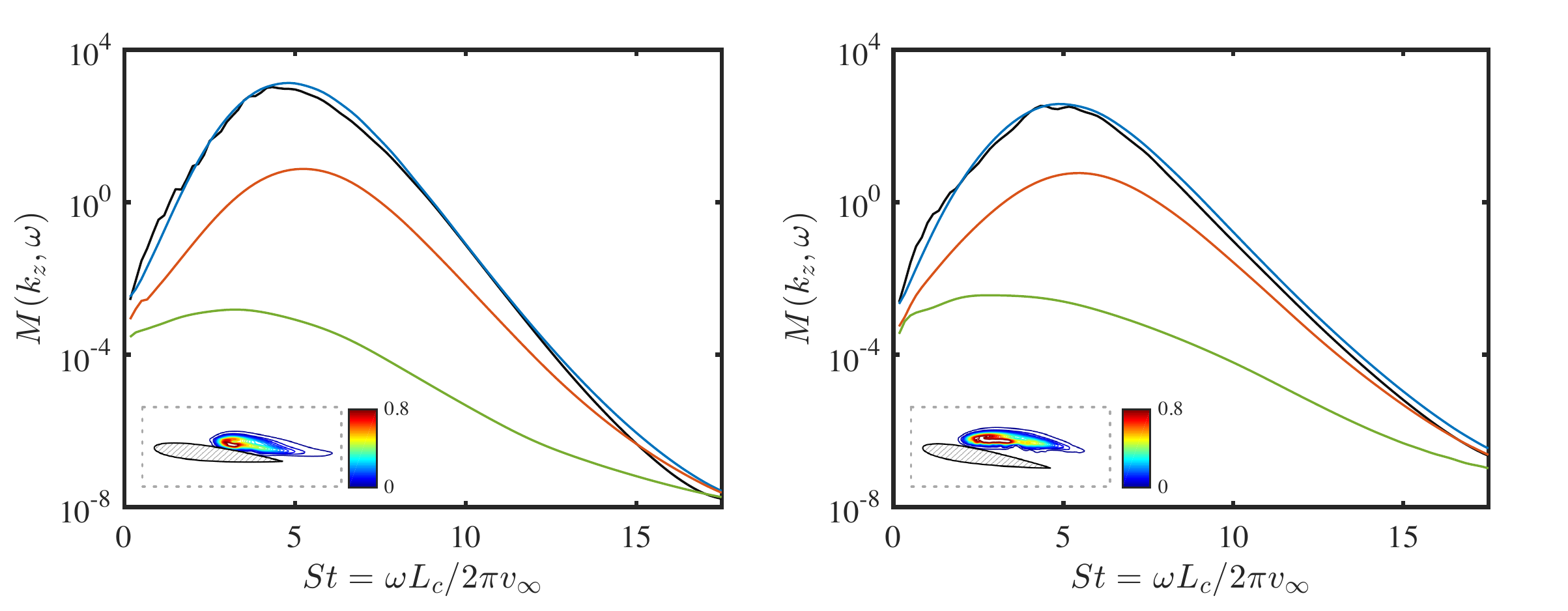}
			\put(0, 35){(a)}
			\put(49, 35){(b)}
			\put(38, 33){\indexsize $\alpha = 6^\circ$}
			\put(87, 33){\indexsize $\alpha = 9^\circ$}
			\put (9, 14) {\scriptsize $w(\boldsymbol{x})/||w(\boldsymbol{x})||_\infty$}
			\put (58, 14) {\scriptsize $w(\boldsymbol{x})/||w(\boldsymbol{x})||_\infty$}
			\put (34, 24){\scriptsize	
							\begin{tabular}{r c} 
								\multicolumn{2}{c}{$k_zL_c$} \vspace{-0.07in}\\ \hline\vspace{-0.15in}\\
								{\color{black}\solid}&{\color{black}$0$} \\
								{\color{blue2}\solid}&{\color{blue2}$10\pi$} \\
								{\color{red2}\solid}&{\color{red2}$20\pi$} \\ 
								{\color{green2}\solid}&{\color{green2}$40\pi$}
							\end{tabular}}
			\put (83, 24){\scriptsize	
							\begin{tabular}{r c} 
								\multicolumn{2}{c}{$k_zL_c$} \vspace{-0.07in}\\ \hline\vspace{-0.15in}\\
								{\color{black}\solid}&{\color{black}$0$} \\
								{\color{blue2}\solid}&{\color{blue2}$10\pi$} \\
								{\color{red2}\solid}&{\color{red2}$20\pi$} \\ 
								{\color{green2}\solid}&{\color{green2}$40\pi$}
							\end{tabular}}
		\end{overpic}
	\end{center}
	\caption{\label{fig:ModalWindow} Spatial integration of the modal Reynolds stress, $M(k_z, \omega)$, for $k_zL_c = 0$, $10\pi$, $20\pi$ and  $40\pi$. (a) $\alpha = 6^\circ$; (b) $\alpha = 9^\circ$. Over the airfoil, the shear-layer window represented by the level-set function in the lower-left corner is used as the weight in the spatial integration performed in equation \ref{eq:MixingFunction}.}
\end{figure}
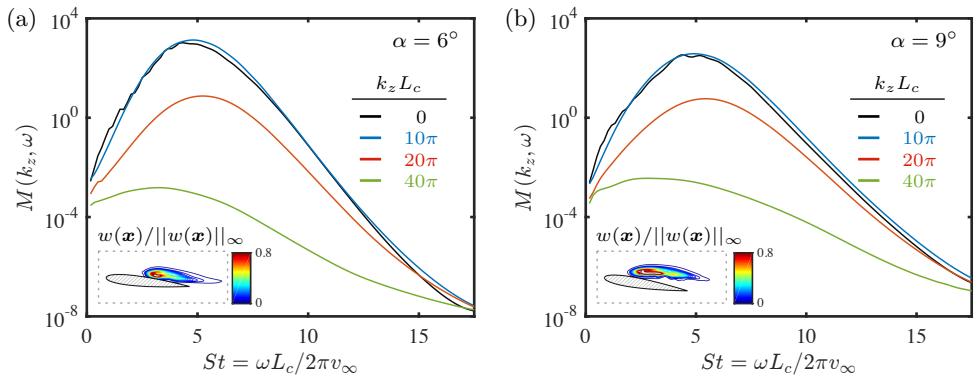

The momentum mixing associated with resolvent response mode can be characterized through performing a spatial integral of the corresponding Reynolds stresses over a region of physical interests \citep{Nakashima:JFM2017}.  Here, we quantitatively assess mixing by introducing a spatial window to perform integration of resolvent Reynolds stress.  We choose a window that covers the shear layer over the airfoil so that only the mixing taking place in this crucial region for suppression of separation is taken into account.  This window $w(\boldsymbol{x})$, shown in figure \ref{fig:ModalWindow}, is designed as a level-set function with $\int_\Omega w(\boldsymbol{x}) {\rm d}\boldsymbol{x} = 1$.   This level-set function is obtained by evaluating $|\hat{v}_x^* \hat{v}_y |$ for the dominant shear-layer eigenmode shown in figure \ref{fig:A93D_BiG}.  In appendix \ref{sec:IntReyWindow}, we also demonstrate that the present assessment is robust with respect to the choice of the window.  The spatial integration for modal Reynolds stress considers $w(\boldsymbol{x})$ as a weighting function and is performed over the entire domain $\Omega$ as
\begin{equation}
	M(k_z, \omega) \equiv \int_\Omega \left[ \sigma^2 (\hat{R}_x^2 + \hat{R}_y^2 + \hat{R}_z^2)^{\frac{1}{2}}\right]_{k_z, \omega} w(\boldsymbol{x}) {\rm d}\boldsymbol{x},
\label{eq:MixingFunction}
\end{equation}
where we also associate the gain $\sigma$ in the integration considering the amplification from a unit energy of forcing.  With this scalar function $M(k_z, \omega)$, the mixing that is favorable for flow control can be evaluated by the integrated Reynolds stresses from the resolvent response mode at $k_z$-$\omega$.   

\begin{figure}
\begin{center}
	\vspace{0.1in}
	\begin{overpic}[width=1.0\textwidth]{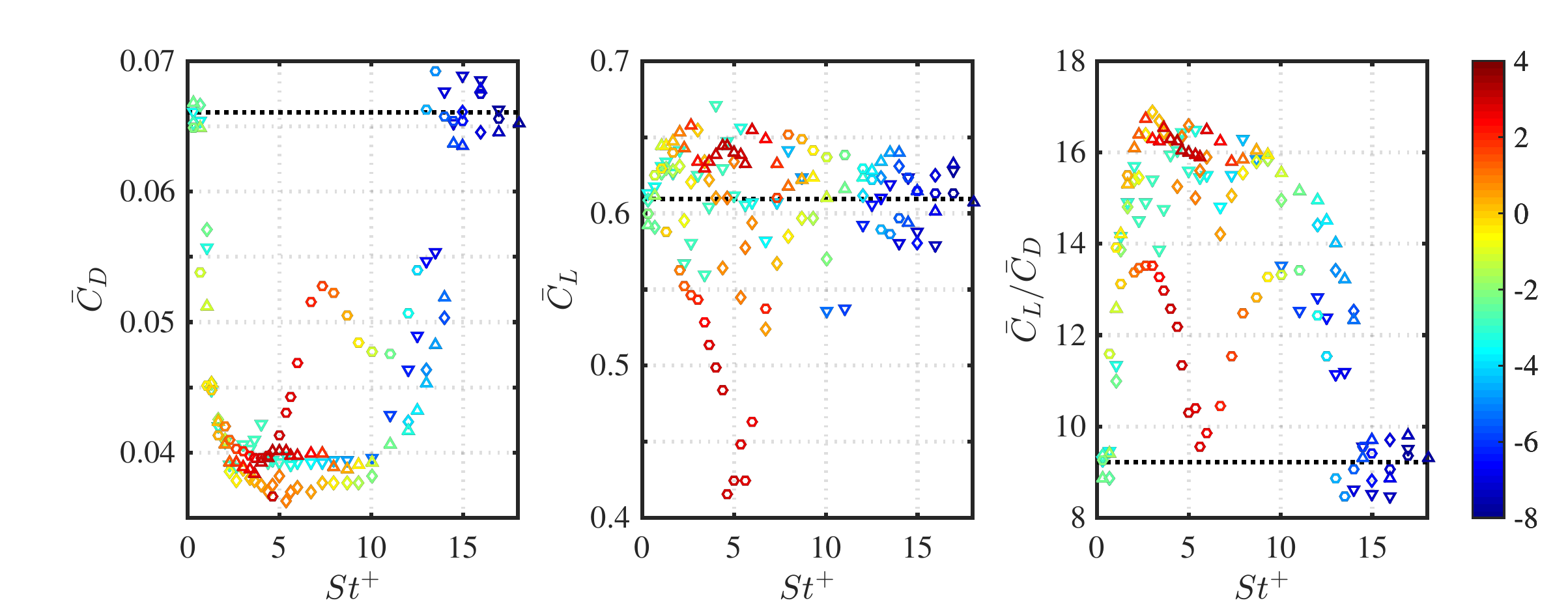}
		\put(1, 17){\rotatebox{90}{$\alpha = 6^\circ$}}
		\put(21.2, 37){(a)}
		\put(50.1, 37){(b)}
		\put(79.3, 37){(c)}
		\put(90.2, 36.8){\indexsize$\log_{10}(M)$}
	\end{overpic}
~\\
	\begin{overpic}[width=1.0\textwidth]{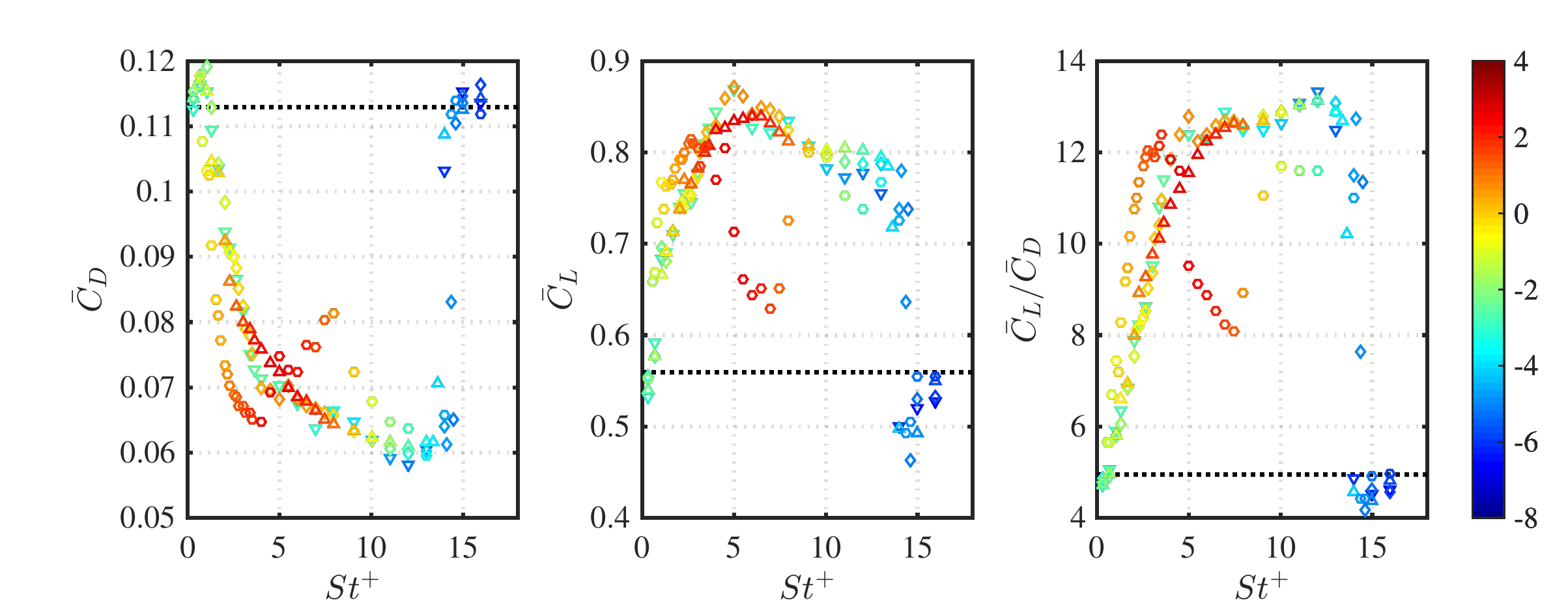}
		\put(1, 17){\rotatebox{90}{$\alpha = 9^\circ$}}
		\put(21.2, 37){(c)}
		\put(50.1, 37){(d)}
		\put(79.3, 37){(e)}
		\put(90.2, 36.8){\indexsize$\log_{10}(M)$}
	\end{overpic}
	\caption{\label{fig:Forces_Rey} Time-average drag, lift, and lift-to-drag ratio colored by the corresponding $M(k_z^+, \omega^+)$ for $\alpha = 6^\circ$ (a-c) and $\alpha = 9^\circ$ (d-f). In each plot, the dashed line corresponds to the baseline level.  Symbols represent different actuation wavenumbers. $\boldsymbol{\circ}$: $k_z^+L_c = 0$; {\scriptsize$\boldsymbol{\triangle}$}: $k_z^+L_c = 10\pi$; $\boldsymbol{\diamond}$: $k_z^+L_c = 20\pi$; {\scriptsize$\boldsymbol{\nabla}$}: $k_z^+L_c = 40\pi$.}
\end{center}
\end{figure}
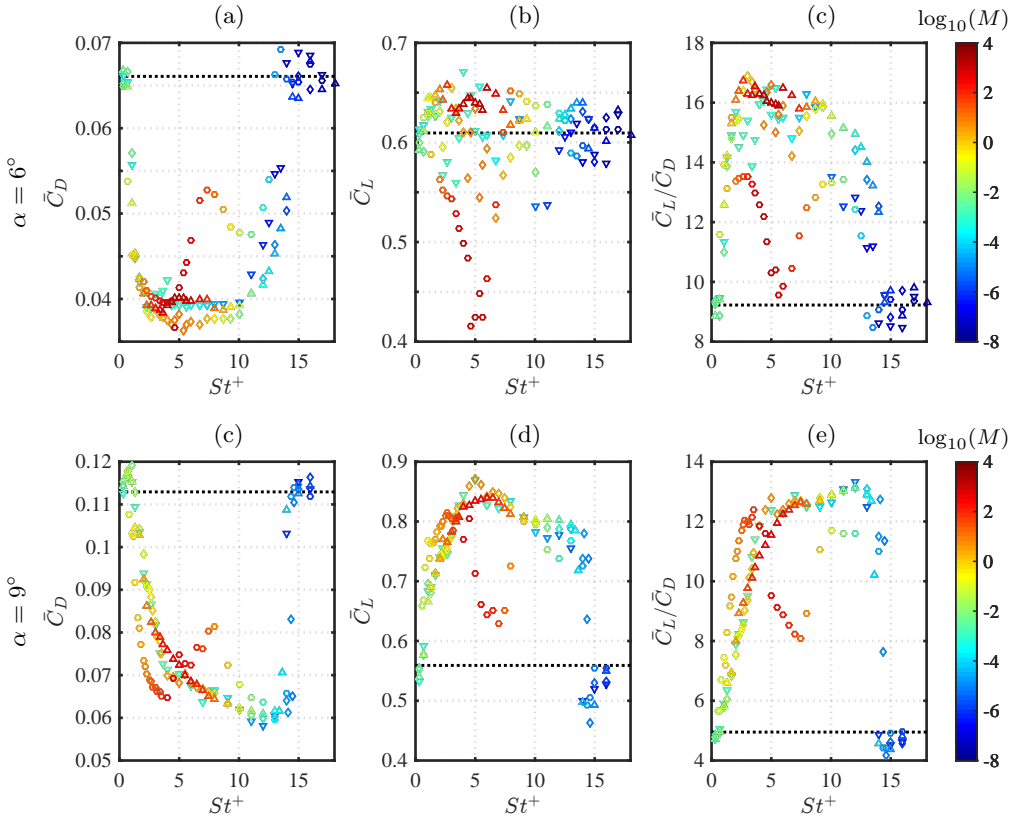

We show the integrated resolvent Reynolds stress $M(k_z, \omega)$ using the shear-layer windows in figure \ref{fig:ModalWindow}.  The trend in $M(k_z, \omega)$ suggests higher mixing is achieved by resolvent response modes over the shear layer near $St \approx 5$ and low $k_zL_c$ for both angles of attacks.  With the mixing quantified by $M(k_z, \omega)$ for resolvent response modes, we color the data points of aerodynamic forces from controlled cases by the corresponding $M(k_z^+, \omega^+)$ for both angles of attack in figures \ref{fig:Forces_Rey}.  In both figures, we show time average drag, lift and lift-to-drag ratio with the level of modal mixing $M(k_z^+, \omega^+)$.  We observe that the drag reduction and lift enhancement achieved by active flow control correlate well with the level of mixing based on $M(k_z^+, \omega^+)$ from resolvent analysis on the mean baseline flow.  Over the actuation frequency range of $3 \lesssim St^+ \lesssim 12$, where most of the effective control cases reside, successful control is characterized by high levels of shear-layer mixing over the airfoil according to resolvent analysis.  Particularly for the lift data of $\alpha = 9^\circ$, the maximum lift agrees well with the high value of $M$.  Similarly for $\alpha = 9^\circ$, the sluggish decrease in drag over $0.3 \lesssim St^+ \lesssim 5$ can also be related to the mixing that takes place in the wake for low frequency modes, as discussed in figure \ref{fig:Ctrl_vs_ReyStress}.  At this stage, we have observed both qualitative and quantitative agreements between resolvent analysis and controlled flows obtained from LES.  The positive correlation between the enhancement of aerodynamic performance and the modal mixing from resolvent analysis suggests its capability of serving as a guiding tool towards selecting effective actuation parameters.

\subsection*{The nonlinear physics beyond resolvent analysis}
With resolvent analysis being a linear technique for the present nonlinear fluid-flow problem, we also observed some limitations of the interpretation in the aerodynamic performance and the prediction of $M(k_z^+, \omega^+)$.  Below, we comment on these limitations and identify the associated nonlinear physics.

For controlled cases with $k_z^+L_c = 0$, drag increases over $4 \lesssim St^+ \lesssim 10$ for both angles of attack.  Such increase in drag is not captured by the value of $M$.  As discussed in the previous section, this drag increase is due to the vortex merging process that causes trailing-edge separation.  Therefore, the difference between the controlled flow results and resolvent analysis can be attributed to the nonlinear nature of the merging process that transfers energy from a fundamental frequency to its subharmonics.   With the energy transfer across frequency space, this nonlinear process not captured by the linear resolvent analysis that deals with a harmonic input-output process.

Another nonlinear process that leads to difference between the LES findings and the results of resolvent analysis is the laminar-turbulent transition following the break-up of spanwise vortices.  In the previous section, the transition process has been shown to be a mechanism responsible for the suppression of separation, in addition to the shear-layer excitation.  Such a mechanism is particularly important for suppressing stall in the control cases with high frequency near $St^+ \approx 10$ and $k_z^+L_c > 0$, leading to the peak drag reduction at $St^+ \approx 12$ for $\alpha = 9^\circ$ and comparable level of force enhancement across three choices of 3D actuation profiles ($k_z^+L_c > 0$).  However, the level of $M(k_z, \omega)$ evaluated from resolvent analysis suggests degraded mixing for $k_zL_c > 0$ and high frequencies.  Therefore, while the aerodynamic forces benefit from the laminar-turbulent transition, this nonlinear process is also beyond the capability of resolvent analysis to predict the force enhancement through transition by using the quantitative level of $M(k_z, \omega)$.

\subsection*{Resolvent analysis as a guiding tool for separation control}

We have demonstrated a design guideline that leverages the knowledge obtained from resolvent analyses performed on mean baseline flows for suppressing stall.  We evaluate the mode-based mixing by combining the knowledge of amplification, modal structure and a shear-layer window over the airfoil, providing a scalar function over the frequency-wavenumber space.  In spite of slight deviations due to the nonlinear physics beyond the present linear modal, the control effect well correlates with the lift enhancement and drag reduction for open-loop controlled flows.  Such a guideline provides quantitative assessment towards selecting actuation frequency and wavenumber for effective unsteady separation control.

\section{Conclusion}
We presented an active flow control effort that capitalizes on large-eddy simulations and resolvent analysis.  This effort considers separated flows over a NACA 0012 airfoil at angles of attack of $\alpha = 6^\circ$ and $9^\circ$ and a chord-based Reynolds number of $23,000$.  The objective of our study was to provide design guidelines for separation control by performing resolvent analysis on the turbulent mean flows.  

The resolvent analysis started by extracting the linear Navier--Stokes operator that governs the perturbations about the statistically stationary turbulent mean flows obtained from the baseline LES.  In the present analysis, the nonlinearity is retained by treating it as an internal forcing in the formulation.  To analyze the unstable linear operators (base states), we considered an extension to the standard approach of resolvent analysis by introducing a temporal filter such that the input-output analysis is performed over a finite-time horizon.  We observed the gain as well as the modal structure physically correlate with the time constant of the temporal filter.  By sweeping through the Fourier space spanned by the frequency and spanwise wavenumber, we observed that the gain distribution scales well with the chord-based Strouhal number between both angles of attack.  This scaling behavior stems from the high nonnormality of the shear-layer modes in the operator spectrum that expands the pseudospectral radius.  Based on these findings, the resolvent analysis revealed a shear-layer dominated mechanism for energy amplification from the input-output process.

The LES of controlled flows were performed with a thermal actuator that introduces time-periodic heat injection with a prescribed spanwise profile.  We swept through different choices of actuation frequency and spanwise wavenumbers to investigate their capability of effects on suppressing stall and enhancing the aerodynamic performance.  In successful controlled cases, the periodic thermal actuation reduces drag by up to $49\%$ and enhance lift by up to $54\%$.  The fluctuation in lift is also reduced by up to $85\%$.  According to the trend of drag reduction over frequencies, we once again observed that the effective frequency for both angles of attack scales well with the chord-based Strouhal number.  Aligning with the literatures are the observations on the shear-layer dominated physics in suppressing separation.  We also examined the control cases in their flow fields and the associated change in the aerodynamic forces.  With the examination,  we concluded that the excitation of shear-layer roll-up and the subsequent laminar-turbulent transition are important mechanisms in enhancing momentum mixing to entrain the free-stream momentum.  Both mechanisms contribute to the enhancement of aerodynamic performances by reducing drag and increasing lift.

The study of controlled flows showed that the mixing over the suction surface plays a key role in suppression of separation.  As such, we evaluated the mixing provided by resolvent response modes obtained from mean baseline flows.  We quantified the modal mixing by integrating the Reynolds stresses associated with the response mode over a shear-layer window.  By comparing the modal mixing to the force data obtained from LES, we observed a good correlation between the higher modal mixing and enhanced control effects for both angles of attack.  Such quantitative agreement assures the utility of resolvent analysis for selecting effective actuation frequencies and wavenumbers, even when the analysis is performed on the mean baseline flow.  Although slight deviations are found in such a correlation, they can be attributed to the nonlinear physics such as vortex merging and laminar-turbulent transition.  These nonlinear processes are beyond the validity of the linear input-output process captured through resolvent analysis.

Through this combined effort, we have demonstrated that resolvent analysis is a valuable tool for providing physics-based guideline for designing separation control.  Such a guideline gives insights on the effective actuation frequencies and wavenumbers for separation control with periodic actuation.  The present analysis was performed on the mean baseline flow to serve as a predictive tool on the choices of actuation frequencies and wavenumbers.  It also provides a quantitative support on the shear-layer dominated physics for separation control.  We believe that this study can provide insights for the use of resolvent analysis in guiding future implementation of active flow control.

~\\
The authors acknowledge the U.S. Office of Naval Research (N00014-16-1-2443, managed by Dr.~Kenneth Iwanski) and Army Research Office (W911NF-14-1-0224, managed by Dr.~Matthew Munson) for supporting this study.  We also thank Prof.~Peter Schmid and Prof.~Mihailo Jovanovi\'c for the insightful discussions on the use of discounted resolvent analysis.  We also acknowledge Dr.~Yiyang Sun for her help with code development and continuous feedback on this study.  The computations were supported by the High Performance Computing Modernization Program at the U.S. Department of Defense and the Research Computing Center at the Florida State University.  We also thank Ms.~Odessa Murray for her help on facilitating the extensive computation in this study.

\appendix
\section{Window of integration on resolvent Reynolds stress}
\label{sec:IntReyWindow}

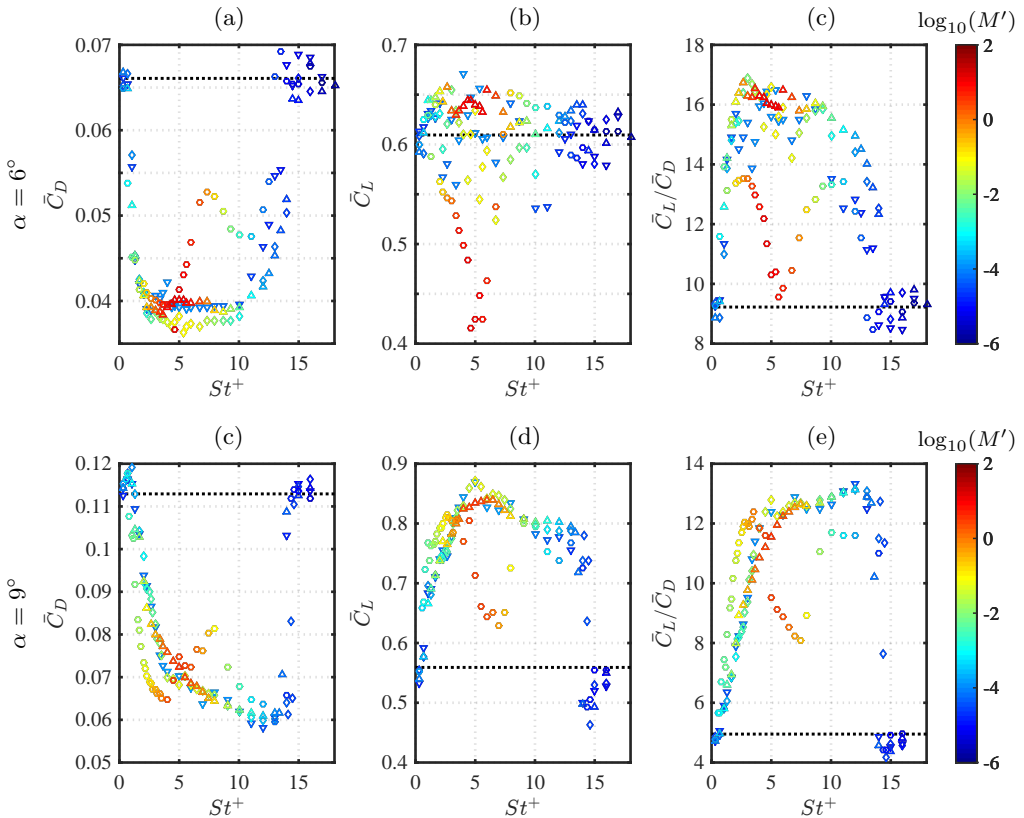
\begin{figure}
\begin{center}
	\vspace{0.1in}
	\begin{overpic}[width=1.0\textwidth]{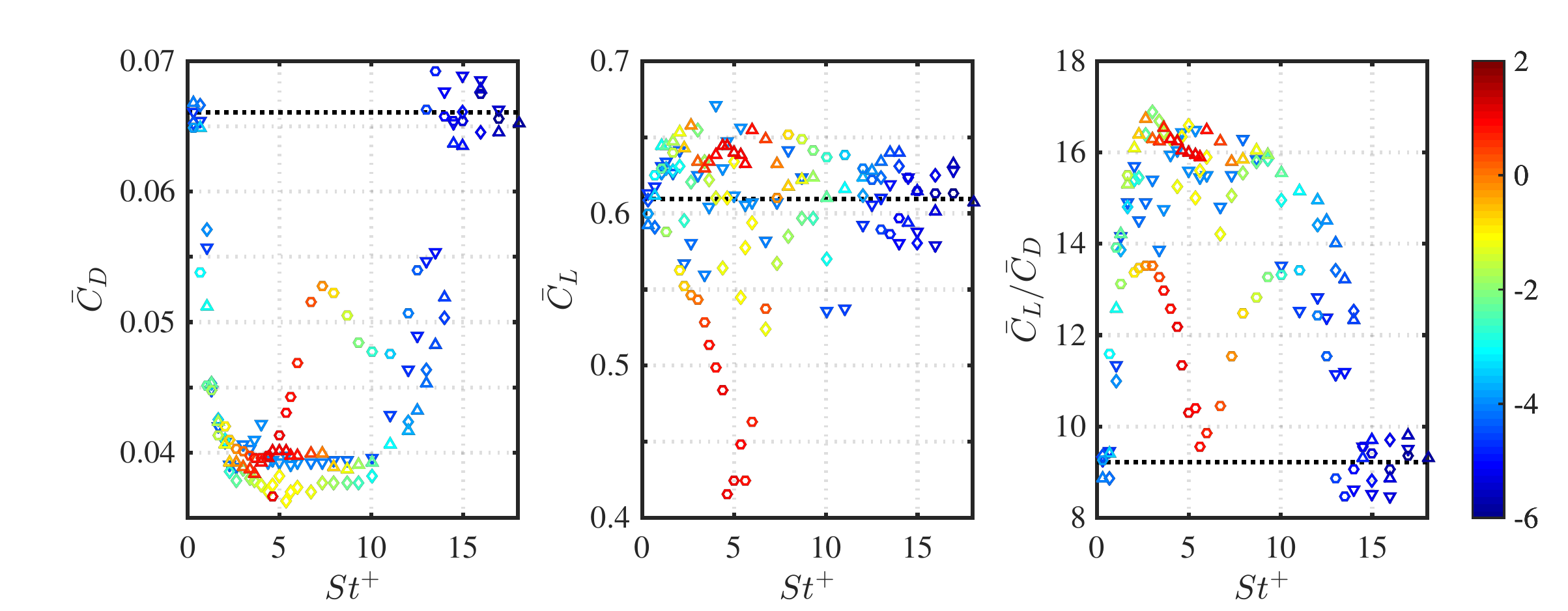}
		\put(1, 17){\rotatebox{90}{$\alpha = 6^\circ$}}
		\put(21.2, 37){(a)}
		\put(50.1, 37){(b)}
		\put(79.3, 37){(c)}
		\put(90.2, 36.8){\indexsize$\log_{10}(M')$}
	\end{overpic}
~\\
	\begin{overpic}[width=1.0\textwidth]{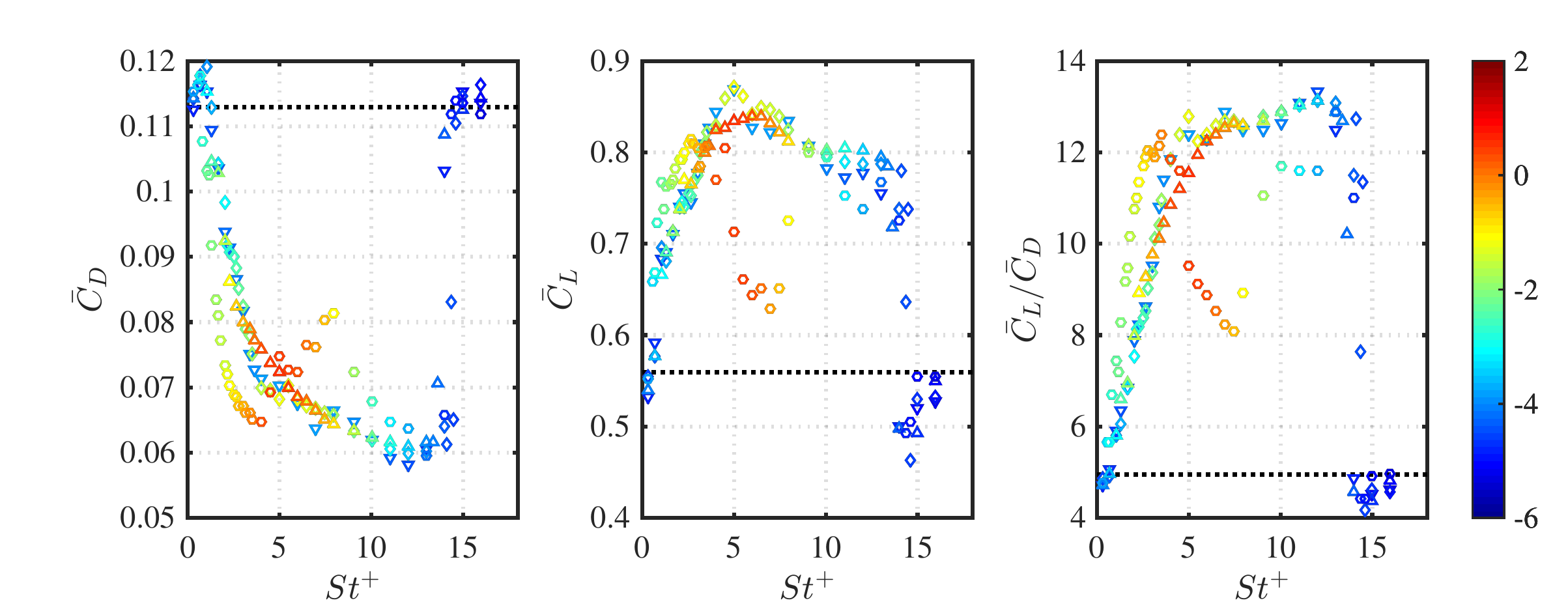}
		\put(1, 17){\rotatebox{90}{$\alpha = 9^\circ$}}
		\put(21.2, 37){(c)}
		\put(50.1, 37){(d)}
		\put(79.3, 37){(e)}
		\put(90.2, 36.8){\indexsize$\log_{10}(M')$}
	\end{overpic}
	\caption{\label{fig:Forces_Rey_WinA} Time-average drag, lift, and lift-to-drag ratio colored by the corresponding $M'(k_z^+, \omega^+)$ for $\alpha = 6^\circ$ (a-c) and $\alpha = 9^\circ$ (d-f). In each plot, the dashed line corresponds to the baseline level. Symbols represent different actuation wavenumbers. $\boldsymbol{\circ}$: $k_z^+L_c = 0$; {\scriptsize$\boldsymbol{\triangle}$}: $k_z^+L_c = 10\pi$; $\boldsymbol{\diamond}$: $k_z^+L_c = 20\pi$; {\scriptsize$\boldsymbol{\nabla}$}: $k_z^+L_c = 40\pi$.}
\end{center}
\end{figure}

The integration of Reynolds stress in equation \ref{eq:MixingFunction} involves a spatial window over which the integration is performed.  Here, we examine another choice for this window and see how it affects the concluding quantitative correlation discussed in figure \ref{fig:Forces_Rey}.  Instead of providing a level-set function according to the dominant shear-layer eigenmode, we integrate the Reynolds stress associated with the response mode over the domain above the airfoil as
\begin{equation}
	M'(k_z, \omega) \equiv \int_{x_{\text{LE}}}^{x_{\text{TE}}} \int_{y_s(x)}^{+\infty} \left[ \sigma^2 (\hat{R}_x^2 + \hat{R}_y^2 + \hat{R}_z^2)^{\frac{1}{2}}\right]_{k_z, \omega} {\rm d}y {\rm d}x,
\label{eq:MixingFunctionWinA}
\end{equation}
where $y_s(x)$ denotes the profile of the suction surface as a function of $x$, $x_{\text{LE}}$ and $x_{\text{TE}}$ respectively denote the streamwise locations of leading and trailing edge.  Using this scalar function $M'(k_z, \omega)$ to quantify modal mixing, we generate similar plots in figure \ref{fig:Forces_Rey_WinA} and compare it to figure \ref{fig:Forces_Rey}.  We observe that the use of the new window in equation \ref{eq:MixingFunctionWinA} provides the same conclusive assessment with the positive correlation between the level of $M'(k_z, \omega)$ and the performance enhancement.  This suggests the developed guideline is robust in the choice of the integration window as long as the window reasonably highlights the shear layer over the suction surface.

\bibliographystyle{jfm}
\bibliography{/Users/DenAprillia/Dropbox/CAYeh_All}

\end{document}